\documentclass[a4paper, 12pt]{article}

\usepackage{graphicx}
\usepackage[utf8]{inputenc}
\usepackage{lineno}
\usepackage{blindtext}
\usepackage{url}
\usepackage[cal=esstix]{mathalfa}
\usepackage{amsmath}
\usepackage{amsthm}
\usepackage{amssymb} 
\usepackage{float, placeins}
\usepackage{caption, subcaption}
\usepackage{dblfloatfix,fixltx2e}
\usepackage[font={it}]{caption}
\usepackage[margin=2.0cm]{geometry}
\usepackage{cancel}
\usepackage{xfrac}
\usepackage{hyperref}
\usepackage{booktabs}
\usepackage[binary-units=true]{siunitx}
\usepackage{accents}
\usepackage{titling}
\usepackage[super]{nth}
\usepackage[toc,page]{appendix}

\newcommand{\whatAmI}{article\xspace}

\newcommand{\vv}[1]{\vec{\mkern0mu#1}}
\newcommand{\kv}[1]{\vec{\mkern5mu#1}}

\newcommand{\pt}{\ensuremath{p_T}\xspace}

\newcommand{\delphes}{\textsc{Delphes}\xspace}

\newcommand{\ttbar}{\ensuremath{t\,\bar{t}}\xspace}
\newcommand{\tautau}{\ensuremath{\tau^+\,\tau^-}\xspace}

\newcommand{\geant}{\textsc{Geant}\xspace}
\newcommand{\xgboost}{\textsc{XGBoost}\xspace}

\newcommand{\mpt}{\ensuremath{\kv{p}_T^{\text{miss}}}\xspace}

\newcommand{\htotautau}{\ensuremath{h \rightarrow {\tautau}}\xspace}

\newcommand{\lumin}{\textsc{Lumin}\xspace}
\newcommand{\pytorch}{\textsc{PyTorch}\xspace}

\newcommand{\sfSmall}{0.49}
\newcommand{\sfWide}{1.0}
\newcommand{\sfMid}{0.8}

\usepackage{algorithm}
\usepackage{algpseudocode}
\usepackage{bm} 
\usepackage{arydshln}
\usepackage{tikz}
\usepackage{tikz-3dplot}
\usepackage{datetime2}

\date{\today}
\title{On the impact of selected modern deep-learning techniques to the performance and celerity of classification models in an experimental high-energy physics use case}
\author{Giles Chatham Strong\\\small{\textit{LIP-Lisbon} and \textit{Università degli Studi di Padova}}\\\small{\texttt{giles.chatham.strong@cern.ch}}}

\begin{document}

\maketitle

\abstract{
Beginning from a basic neural-network architecture, we test the potential benefits offered by a range of advanced techniques for machine learning, in particular deep learning, in the context of a typical classification problem encountered in the domain of high-energy physics, using a well-studied dataset: the 2014 Higgs ML Kaggle dataset. The advantages are evaluated in terms of both performance metrics and the time required to train and apply the resulting models. Techniques examined include domain-specific data-augmentation, learning rate and momentum scheduling, (advanced) ensembling in both model-space and weight-space, and alternative architectures and connection methods.

Following the investigation, we arrive at a model which achieves equal performance to the winning solution of the original Kaggle challenge, whilst being significantly quicker to train and apply, and being suitable for use with both GPU and CPU hardware setups. These reductions in timing and hardware requirements potentially allow the use of more powerful algorithms in HEP analyses, where models must be retrained frequently, sometimes at short notice, by small groups of researchers with limited hardware resources. Additionally, a new wrapper library for \pytorch called \lumin is presented, which incorporates all of the techniques studied.
}

\clearpage

\begin{center}
	\hrule
\end{center}

\tableofcontents

\begin{center}
	\hrule
\end{center}

\clearpage


\FloatBarrier
	\section{Introduction}\label{sec:intro}
	The rise of machine-learning (ML) applications has had a remarkable impact on many areas of industry and in a few academic disciplines. The field of experimental high-energy physics (HEP), however, was comparatively slower to adopt these new approaches, with only the occasional and isolated use of basic \textit{multivariate analyses} (MVAs), such as $b$-jet tagging at LEP e.g.~\cite{lep_tagging} in 1995, and D{\o}'s observation of single-top-quark production~\cite{dzero_singletop} in 2001; in the latter, the authors explicitly noted the importance of their MVA (a neural network) in fully utilising all available data to make the observation. Indeed, the importance and potential benefits of these methods was well recognised within the HEP community, as indicated by Ref.~\cite{runII_tevatron_mva} (2003), which notes "The best use of data is ensured only with a multivariate treatment" when discussing approaches for data-analysis by the CDF~\cite{cdf} and D{\o}~\cite{d0} collaborations during Run-II of the Tevatron (2001-2011).
	
	This process of adoption continued at the Large Hadron Collider (LHC) with analyses such as Ref.~\cite{first_cms_bdt} (2011), which performed two complimentary measurements: one using physics-derived variables, and the other using an MVA (a boosted decision tree (BDT)). ML continued to became more and more integrated into the general HEP framework, with multiple MVA tools being used for different tasks within analyses. A notable example of this came in 2012 with the concerted use of no less than four MVAs (BDTs) in a single analysis: the search for Higgs boson decays to pairs of photons performed by the CMS collaboration at the LHC \cite{cms_diphoton} - which contributed significantly to the discovery of the Higgs boson by ATLAS~\cite{atlas_higgs} and CMS~\cite{cms_higgs}. This process of MVA integration accompanied a shift in the community's trust of approaches which relied less on expert knowledge.
	
	This degree of trust had yet to extend fully to neural networks, and although they had been used to some extent at both LEP and the Tevatron, they were still perceived as ``black-box" methods; whilst researchers may be able to achieve better results by using them, they might be unable to explain how the results were achieved to sufficiently satisfy internal collaboration reviewers. In more recent years, the advancement of deep-learning approaches has led to improvements in model interpretability and greater knowledge of neural networks in the scientific community; both of which have helped foster a climate more amenable to their use \cite{hepml_cwp, dl_for_hep}.

	Whilst the improvements in performance that more advanced algorithms can bring is often the primary driving force in optimisation, it is also necessary to appreciate the run-time environment and time-requirements of the tasks they intend to solve. Algorithms used during data-collection at particle colliders must be extremely quick in order to not cause delays, which could lead to data not being recorded. Algorithms run after data are recorded have slightly relaxed time-requirements, however given the quantity of data that must be processed, which will increase considerably at the High Luminosity LHC, these algorithms must be sufficiently quick so as not to cause delays in the analyses. These two types of algorithms normally require infrequent training, i.e. they are trained once and used for a long period of data taking and processing by a large number of researchers. Therefore the training time is not so much an issue as their application time. Algorithms used at analysis-level, however, are normally developed by a small group of researchers, and tailor-made for a specific analysis. As the analysis evolves, these algorithms will need to be adapted and re-optimised several times, occasionally at short notice. It is therefore necessary that they are both quick to train and reasonably quick to apply. It is also important to consider the hardware available on demand to analysis-level researchers; while some institutes or individual researchers may have their own private GPUs, this cannot be assumed and algorithms must work sufficiently well on CPU as well.
	
	Nowadays, deep neural networks (DNNs) are one of the main drivers of solutions to the domain-specific problems in HEP. The benefit of these approaches is that although we have theories and models to describe particle collisions, these do not automatically translate into optimal treatments of the data that are recorded. Previously researchers would have to handcraft solutions to reconstruct or represent the data in a more useful way. Deep learning instead allows the vast amount of data (both recorded and simulated) available to be used not only to reduce the statistical uncertainties on measurements, but to learn also the optimal treatment of the data starting from comparatively low-level representations. Such tasks faced in HEP are: object reconstruction\cite{end2endDL, qcd_rnn}, collision and detector simulation \cite{CaloGAN, fast_gans_hep, hep_gans}, object identification \cite{deepcsv, dnn_toptagging, jedi_net}, event triggering \cite{lhcb_trigger}, likelihood-free inference \cite{mining_for_gold, inferno, cross_entropy_imporvement}, parameterised learning for multiple search hypotheses \cite{parameterised_nn}, unsupervised searches for new physics (anomaly detection) \cite{unsupervised_searches}, and the optimal treatment of systematic uncertainties \cite{learning_to_pivot, ml_systs}. Although these tasks are mostly peculiar to particle physics, their solutions normally rely on applying and adapting techniques developed outside of HEP.
	
	Those techniques are continually being refreshed and updated, and are normally presented on benchmark
	datasets for some specific task, such as image recognition on ImageNet
	\cite{ImageNet}. However, it is not always obvious whether improvements on such
	datasets would be reflected by similar success when applied in other domains. We
	consider this a strong motivation to study how large an improvement these new techniques can bring. To do so we will use a very well-studied, HEP-specific benchmark dataset: that of the Higgs ML Kaggle challenge \cite{higgsml_description}, in which participants were tasked with developing classifiers to help search for the Higgs boson amongst a set of common background processes.
	
	In this study we examine the benefits and cross-domain applicability of several recent advances in deep-learning, including: a method to quickly optimise the learning rate; newer activation functions; learning rate scheduling; data augmentation; alternative ensembling techniques. Following these experiments we present a solution which is able to achieve consistently the performance of the winning solution, whilst being significantly quicker to train and apply, and being suitable for use with both GPU and CPU hardware setups. These reductions in timing and hardware requirements potentially allow the use of more powerful algorithms in HEP analyses, where models must be retrained frequently, sometimes at short notice, by small groups of researchers with limited hardware resources.
	
	We begin with a more detailed description of both the general problem (\autoref{sec:ml_in_hep}) and the specific challenge (\autoref{sec:higgsml}). In \autoref{sec:solution_development} we describe the baseline model and the various improvements, reporting performances in terms of a range of optimisation metrics. In \autoref{sec:testing} we report the final performance on the testing data, expanding to a fuller comparison with the winning solutions. Section~\ref{sec:ablation} studies the individual benefits of each method on the final model. The investigation and the main results are summarised in \autoref{sec:conclusion}.
	
    The study presented here was preceded by an initial investigation documented in Ref.~\cite{amva_d1.4}. The study here uses a more robust approach of multiple trainings to compute average performance changes, and also explores some new methods. The framework, notebook experiments, and raw paper results are made available at Ref.~\cite{higgsml_lumin_git}.

\FloatBarrier
	\section{The general search-strategy at the LHC}\label{sec:ml_in_hep}
    At particle colliders such as the LHC \cite{lhc}, sub-atomic particles or ions are accelerated to high energies and made to collide with one another. The resulting collisions are attributable to fundamental-physics interactions between the particles, and their study can be used to compare theoretical models to reality in order to better understand the nature of the universe. The energy of these collisions gives rise to particles which are recorded with specialised instruments called \textit{particle detectors}. Since there are many possible processes which could have given rise to a particular final state, and because the detectors only capture information of the products of the collision, there is no way to tell exactly how the final state was created. However, each of the contributing processes is likely to have some specific signature, which is reflected in its outgoing products, e.g. a process which gives rise to an unstable particle will have decay products which can be used to reconstruct its mass, but due to experimental limits, such as the finite resolution of the detector, other processes can still give rise to collisions which reproduce the signature of desired process. Effectively, the data are unlabelled with respect to the classes of process that gave rise to them.
            
    When looking to test some theory, such as the presence of a new \textit{signal} process \footnote{Signal refers to a process or particle which is being searched for. An example of this is the presence of the Higgs boson as in Ref.~\cite{cms_diphoton} and the HiggsML Kaggle challenge studied here, where the Higgs boson was treated as a new particle and other processes that had already been discovered formed a background (noise) to its discovery.}, one normally defines some signal-enriched region, in which the process being searched for is expected to contribute appreciably. One can then examine the rate at which events populate this region and compare it to the rate expected if the new process were not occurring and the region were only being populated by known processes (background). This is done via a test of the signal+background hypothesis against the null hypothesis of background only, e.g. using the $CL_s$ method~\cite{cls0,cls1}. The catch is that, due to the data being unlabelled, the background and signal rates are unknown, and only their sum is known. In some situations the signal signature (e.g. a particle mass with some determined resolution) is well defined, and the background rate can be extrapolated from, or interpolated between, signal-depleted regions, but this is not always the case.

    Instead, what can be done is to simulate particle collisions and the experimental equipment using a combination of analytical and stochastic modelling (see e.g. Refs.~\cite{mc_gen, geant0, delphes0}), a process referred to as Monte Carlo generation and detector simulation. Since one is able to generate collisions for specified processes, one now has a labelled approximation of the collider data and can estimate the signal and background rates in the signal-enriched region, applying correction factors if need be to improve the modelling and marginalising over any remaining unknown (nuisance) parameters in the hypothesis test.

    The exact specification of the signal-enriched region is an important task, which can have a strong influence on the outcome of a search; if the region is too wide the included signal becomes washed out by background, while if it is too narrow it may end up not being sufficiently populated. The main problem is the high dimensionality of the data; due to the high energy of the collisions, each collision can result in hundreds to thousands of particles, and each particle can produce multiple `hits' and energy deposits in the detector. Traditionally, the first step is to reduce the dimensionality by \textit{reconstructing} the hits and deposits back into known physics objects (fundamental particles, jets, missing energy, et cetera). The next stage is to select objects from the data which correspond to the expected final states of the signal process (and cut away events which fail to produce such objects). The signal region can then be defined using some theoretically or phenomenologically motivated function of the properties of these final states and other objects in the event.

    As mentioned in \autoref{sec:intro}, machine learning techniques are becoming more and more used in high energy physics analyses, due to the ease with which they can discover high-dimensional patterns in data and use them to learn to solve some problem, such as learning to separate particle collisions by class (e.g. signal or background). This ability, however, has limits and it is currently beyond our computational capacity to run such event-level classification algorithms directly on the detector recordings. The contemporary compromise is to still perform the particle reconstruction and event selection, but then to use features based on the properties of the selected particles and the event as a whole as inputs to a machine-learning based classifier and use its response to help define the signal-enriched region. The development, training, and inference of such algorithms is still a difficult task and a source of experimentation in its own right. 

\FloatBarrier
	\section{Higgs ML challenge}\label{sec:higgsml}
	In 2014 the Higgs ML challenge was launched on Kaggle (\url{https://www.kaggle.com/c/higgs-boson}). The challenge was designed to help stimulate outside interest and expertise in solving high-energy physics problems. Participants were tasked with working on a specific kind of problem one often faces in HEP: the search for a rare signal by classification of data based on its features. This signal process was the production of a Higgs boson decaying to a pair of tau leptons, against a background comprised of several more common processes.
	
	The competition was highly successful with \num{1785} teams competing \footnote{Typical Kaggle competitions can attract anywhere between \num{50} and \num{9000} teams. \num{2000} teams puts this competition in approximately the top \SI{20}{\%} of the most popular competitions. What is worth noting, though, is that this competition required participants to become familiar with a complex field, which could have potentially discouraged people from joining, and yet it still managed to attract so many participants.}, and helped to introduce many new methods to HEP, as well as produce more widely used tools, such as \xgboost \cite{XGBoost}. Given the level of expertise and effort that went into the solutions, the challenge forms a viable method to benchmark models and quantitatively measure the impact of new methods and approaches. The challenge is presented in Ref.~\cite{higgsml_description}, and the results of the solutions are discussed in Ref.~\cite{higgsml_discussion}.

	\subsection{Challenge description}\label{sec:challenge}
		\subsubsection{Overview}\label{sec:overview}
			The data used in the challenge consist of simulated particle collisions, generated to mimic those expected at the LHC during its 2012 running. These are fed through a detailed \geant~4 \cite{geant0, geant1}-based simulation of the ATLAS detector \cite{ATLASExperiment}, and finally through the ATLAS reconstruction algorithms, resulting in a set of physics objects per simulated collision. These \textit{events} are then filtered to select those compatible with containing the semi-leptonic decay of a pair of tau leptons, i.e. events which contain a hadronically decaying tau lepton and either an electron or a muon.
			
			The properties of the reconstructed physics objects are then recorded in columnar-data format, with each row corresponding to a separate collision event, along with the process label and a weight. The event labels are: ``0" for background and ``1" for signal. The weight of an event is the product of the production cross-section of the particular physical process that gave rise to the collision (effectively how likely the process was to occur), the probability of the event then meeting the requirements to be included in the final dataset (in this case containing a hadronically decaying tau and either as electron or muon), and the amount of data collected for which the sample was simulated. It is used to normalise the contributions of the events. Both the labels and the weights are only known due to the events being simulated. 
			
			The top solutions to the challenge made heavy use of ensembling techniques, combining tens to hundreds of models. The bases of the models were mostly either decision trees/forests, or neural networks. A lot of work seemed to go into feature engineering and selection, with a new fit-based high-level feature (CAKE\footnote{CAKE discussion: \url{https://www.kaggle.com/c/higgs-boson/discussion/10329}\\CAKE code: \url{https://bitbucket.org/tpgillam/lesterhome/src/master/}}) being developed towards the end of the competition. This appeared to improve the results for tree-based models, but gave no significant improvements to DNN-based models, indicating that sufficiently well trained networks were able to learn their own versions of it. The other focus appeared to be optimising the way in which models were ensembled, e.g regularised greedy forests \cite{rgf}.
		
		\subsubsection{Scoring metric}\label{sec:scoring}
			The performance of a solution is measured using the Approximate Median Significance \cite{ams}, as computed in \autoref{eq:ams}:
			\begin{equation}
				\mathrm{AMS}=\sqrt{2\left(\left(s+b+b_\text{reg.}\right)\ln\left(1+\frac{s}{b+b_\text{reg.}}\right)-s\right)},
				\label{eq:ams}
			\end{equation}
			in which $s$ is the sum of weights of true positive events (signal events determined by the solution to be signal), $b$ is the sum of weights of false positive events (background events determined by the solution to be signal), and $b_\text{reg.}$ is a regularisation term that was set to 10 by the competition organisers. This term effectively ensures that a minimum number of background events are always present in the signal region, which reduces the variance of the AMS (see Sec.~4.3.2 of Ref.~\cite{higgsml_discussion}). The AMS provides a quick, analytical value which is an approximation of the expected significance one would obtain after a full hypothesis test of signal+background versus background only. The significance is a $Z$ score corresponding to the minimum number of standard deviations, $\sigma$, a Gaussian distributed variable would have to fluctuate in order to produce the observed data, see e.g. Ref.~\cite{asymptotic_limit} for a compact introduction to hypothesis testing in new-physics searches. In High Energy Physics, an observed significance of at least five sigma is commonly required to claim observation of a new particle or process; for a discussion on the history and appropriateness of this convention one may read Ref.~\cite{extraordinary_claims}.
			
			The common practice for these kinds of problems is to cut events from the data in order to remove preferentially the background events whilst retaining the signal events, in order to improve the AMS of the remaining data. Either a single cut can be used on some highly discriminating feature of the data, or multiple cuts can be used over several features. The common ML approach is to adapt the former procedure by using the features of the data to learn a new single highly discriminating feature, place a threshold on it, and then only accept events which pass the threshold. The feature learnt in this approach is simply the predicted class distribution of events, in which background events will be clustered towards zero, and signal events towards one. The AMS can then be optimised by only accepting events with a class prediction greater than some value.
			
			In a full HEP analysis, the sensitivity is usually improved further by fitting histograms of the data along one or two discriminating features, and computing the significance using multiple bins of the data. This can either involve cutting on the ML prediction and then fitting to a different feature, or not cutting on the prediction and instead directly fitting to the histogram of the prediction. Additionally, the data may be split into analysis-specific sub-categories to further increase sensitivity, due to varying background contributions in each category. Examples of such sub-categories for this search might be: channel-wise ``tau+electron'' and ``tau+muon''; and jet-wise ``zero jets'', ``one jet'', ``two or more jets''. However, for simplicity the challenge here requires only a single cut on the ML prediction.
			
			The threshold cut can easily be optimised by scanning across the range of possible cuts (either at fixed steps, or checking at each data point) and picking the value which maximises the AMS. This is likely, however, to be optimal only for the dataset on which it is optimised, and performance can be expected to drop when applying the cut to unseen data. This is reflected in the challenge by requiring the solutions to predict on test data for which the class labels are not provided. It is important then that one is more careful when choosing a cut, in order to generalise better to unseen data. The approach adopted here is to consider the top \SI{10}{\%} of events as ranked by their corresponding AMS, and compute the arithmetic mean of their class predictions and use this as the cut. This reduces the influence of statistical fluctuations in the AMS, resulting in a more generalising cut.
			
	\subsection{Current state of the art}\label{sec:higgsml_sota}
		It is our understanding that the winning solution to the Higgs ML Challenge, with an AMS of \num{3.80581}, has yet to be outperformed under challenge conditions and so will be used as a comparison to the solution developed in this \whatAmI. Whilst Ref.~\cite{super_tml} claims to achieve an AMS of 3.979 using a method of converting the tabular data to images, we find that this result is likely a mistake in the authors' handling of the testing data which led them to compute the AMS on double the integrated luminosity, thereby increasing their score by a factor of approximately $\sqrt{2}$. We assert that their method actually achieves an AMS of around 2.8 and document our study in App.~\ref{sec:super_tml}.

	\subsection{ATLAS geometry and coordinate system}
		The ATLAS detector \cite{ATLASExperiment} is cylindrical in shape, displaying rotational and forwards-backwards symmetry in the transverse and longitudinal planes, respectively, with respect to the beam-axis. It is worth noting that such a configuration is common amongst other particle detectors which are also designed to be suitable for searches for, and studies of, the Higgs boson, such as the CMS detector.
		
		Two coordinate systems are normally used in particle physics: the Cartesian system of $\left(x,y,z\right)$, which can be used to express geometric position and components of particle momenta; and cylindrical coordinates $\left(r,\eta,\phi\right)$, again used for geometric position and components of particle momenta (in which case radius $r$ is the particles' transverse momenta $\pt$).

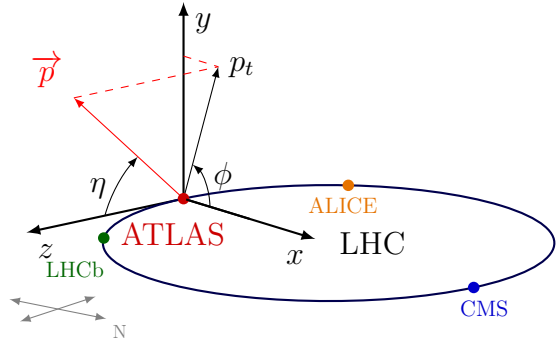
\begin{figure}[ht]
			\begin{center}

\tikzset{>=latex} 

\tdplotsetmaincoords{75}{50} 
    \begin{tikzpicture}[scale=2.7,tdplot_main_coords,rotate around x=90]
    
    \def\rvec{1.2}
    \def\thetavec{40}
    \def\phivec{70}
    \def\R{1.1}
    \def\w{0.3}
    
    \coordinate (O) at (0,0,0);
    \draw[thick,->] (0,0,0) -- (1,0,0) node[below left]{$x$};
    \draw[thick,->] (0,0,0) -- (0,1,0) node[below right]{$y$};
    \draw[thick,->] (0,0,0) -- (0,0,1) node[below right]{$z$};
    \tdplotsetcoord{P}{\rvec}{\thetavec}{\phivec}
    
    \draw[->,red] (O) -- (P) node[above left] {$\overrightarrow{p}$};
    \draw[->] (O)  -- (Pxy) node[right] {$p_t$};
    \draw[dashed,red] (P)  -- (Pxy);
    \draw[dashed,red] (Py) -- (Pxy);
    
    \tdplotdrawarc[thick,rotate around x=90,black!70!blue]{(\R,0,0)}{\R}{0}{360}{}{}
    
    \begin{scope}[shift={(1.1*\R,0,1.65*\R)},rotate around y=12]
        \draw[<->,black!50] (-\w,0,0) -- (\w,0,0);
        \draw[<->,black!50] (0,0,-\w) -- (0,0,\w);
        \node[below right,black!50,scale=0.6] at (\w,0,0) {N};
    \end{scope}
    
    \node[right] at (\R,0,0) {LHC};
    \fill[radius=0.8pt,black!20!red]
        (O) circle node[left=4pt,below=5pt] {ATLAS};
    \draw[thick] (0.02,0,0) -- (0.5,0,0); 
    \fill[radius=0.8pt,black!20!blue]
        (2*\R,0,0) circle
        node[right=4pt,below=2pt,scale=0.7] {CMS};
    \fill[radius=0.8pt,black!10!orange]
        ({-\R*sqrt(2)/2+\R},0,{-\R*sqrt(2)/2}) circle 
        node[left=2pt,below=2pt,scale=0.7] {ALICE};
    \fill[radius=0.8pt,black!60!green]
        ({-\R*sqrt(2)/2+\R},0,{\R*sqrt(2)/2}) circle 
        node[below=6pt,right=3pt,scale=0.7,anchor=north east] {LHCb};
    
    \tdplotdrawarc[->]{(O)}{0.2}{0}{\phivec}
        {above=2pt,right=-1pt,anchor=mid west}{$\phi$}
    \tdplotdrawarc[->,rotate around z=\phivec-90,rotate around y=-90]{(0,0,0)}{0.5}{0}{\thetavec}
        {anchor=mid east}{$\eta$}
\end{tikzpicture}
				\caption{Illustration of the coordinate systems used at the ATLAS experiment in the geographical context of the LHC. Image notice \cite{atlas_coord_system}.}
				\label{fig:coord_system}
			\end{center}
		\end{figure}

		At ATLAS the Cartesian system is defined as shown in \autoref{fig:coord_system} such that the origin is the nominal collision-point of the experiment, the $z$-axis points along the beam-axis, the $x$ axis points towards the centre of the LHC ring, and the $y$ axis points perpendicularly upwards.

		The cylindrical coordinates for momenta are defined in terms of transverse momentum $\pt$, azimuthal angle $\phi$, and pseudorapidity $\eta$, where for a particle with momentum $p$:
		\begin{equation}
			\eta=\tanh^{-1}\left(\frac{p_z}{\left|\bar{p}\right|} \right).
		\end{equation}

		In this \whatAmI the $\Delta R$ separation between two objects refers to the Euclidean distance between two points in $(\phi,\eta)$ space:
		\begin{equation}
			\Delta R=\sqrt{\Delta\phi^2+\Delta\eta^2},
		\end{equation}
		where $\Delta\phi$ and $\Delta\eta$ are the angular separations between the points in terms of azimuthal angle and pseudorapidity, remembering that $\phi$ is cyclical within $[-\pi,\pi)$.
	
	\subsection{Data and processing}\label{sec:data_processing}
		Although the full dataset created for the challenge has now been made public \cite{higgsml_data}, in order to provide an accurate comparison to the previous scores only the subset that was used during the challenge is used here, both for training and testing. The training and testing sets consist of \num{250000} and \num{550000} events, respectively. Both contain a mixture of both classes: signal (\htotautau); and background (\ttbar, $Z\rightarrow\tautau$, and $W$ boson decay). The full dataset is only slightly larger, containing an extra \num{18238} events (an increase of \SI{2.3}{\%}), and so it is unlikely to offer much in the way of benefits to training or evaluation. Use of the full dataset was therefore not thought to be worth the cost of being unable to compare to the baseline results on a like-for-like basis.
		
		Each event is characterised by two sets of training features: primary - lower level information such as tau \pt; and derived - higher-level information calculated via (non)-linear combinations of the low-level features or from hypothesis-based fitting procedures. Tables~\ref{tab:der_features} and \ref{tab:pri_features} list and describe the features, and further details are available in Ref.~\cite{higgsml_description}. The coordinate system of the data is originally provided in terms of $\left(\pt, \eta, \phi\right)$, however initial tests and past experience dictates that NN-based models perform better when vectors are mapped into a Cartesian coordinate system, due to the cyclical nature of the azimuthal angle, and the non-linear nature of the pseudorapidity, $\eta$.
		
		The dataset also contains information about the hardest two jets in each event. For cases in which there were not enough jets reconstructed, the features are set to default values of \num{-999}. In order to prevent these values from having an adverse effect on the normalisation and standardisation transformations that will later be applied, these values were replaced with \textit{NaN} (Not a Number - a special value that is ignored by the pre-processing methods), which prevents them influencing the transformations during fitting, and being altered by the transformations during application. Additionally, one of the features, \texttt{DER\_mass\_MMC}, which is the mass of the Higgs boson estimated by a hypothesis based fitting technique, sometimes does not converge, in which case the value is set to \textit{NaN}.
		
		Since scoring the testing data requires optimising a threshold on some data, and also to provide some way to compare architectures without relying on too much on the public score of the testing data, an $80:20$ random, stratified split \footnote{Stratified random splitting involves randomly splitting a dataset whilst ensuring that the fractional composition of events in the resulting datasets with some \textit{stratification key} are as close as possible to those of the starting dataset. Initially this key is the event class, i.e. the ratio of signal to background in the resulting datasets matches that of the full dataset. From \autoref{sec:cat_embed} onwards, the stratification key is based on the event class within categories of jet multiplicity.} into training and validation sets is performed on the original training data. Since it was found that the validation set had a large impact on the test scores, due to the cut optimisation, experiments are repeated multiple times using different random seeds for the splitting.
		
		After processing the data consist of 31 training features. The training data features are transformed to have a mean of zero, and a standard deviation of one. The exact same transformation is then applied to both the validation and testing data. Following the pre-processing stage, all \textit{NaN} values are replaced with zeros. The training and validation datasets are then each split into ten folds via random stratified splitting on the event class (signal or background). The testing data is also split into ten folds, but via simple random splitting. This was done to allow for easy training of ensembles and to make the process of augmenting the data easier. Additionally, we henceforth redefine one epoch as being a complete use of one of these folds, and during training will load each fold sequentially.
		
		As mentioned in \autoref{sec:overview}, each event in the training and validation data also contains a \textit{weight}. This is normalised to be used to evaluate the AMS, but also reflects the relative importance of the particular event. It is a product of the production cross-section of the underlying process and the acceptance efficiency of the initial skimming procedure. Since the model is unlikely to achieve perfect classification, it is important that it focuses on learning to classify the higher weighted events better than other, less important events. This can be achieved by applying the weights as multiplicative factors during the evaluation of the loss:
		\begin{equation}
			\mathcal{L}=\frac{1}{N}\sum^N_{n=1}w_n\times\epsilon\!\left(y_n,\hat{y}_n\right),
		\end{equation}
		where $w_n$ is the weight associated with event $n$ in a mini-batch of $N$ events. This method of loss weighting is also used to account for class imbalances in the data by normalising the event weights in training data such that the sum of weights for each class are equal:
		\begin{equation}
			\sum w_{\text{signal}} = \sum w_{\text{background}} = 1.
		\end{equation}
		This has the advantage of retaining the relative importance of different events within each class, whilst balancing the importance of each class overall.
			
	\subsection{Spirit of the investigation }
		The HiggsML dataset is used due to it being a publicly accessible dataset which is representative of a typical HEP problem in both size and scoring metric, and already has strong baseline solutions from the 2014 Kaggle challenge.
		
		The public release of the dataset now includes the truth labels of the original Kaggle test set meaning that solutions can be fully scored without requiring submissions to Kaggle. This also means that the test set can be further subdivided and mean scores computed allowing both a score and an uncertainty to be reported.
		
		The preceding version of this investigation, available in Ref.~\cite{amva_d1.4}, attempted to reproduce the challenge conditions by only evaluating the public score during development and then checking the private score at the end. Additionally, that investigation only considered single trainings with a fixed seed for the train/validation splitting. The investigation presented here instead attempts to evaluate the general effects that new techniques have on model performance, and so will report the mean scores from multiple trainings, each with unique random seeds for train/validation splitting. Additionally, the mean public score over ten stratified folds will be computed.
		
		Whilst this metric would be unavailable in true challenge conditions, the final score is a product of both model optimisation and cut optimisation. Since the method to pick the cut is fixed (as per \ref{sec:scoring}), the problem of \textit{cut optimisation} reduces to \textit{cut compatibility}. Even so, the validation AMS alone is not representative of the true performance of the method; the score on some withheld data at the chosen cut is required. Data limitations mean that  it is necessary to rely on the public scores to provide some indication of true model performance. However in order to avoid overfitting to the public score, the mean score will be used, and this will be averaged over repeated trainings.
		
		This all is to say that, although the private scores will be kept blind until the end, this investigation does not fully reproduce challenge conditions and instead aims to demonstrate the general effect of different architectures and training choices. Additionally, the solutions developed will consider the train and inference time requirements, not just performance, and hopes to produce solutions and approaches which are appropriate for use in other HEP applications.

	\subsection{Reported metrics}
		Several different metrics will be reported during the investigation and are averaged over six repeated trainings, each of which uses a unique train/validation splitting. The metrics are:
		\begin{itemize}
			\item The Mean Maximal Validation AMS (MMVA); the maximum AMS achievable on the validation data. MMVA can be thought of as representing an optimistic upper-limit of solution performance.
			\item The Mean Validation AMS at Cut (MVAC); the AMS  on the validation data at the cut chosen according to the technique described in \autoref{sec:scoring}. MVAC offers a more general measure of performance, but assumes that the cut used generalises well to unseen data.
			\item The Mean Averaged Public AMS (MAPA); the value of the AMS on the public part of the testing data at the chosen cut averaged over ten sub-splittings. MAPA represents a more realistic indication of performance since it accounts for cut generalisation.
		\end{itemize}
		MAPA will be the main metric for determining improvements, with MMVA and MVAC available to provide further indications of performance. Uncertainties computed for these metrics only account for statistical fluctuations in performance as systematic uncertainties due to theoretical modelling and experimental precision are not accounted for in the challenge.
		
		Additionally, the time taken to train and apply each setup will be reported as the fractional increase in time compared to the baseline setup (see \autoref{sec:baseline}), and this will be averaged over the six hardware setups used. A more complete set of timings will be reported towards the end of this \whatAmI.
		

\FloatBarrier
	\section{Solution development}\label{sec:solution_development}
		\subsection{Machinery}\label{sec:machinary}
    Since this paper will report the timings of the solutions, we introduce here the machinery used. There are five machines, two of which have the same configuration, and one of which is used in two different hardware configurations (CPU or GPU). Each solution is trained once per hardware configuration, meaning that six trainings are performed. Whilst RAM and VRAM capacity will be stated, we emphasise that the memory requirements of the solutions are minimal and less than \SI{1}{\giga\byte}.

    \paragraph{Desktop}
        \begin{itemize}
            \item CPU: Intel Core i7-8700K, 6 cores, 2 threads per core, clock-speed \SI{3.7}{\giga\hertz}
            \item GPU: Nvidia GeForce GTX 1080 Ti (\SI{11}{\giga\byte} VRAM)
            \item RAM: \SI{32}{\giga\byte}, DDR4, clock-speed \SI{2666}{\mega\hertz}
            \item Data loaded from SSD hard-drive (SATA connection), read speed \SI{550}{\mega\byte\per\second}
            \item Used to provide two hardware configurations: CPU or GPU
        \end{itemize}

    \paragraph{MacBook Pro (2018, 13")}
        \begin{itemize}
            \item CPU: Intel Core i7-8559U, 4 cores, 2 threads per core, clock-speed \SI{2.7}{\giga\hertz}
            \item GPU: N/A
            \item RAM: \SI{16}{\giga\byte}, LPDDR3, clock-speed \SI{2133}{\mega\hertz}
            \item Data loaded from internal SSD hard-drive, read speed \SI{2520}{\mega\byte\per\second}
        \end{itemize}

    \paragraph{Virtual Machine (m2.medium)}
        \begin{itemize}
            \item CPU: Intel Xenon Skylake CPU, 2 virtual single-thread cores, clock-speed \SI{2.2}{\giga\hertz}
            \item GPU: N/A
            \item RAM: \SI{3.7}{\giga\byte}, unknown clock-speed
            \item Unknown hard-drive
        \end{itemize}

    \paragraph{2 X Virtual Machine (m2.large)}
        \begin{itemize}
            \item CPU: Intel Xenon Skylake CPU, 4 virtual single-thread cores, clock-speed \SI{2.2}{\giga\hertz}
            \item GPU: N/A
            \item RAM: \SI{7.3}{\giga\byte}, unknown clock-speed
            \item Unknown hard-drive
        \end{itemize}

		\subsection{Feature selection}\label{sec:feature_selection}
	\textit{The definitions of the features available, and selection of the training features to use. All training features are found to be useful.}\\
	
\begin{figure}[ht]
		\begin{center}
			\begin{subfigure}[t]{\sfMid\textwidth}
				\begin{center}
					\includegraphics[width=\textwidth]{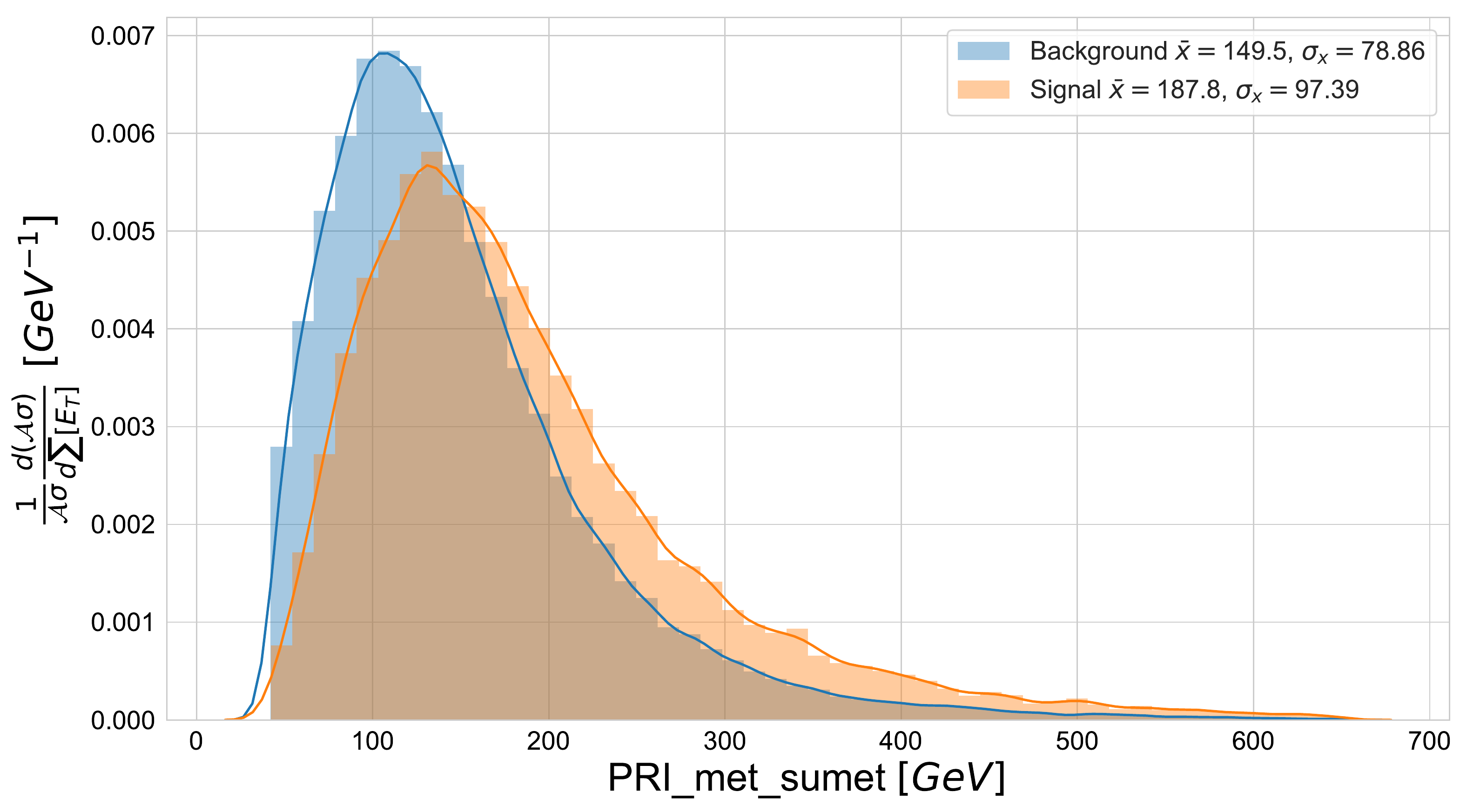}
					\caption{Example low-level feature: the sum of missing transverse energy $\left(\sum{E_T}\right)$}
				\end{center}
			\end{subfigure}
			\begin{subfigure}[t]{\sfMid\textwidth}
				\begin{center}
					\includegraphics[width=\textwidth]{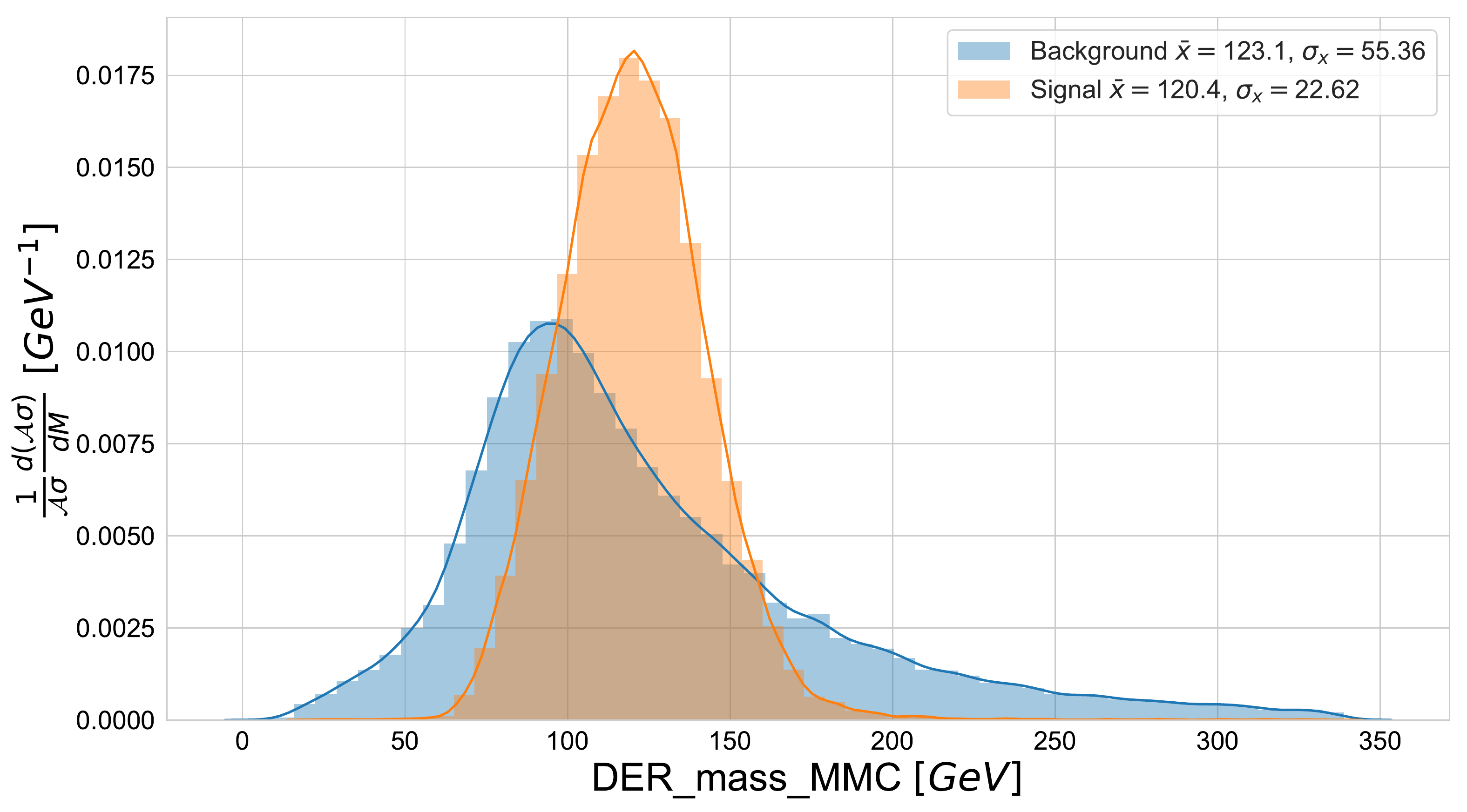}
					\caption{Example high-level feature: the masses of the Higgs candidates as computed by a hypothesis-based fitting algorithm $\left(M_{\tau\,\tau, \text{MMC}}\right)$}
				\end{center}
			\end{subfigure}
			\caption{Example distributions for low-level and high-level features in the data.}
			\label{fig:ll_hl_feats}
		\end{center}
	\end{figure}

	As discussed in \autoref{sec:data_processing}, the dataset consists of both low-level and high-level features: 18 low-level features consisting of 3-momenta $(\pt,\eta,\phi)$ \footnote{The 3-momentum of a particle is a vector of its momentum in terms of three spatial components. The 4-momentum of a particle contains its 3-momentum and its energy, and fully describes the particle's kinematic properties.}, missing transverse momentum \footnote{Due to momentum conservation, the vectorial sum of the final-state momenta in the transverse plane should be zero, however particles such as neutrinos resulting from the tau decays cannot be detected and so lead to a loss of visible momentum. The missing transverse momentum is a vector computed to balance the visible momenta to zero and can be used to infer the kinematics of neutrinos.}, and sum of transverse momenta of all jets; and 13 physics-inspired high-level features such as invariant masses \footnote{The invariant mass of a particle is its mass in its rest-frame ($m_0=\sqrt{E^2-|\vv{p}|^2}$). It is a useful quantity for identifying unstable particles, such as the Higgs boson: given conservation of energy, the final-state particles from the Higgs decay should sum to form a vector with an invariant mass close to that of the Higgs boson (\SI{125}{\giga\electronvolt}), where as particles from background processes are expected to form either masses at other values (such as \SI{91}{\giga\electronvolt} for the $Z\rightarrow\tautau$ background), or a broad ranges of masses (such as for the \ttbar background).} and angles between final-state objects. Tables~\ref{tab:der_features} and \ref{tab:pri_features} describe the features used along with the associated name that will be used throughout this \whatAmI. Figure~\ref{fig:ll_hl_feats} illustrates the density distributions of two example features, one high-level and the other low-level. It should be borne in mind that 3-momenta of final-states are the result of complex reconstruction techniques and so are not strictly ``low-level'' features, these would be the tracking hits and energy deposits in the particle detector, however they are the lowest-level of information provided in the dataset.
	
	The mass of the Higgs boson candidate is expected to be a particularly useful feature: the $\tautau$ final-state candidates should form an invariant masses around \SI{125}{\giga\electronvolt} for signal, \SI{91}{\giga\electronvolt} the $Z\rightarrow\tautau$ background, and a broad range of masses for the $W$-decay (with perhaps a peak around \SI{80}{\giga\electronvolt}) and \ttbar backgrounds. The tau decays, however, involve neutrinos, which cannot be detected at the LHC detectors leading to a loss of kinematical information. \texttt{DER\_mass\_vis} is an estimation of the Higgs mass using the visible decay products, and underestimates the Higgs boson mass. \texttt{DER\_mass\_MMC} instead attempts to account for the missing energy and produce a more accurate estimate for the Higgs mass by performing a kinematic fit using the visible decay products and the missing transverse momentum under the hypothesis of a di-tau decay \cite{mmc}.
	
	\begin{table}
		\begin{center}
			\begin{tabular}{lll}
				\toprule
				Feature name & Description & Grouping\\
				\midrule
				\texttt{DER\_mass\_MMC} & Mass of the Higgs boson estimated by &Mass,\\& a hypothesis based fitting technique &  Higgs\\\hdashline
				\texttt{DER\_mass\_transverse\_met\_lep} & Transverse mass of the lepton and \mpt & Mass Higgs\\\hdashline
				\texttt{DER\_mass\_vis} & Invariant mass of the lepton and the tau &Mass,\\& &Higgs\\\hdashline
				\texttt{DER\_pt\_h} & Transverse momenta of the vector sum  &3-momenta,\\& of the lepton, tau, and \mpt & Higgs\\\hdashline
				\texttt{DER\_deltaeta\_jet\_jet} & Absolute difference in pseudorapidity &Angular\\& of the leading and subleading jets &Jet\\& (undefined for less than two jets) &\\\hdashline
				\texttt{DER\_mass\_jet\_jet} & Invariant mass of the leading and &Mass,\\& subleading jets &Jet\\& (undefined for less than two jets) &\\\hdashline
				\texttt{DER\_prodeta\_jet\_jet} & Product of the pseudorapidities of the &3-momenta,\\& leading  and subleading jets &Jet\\& (undefined for less than two jets) &\\\hdashline
				\texttt{DER\_deltar\_tau\_lep} & Separation in $\eta-\phi$ space of the lepton &Angular,\\& and the tau & Final-state\\\hdashline
				\texttt{DER\_pt\_tot} & Transverse momentum of the vector sum &Sum,\\& of the transverse momenta of the lepton, &Final-state\\& tau, the leading and subleading jets &\\& (if present),  and \mpt &\\\hdashline
				\texttt{DER\_sum\_pt} & Sum of the transverse momenta of &Sum,\\& the lepton, tau, and all jets & global event\\\hdashline
				\texttt{DER\_pt\_ratio\_lep\_tau} & Transverse momenta of the lepton divided &3-momenta,\\& by the transverse momenta of the tau & Final-state\\\hdashline
				\texttt{DER\_met\_phi\_centrality} & Centrality of the azimuthal angle of \mpt &Angular,\\& relative to the lepton and the tau & Final-state\\\hdashline
				\texttt{DER\_lep\_eta\_centrality} & Centrality of the pseudorapidity of the &Angular,\\& lepton relative to the leading and &Jet\\& subleading jets &\\& (undefined for less than two jets)  & \\
				\bottomrule
			\end{tabular}
		\end{center}
		\caption{Glossary of high-level (derived) features used for training, along with their grouping.}
		\label{tab:der_features}
	\end{table}

	\begin{table}
		\begin{center}
			\begin{tabular}{lll}
				\toprule
				Feature name & Description & Grouping\\
				\midrule
				\texttt{PRI\_tau\_[px/py/pz]} & 3-momenta of the tau in &3-momenta,\\& Cartesian coordinates & Final-state\\\hdashline
				\texttt{PRI\_lep\_[px/py/pz]} & 3-momenta of the lepton in &3-momenta,\\& Cartesian coordinates & Final-state\\\hdashline
				\texttt{PRI\_met\_[px/py]} & Components of the vector of &3-momenta,\\& missing transverse momentum & Final-state\\& in Cartesian coordinates &\\\hdashline
				\texttt{PRI\_met} & Modulus of the vector of &3-momenta,\\& missing transverse  momentum & Final-state\\& in Cartesian coordinates &\\\hdashline
				\texttt{PRI\_met\_sumet} & Sum of all transverse energy & Energy, \\&& Final-state\\\hdashline
				\texttt{PRI\_jet\_num} & Number of jets in event & Multiplicity, \\&& Jet\\\hdashline
				\texttt{PRI\_jet\_leading\_[px/py/pz]} & 3-momenta of the leading jet in &3-momenta,\\& Cartesian coordinates & Jet\\& (undefined if no jets present ) &\\\hdashline
				\texttt{PRI\_jet\_subleading\_[px/py/pz]} & 3-momenta of the subleading jet in &3-momenta,\\& Cartesian coordinates & Jet\\& (undefined if less than two jets present) & \\\hdashline
				\texttt{PRI\_jet\_all\_pt} & Sum of transverse momenta of all jets &3-momenta,\\& Cartesian coordinates & Jet\\
				\bottomrule
			\end{tabular}
		\end{center}
		\caption{Glossary of low-level (primary) features used for training, along with their grouping. Note that \texttt{PRI\_jet\_all\_pt} can be expected to be different from the sum of \pt of leading and subleading jets, since there can be events which contain more than two jets, however the dataset only contains details of the two jets with the highest \pt.}
		\label{tab:pri_features}
	\end{table}
	
	Our hope is that the networks will be able to exploit the low-level information, but without this exploitation the low-level features risk being removed during feature selection. We are therefore more interested in seeing whether all of the high-level features are necessary. Testing will use a simplified type classification model called a Random Forest \cite{random_forests}. We use these as they are quick and easy to train, provide decent performance, and do not require the data to be transformed in any particular fashion prior to training.

	Quantification of a feature's importance is done using its \textit{permutation importance}. This aims to quantify how much a given feature influences the model's output. Features with a higher importance will have a greater influence overall that those of lesser importance. The permutation importance (PI) is computed by first training the model as normal and evaluating its performance on the training set. Next, a copy of the training data is made and one of the features is randomly shuffled datapoint-wise, thereby destroying any information carried by that feature for each datapoint, whilst keeping its values at the same scale as seen during training. The performance of the model on the shuffled data is computed and the difference in performance compared to the original performance is the permutation importance of the feature that was shuffled. If shuffling the feature caused a large change in performance, then that feature was important to the model. If the change was low, then the feature was less important. By then dividing the PI by the original performance we arrive at the \textit{fractional} permutation importance.
    
\begin{figure}[ht]
		\begin{center}
			\includegraphics[width=\sfWide\textwidth]{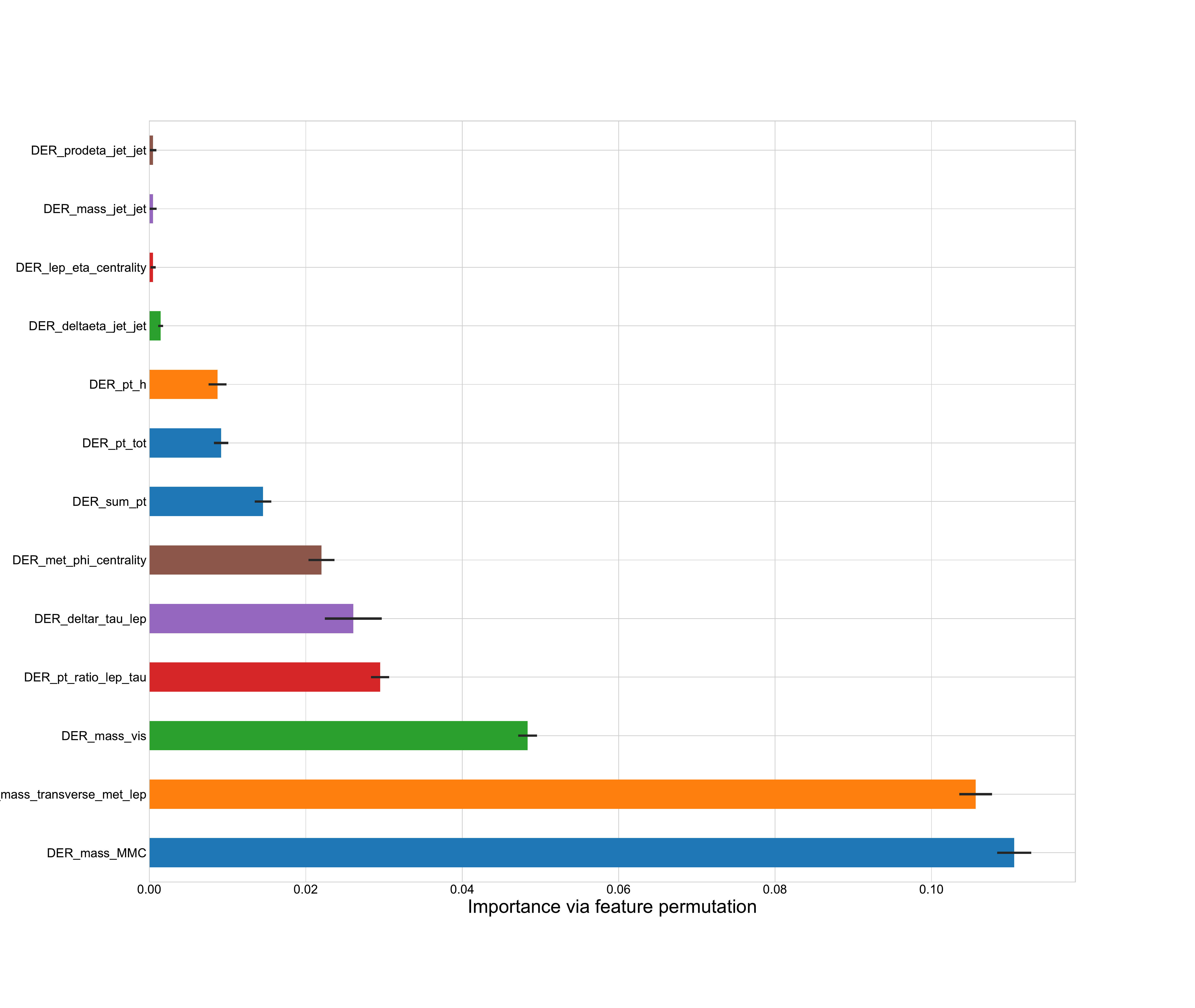}
			\caption{Illustration of the fractional permutation importance of the high-level features as estimated using the average importance over five Random Forest models.}
			\label{fig:hl_imp}
		\end{center}
	\end{figure}

	This process of copying and shuffling is repeated for every feature. By retraining new models, an average PI can be computed for each feature. Although the dataset is the same each time, the feature subsampling at each split acts as a source of stochasticity. Here we compute PIs using an average of five retrainings. A threshold on the PI can then be used to select features by minimum importance. We use a threshold of \num{0.005} as to identify features as ``important". From \autoref{fig:hl_imp}, we can see that all features except the four nearest the top of the plot, i.e. those associated with jet pairs, appear to be important.
	
	The lack of importance of jet-pair features is to be expected: since the majority of events do not contain two or more jets, the values for these features are often the default value of zero. When the feature is permuted, most of the zeros will be replaced with another zero, and the model response remains mostly unaffected. Effectively the di-jet features are only important for a subset of events, but their importance is smeared out across all events. In actual application, the jet number will be encoded in a way such that the model will be able to more easily target these di-jet events and make use of their features. The di-jet features will therefore be kept. As additional motivation, training a new Random Forest without these features showed a drop in performance on unseen data.
	
\begin{figure}[ht]
		\begin{center}
			\includegraphics[width=\sfMid\textwidth]{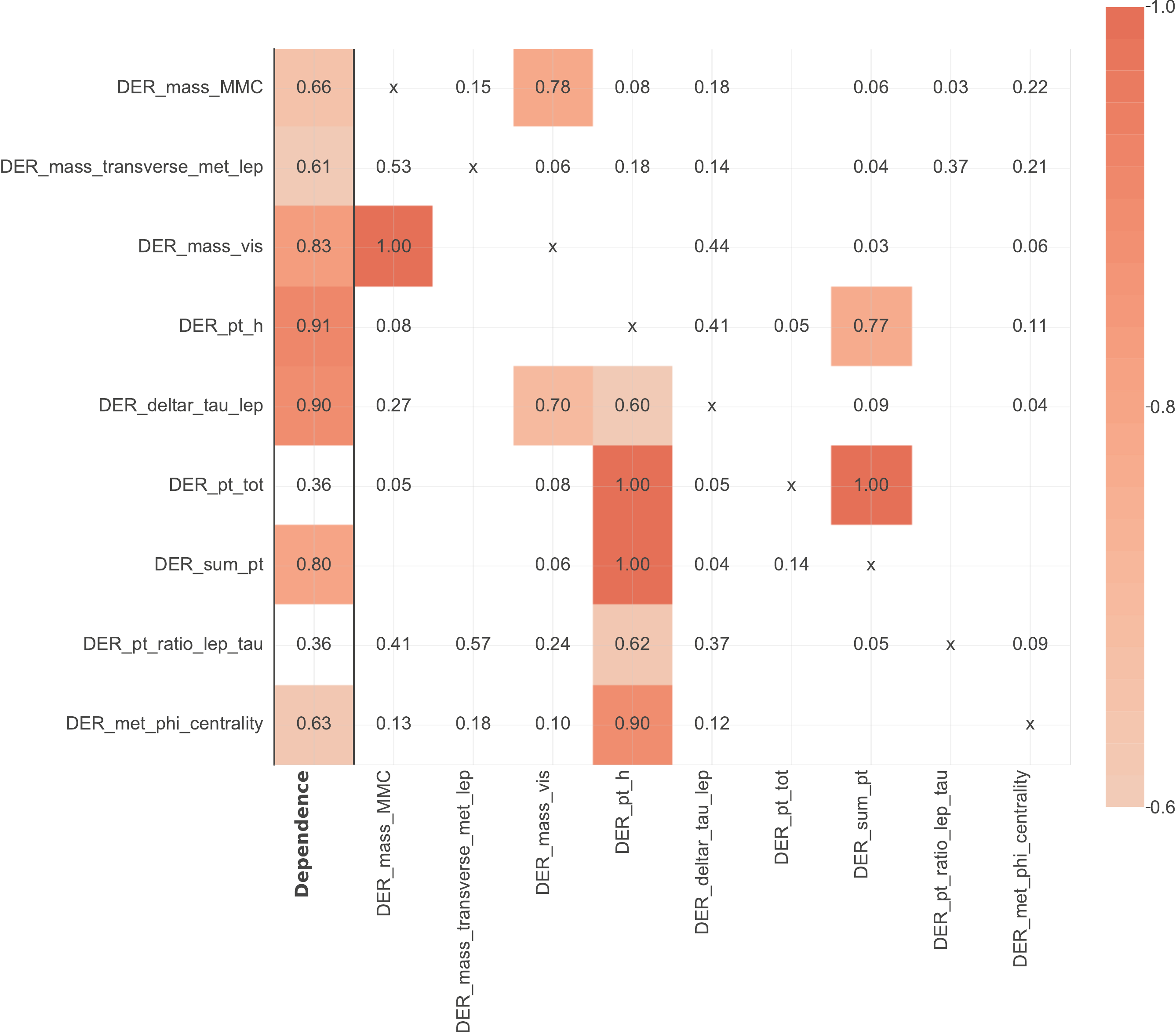}
			\caption{Illustration of feature dependence and associated feature importance using Random Forest regression. Considering each row separately, the feature on the $y$ axis is one being regressed to. The value in the ``Dependence" column is the coefficient of determination ($R^2$); higher means easier to predict the feature values. The remaining columns indicate how much each feature on the $x$ axis contributes to the performance of the regressor; higher means more important. As an example: \texttt{DER\_pt\_h} is the easiest feature to predict (dependence of 0.91), and \texttt{DER\_sum\_pt} is moderately important to the regressor (importance of 0.77).}
			\label{fig:hl_checks}
		\end{center}
	\end{figure}

	In order to check for monotonic relationships between features we then compute the Spearman's rank-order correlation coefficients \cite{spearman} for every pair of high-level features and do not find any pairs that are fully (anti-)correlated with one another. As a final check we examine the mutual dependencies of the features, i.e. how easy it is to predict the value of a feature using the other features as inputs to a Random Forest regressor. As shown in \autoref{fig:hl_checks}, none of the non-di-jet high-level features are completely predictable using the other features (dependencies are never equal to one). We can therefore expect that each high-level feature is bringing at least some unique information not carried by the others.
	
	\FloatBarrier
		\subsection{Baseline model}\label{sec:baseline}
	\textit{The basic architecture for later comparisons, and also a demonstration of the benefits of ensembling models.}\\
	
	Training and inference of all solutions developed in this \whatAmI are performed in \lumin \cite{lumin}, a wrapper library for \pytorch~\cite{pytorch}, which so far has been primarily developed by the author of this \whatAmI.

	\subsubsection{Architecture}
		The baseline model consists of a simple neural network architecture. Specifically, it contains four fully-connected layers, each of 100 neurons, with weights initialised according the He-normal \cite{He} scheme. ReLU activation functions \cite{relu} follow each fully-connected layer. The output layer consists of a single neuron, whose weights are initialised according to the Glorot-normal \cite{Glorot} scheme. A sigmoid activation function is applied to the output of the network. The starting biases for all layers are set to zero.

		As usual, the parameters (weights and biases) of the network are optimised though backpropagation \cite{backprop} of a loss-gradient estimated by averaging batches of data points (here we use a minibatch size of 256 events). Given that the problem involves binary classification, the loss function is chosen to be the binary cross-entropy of network predictions. The update steps are determined using ADAM \cite{Adam} with relatively standard parameter settings of $\beta_1=0.9$, $\beta_2=0.999$, and $\epsilon=\num{1e-8}$. The learning rate (LR) is kept constant during training at \num{2e-3}, which was found using a LR Range Test  as described in Ref.~\cite{Smith_2015}. Training continues until the loss on a validation fold of the training data has not decreased in the last 50 epochs (a \textit{patience} of 50), after which the model state corresponding to the lowest validation loss is loaded.
		
	\subsubsection{Performance}
		\paragraph{Single model}
			A single model is trained on nine out of the ten training folds, with the tenth fold acting as validation data for loss monitoring. Once finished, the model is applied to both the validation and testing data, and the required metrics recorded. This process is run once per hardware configuration, as defined in \autoref{sec:machinary} (six times in total), each using a unique random seed for the train/validation splitting as described in \autoref{sec:data_processing}. Performance metrics for the single model are reported in \autoref{tab:baseline_performance}.
		
		\paragraph{Ensemble}\label{sec:weighted_ensemble}
			An ensemble of ten models is trained, again using nine out of the ten training folds, with the tenth fold acting as validation data. Each model uses a different fold of the training data for validation. The resulting models are ensembled by weighting their predictions proportionally to the inverse of their associated loss on their validation folds. The prediction of the ensemble is then:
			\begin{equation}
				p_{\mathrm{ensemble}}(x)=\sum_{i}^{10} w_ip_i(x),
			\end{equation}
			where $p_{\mathrm{ensemble}}(x)$ is the prediction of the ensemble for datapoint $x$, $p_i(x)$ is the prediction of model $i$ for datapoint $x$, and $w_i$ in the importance weight of model $i$. Importance weights are normalised such that they sum to one. This means that models that reached a lower validation loss  during training have a higher importance weight and consequent a greater influence over the ensemble prediction than those with a higher loss.
			
			Ensembling is seen to provide improvements in all metrics compared to the single model setup, with expected increases in train and inference time. Note that the inference-time increase appears to be sub-linear due to the fact that the data are kept in memory between model evaluations. Based on the improved performance, we proceed to adopt the ensembled version as the baseline model for later comparisons.
			
			\begin{table}
				\begin{center}
					\begin{tabular}{lccccc}
						\toprule
						Setup & MMVA & MVAC & MAPA & \multicolumn{2}{c}{Fractional time-increase}\\
						& & & & Training & Inference\\
						\midrule
						Single model & $3.43\pm0.04$ & $3.38\pm0.04$ & $3.44\pm0.02$ & \textbf{-} & \textbf{-}\\
						\textbf{Ensemble} & $\mathbf{3.70\pm0.05}$ & $\mathbf{3.64\pm0.04}$ & $\mathbf{3.664\pm0.007}$ & $10.4\pm0.6$ & $5.9\pm0.9$\\
						\bottomrule
					\end{tabular}
				\end{center}
				\caption{Validation-performance comparison between using the single model and weighted ensemble setups. Higher values of MMVA, MVAC, and MAPA indicate better performance, and higher values of fractional time-increase indicate that the solution takes longer to train and/or apply. The best values for each metric are shown in bold, and the setup chosen is also indicated in bold.}
				\label{tab:baseline_performance}
			\end{table}
		
	\FloatBarrier
		\subsection{Categorical entity embedding}\label{sec:cat_embed}
	\textit{The testing of an efficient way to encode categorical information, which results in mild improvements.}\\
	
	In \autoref{sec:baseline} we considered all features in the data to be continuous variables, however the number of jets in an event (\texttt{PRI\_jet\_num}) could instead be thought of as a categorical feature. Although two jets are more than one jet, this can also be treated as defining subcategories of particle interactions, e.g. a two-jet event and a one-jet event, in which case there is no longer a numerical comparison. Optimal treatment of categorical features requires encoding the different categories in a way such that the ML algorithm can easily access them individually. A common way to do this is to assign each category an integer code which begins at zero and sequentially increases, and then treat these codes as row indices of an \textit{embedding} matrix. The elements in the indexed row are then used to represent the corresponding category.
	
	Under the \textit{1-hot encoding} scheme, the embedding matrix is  a $N\times N$  identity matrix (where $N$ is the number of categories in a feature; its \textit{cardinality}), i.e. each category is represented by a vector of length $N$ in which only one value is non-zero (one element is \textit{hot}). Whilst this approach provides unique access to each category, the number of input variables is now $N$. For high cardinality features, or many low-cardinality features, this can quickly lead to a huge increase in the number of model parameters which must be learnt.
	
	Reference~\cite{cat_embed} instead suggests for neural networks to use an $N\times M$ weight matrix for the embedding, where $M<N$ is a hyper-parameter which must be chosen. Each category is now represented by a real-valued vector of length $M$, i.e. in a more compact fashion than 1-hot encoding would provide. The values of the weight matrix are then learnt via backpropagation during training. This approach of \textit{categorical entity embedding} works because, for example, it is likely that it is more important to know whether an event contains zero, one, several, or many jets, rather than exactly how many jets, but rather than manually compacting the categories (and having to define a priori ``several" and ``many"), a compact, information rich, and task specific representation can be learnt. The additional hyper-parameter $M$ can be estimated using domain knowledge or by using some rule-of-thumb: the original paper suggests to use $M=N-1$, and \textsc{FastAI} suggests to use half the feature cardinality or 50, which ever is smaller ($M=\text{min}\left(50, \left(N+1\right)//2\right)$ \cite{fastai_embed})
	
\begin{figure}[ht]
		\begin{center}
			\includegraphics[width=\sfSmall\textwidth]{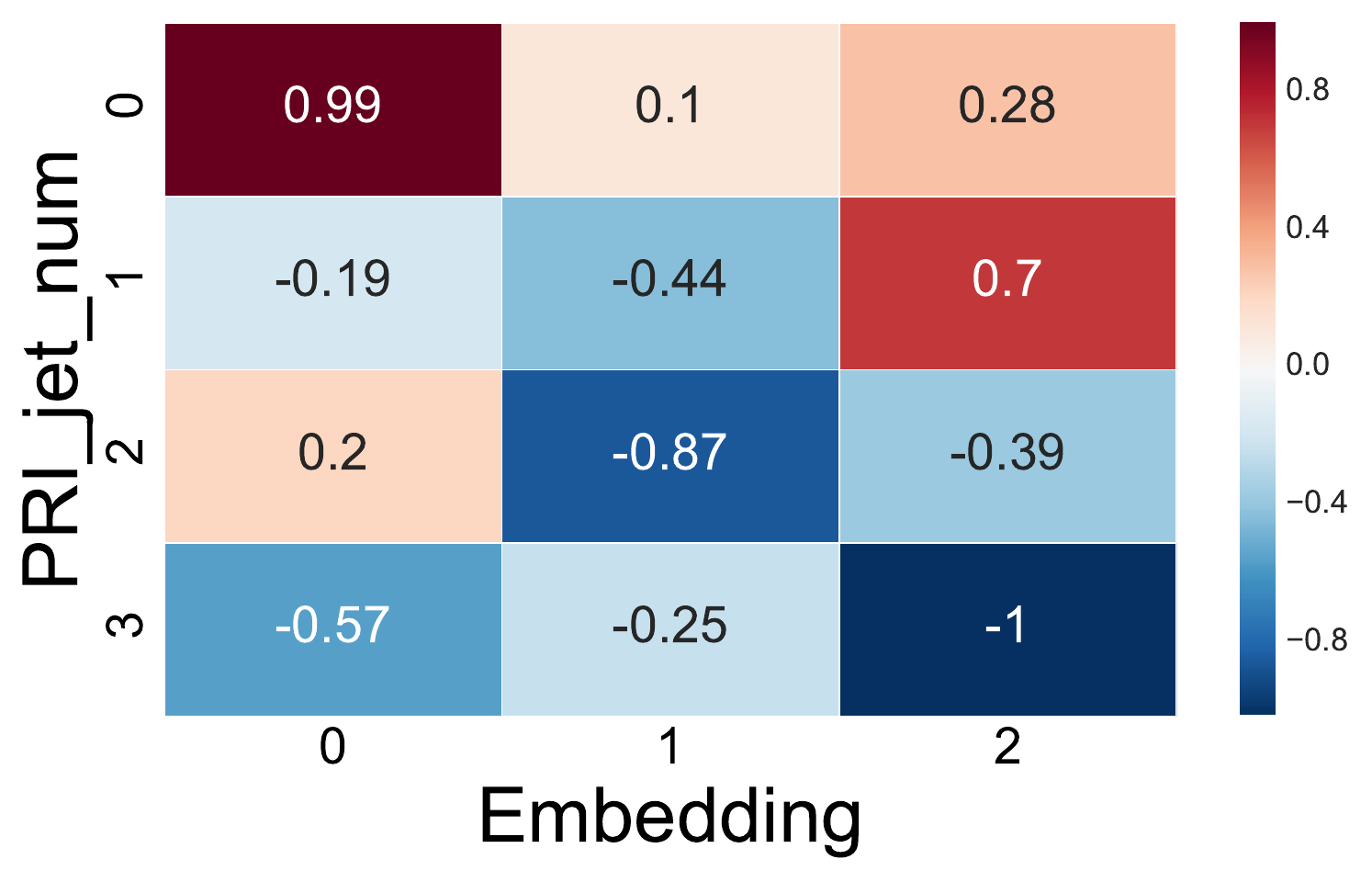}
			\caption{Illustration of an example of a learned embedding matrix for the number of jets. Note that each row is unique, indicating that the NN found it useful to learn separate embeddings for each category, and was able to transform the feature into a lower dimensional representation.}
			\label{fig:embed}
		\end{center}
	\end{figure}

	For embedding \texttt{PRI\_jet\_num}, $M$ is chosen to be three and initial weights for the $4\times3$ matrix are sampled from a unit-Gaussian distribution. As shown in \autoref{tab:solution_evolution}, this is found to provide metric improvements across the board, for only minor increases in train and inference time. Figure~\ref{fig:embed} shows an example of one of the learned embeddings. Please be aware that \autoref{tab:solution_evolution} includes all improvements that are discussed in this \whatAmI, and so contains solutions which may not have been introduced at time of reading.

	Whilst not required for this particular study, as discussed in \autoref{sec:scoring}, HEP analyses potentially categorise data in order to further improve the sensitivity of their search. These inference categories are search channel-dependent, such as di-tau decay-channel, as well as categorisation by jet-multiplicity, but category flags can potentially be encoded as categorical features and embedded. This then allows the classifier to better parameterise its responses to each inference category, whilst benefitting from a larger training sample (as opposed to training separate classifiers for each category) and learning information common to all categories.
	
	\begin{table}
		\begin{center}
			\begin{tabular}{lccccc}
				\toprule
				Setup & MMVA & MVAC & MAPA & \multicolumn{2}{c}{Fractional time-increase}\\
				& & & & Training & Inference\\
				\midrule
				Baseline & $3.70\pm0.05$ & $3.64\pm0.04$ & $3.664\pm0.007$ & \textbf{-} & \textbf{-}\\
				+Embedding & $3.78\pm0.05$ & $3.70\pm0.06$ & $3.71\pm0.02$ & $0.13\pm0.03$ & $0.08\pm0.04$\\
				+Augmentation & $3.96\pm0.04$ & $3.89\pm0.05$ & $3.79\pm0.01$ & $1.45\pm0.08$ & $14\pm5$\\
				+SGDR & $3.94\pm0.06$ & $\mathbf{3.90\pm0.04}$ & $3.80\pm0.02$ & $2.12\pm0.07$ & $14\pm5$\\
				+Swish & $3.96\pm0.06$ & $3.86\pm0.05$ & $3.81\pm0.02$ & $2.32\pm0.07$ & $17\pm4$\\
				-SGDR + SWA & $3.96\pm0.06$ & $3.88\pm0.05$ & $3.81\pm0.02$ & $2.03\pm0.07$ & $17\pm4$\\
				-SWA + 1cycle & $\mathbf{4.04\pm0.06}$ & $3.85\pm0.04$ & $3.81\pm0.02$ & $0.66\pm0.04$ & $17\pm4$\\
				\textbf{+Dense} & $3.95\pm0.04$ & $3.89\pm0.05$ & $\mathbf{3.82\pm0.02}$ & $0.81\pm0.08$ & $17\pm5$\\
				\bottomrule
			\end{tabular}
		\end{center}
		\caption{A summary of solution evolution in terms of the optimisation metrics and the time impact. Fractional time-increases are computed with respect to the baseline model, which is the ensembled solution from \autoref{sec:baseline}. A `+' indicates the addition of a new method to the solution, and `-' indicates the removal of a previously added method. The best values for each metric are shown in bold, and the setup chosen is also indicated in bold.}
		\label{tab:solution_evolution}
	\end{table}
		
	\FloatBarrier
		\subsection{Data symmetry exploitations}\label{sec:data_symmetry}
	\textit{Discussion and application of HEP-specific data augmentation, which result in large improvements.}\\
	
	\subsubsection{Overview of data symmetries}		
		At the LHC \cite{lhc}, beams of protons are collided head-on with approximately zero transverse momentum. Because of this the resulting final states can be expected to be produced isotropically in both the transverse plane ($x,y$) and the longitudinal axis ($z$). Particle detectors such as CMS \cite{CMSExperiment} and ATLAS \cite{ATLASExperiment} are built to account for these isotropies by being symmetric in both azimuthal angle ($\phi$) and pseudorapidity ($\eta$). All this is to say that the class of physical process that gave rise to the collision event is only related to the relative separation between the final states, and not the absolute position of the event.
		
\begin{figure}[ht]
			\begin{center}
				\begin{subfigure}[t]{\sfMid\textwidth}
					\begin{center}
						\includegraphics[width=\textwidth]{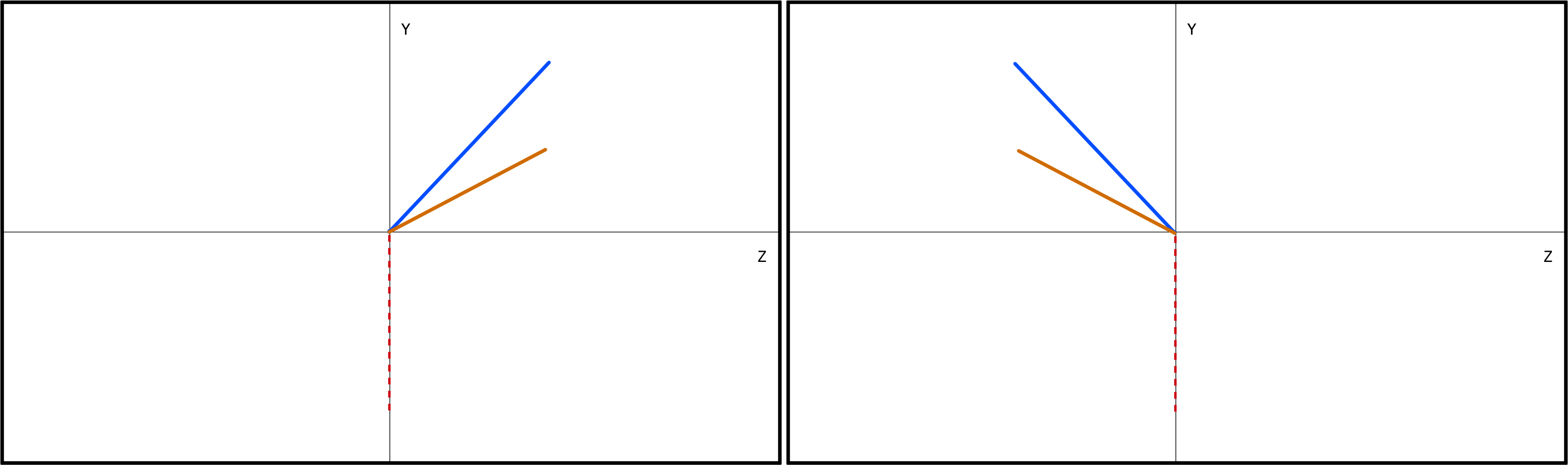}
						\caption{Left: an example event in the $y-z$ plane, with blue and orange lines representing arbitrary final-state vectors and the dashed line representing missing transverse momentum. Right: the same event flipped in the longitudinal axis.}
					\end{center}
				\end{subfigure}
				\begin{subfigure}[t]{\sfMid\textwidth}
					\begin{center}
						\includegraphics[width=\textwidth]{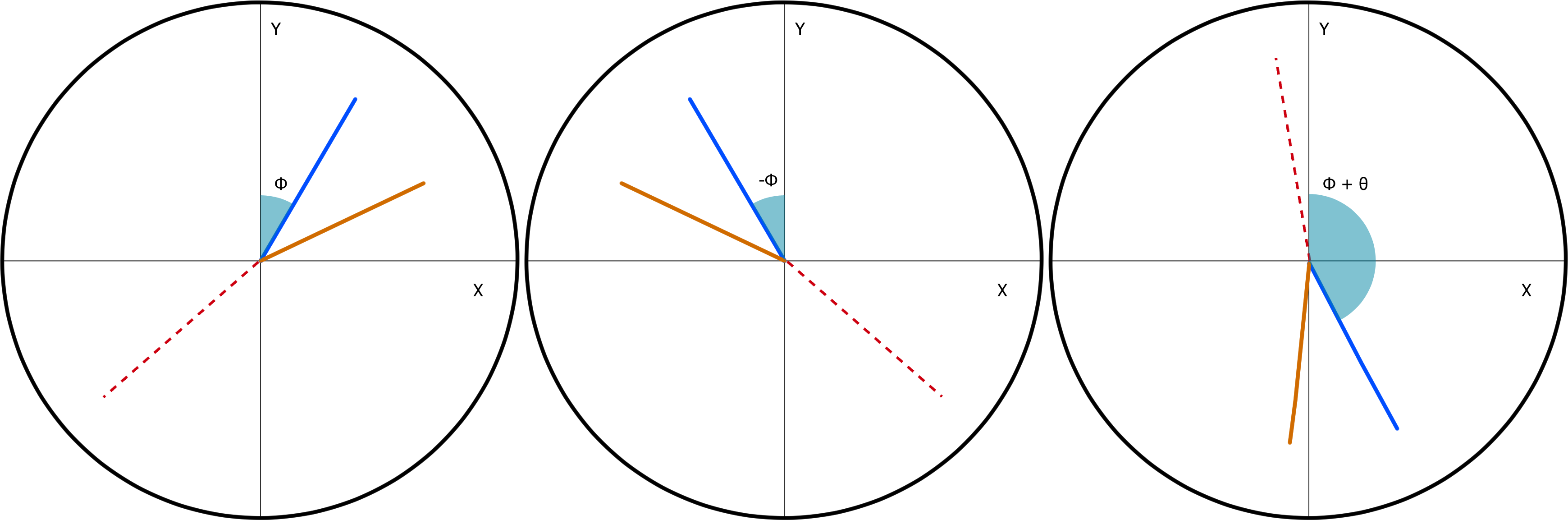}
						\caption{Left: an example event in the transverse plane, with blue and orange lines representing arbitrary final-state vectors and the dashed line representing missing transverse momentum. Centre: the same event rotated in the azimuthal angle ($\phi$). Right: the same event flipped in the $y$ axis.}
					\end{center}
				\end{subfigure}
				\caption{Illustrations of class-preserving transformations for particle collisions.}
				\label{fig:data_aug}
			\end{center}
		\end{figure}

		Since the data used in this problem are simulated for such a collider and detector combination, the class is invariant to flips in the transverse or longitudinal planes, and to rotations in the azimuthal angle, as illustrated in \autoref{fig:data_aug}. Note, it is important to reconsider the set of class-preserving transformations if intending to use the following techniques to analyse data for a different experimental configuration, such as: colliding beams of different particles, using a fixed-target experiment, or an asymmetrical detector.
		
	\subsubsection{Data fixing}
		The existence of class-invariant symmetries in the data acts to increase the data complexity; a classifier must learn to become invariant to the absolute orientation of events and instead focus on the positions of the particles relative to one-another.  This complexity can easily be removed by rotating all events to have a common alignment.
		
		We refer to this pre-processing step as \textit{data fixing}, and apply it by first rotating events in the transverse plane and reflecting in the longitudinal plane such that the light-leptons (\texttt{PRI\_lep}) are at $\phi=0$ and $\eta\geq0$, and then reflecting the event in the transverse plane such that the tau leptons (\texttt{PRI\_tau}) are in the positive $\phi$ region. Since the $y$ component of the light-lepton's 3-momentum is now constant (zero) it can be dropped as an input feature during training. Note that rotation in the transverse plane is effectively computing the $\Delta\phi$ angles of other final-states with respect to the light-leptons, and indeed this is how the fixing is implemented in code.
			
	\subsubsection{Data augmentation}
		An alternative approach to dealing with the data symmetry is instead to exploit it as a means of generating augmented copies of events. Data augmentation is a common technique for improving the generalisation power of a model. It involves applying class-preserving transformations to the data in order to exploit or highlight certain invariances between the input and target features. In the field of image classification, these transformations could be small rotations, flips, zooms, or colour and lighting adjustments, e.g. Ref.~\cite{ImageNet_data_aug}. The data augmentation applied here consists of event flips in the $y$ and $z$ planes and rotations in the transverse plane. Application of the augmentations may be done during training in order to artificially inflate the size of the training data (train-time augmentation), or during inference by taking the mean prediction for a set of augmentations (inference-time augmentation).
		
		Train-time augmentation was performed by applying a random set of augmentations before each data point is used (implemented by augmenting each fold when loaded). Inference-time augmentation was performed by taking the mean prediction for each data point over a set of eight transformations: each possible combination of flips in $y$ and $z$ for two different $\phi$ orientations (original $\phi$ and $\phi+\pi$).
	
	\subsubsection{Comparison}
		Table~\ref{tab:data_symmetry} compares the two approaches in terms of the optimisation metrics. Data augmentation produces more favourable scores, but at the expense of a large increase in inference time. Nevertheless, the absolute inference time is still acceptable (\SI{15}{\second} on a GPU (Nvidia GeForce GTX 1080 Ti) and \SI{2}{\minute} on the slowest CPU (Intel Xenon Skylake CPU, 2 virtual single-thread cores, clock-speed \SI{2.2}{\giga\hertz})), so the augmentation approach was chosen going forwards. Table~\ref{tab:solution_evolution} compares the improvements against the baseline solution.
		\begin{table}
			\begin{center}
				\begin{tabular}{lccccc}
					\toprule
					Setup & MMVA & MVAC & MAPA & \multicolumn{2}{c}{Fractional time-increase}\\
					& & & & Training & Inference\\
					\midrule
					Fixing & $3.90\pm0.04$ & $3.83\pm0.05$ & $3.76\pm0.01$ & \textbf{-} &  \textbf{-}\\
					\textbf{Augmentation} & $\mathbf{3.96\pm0.04}$ & $\mathbf{3.89\pm0.05}$ & $\mathbf{3.79\pm0.01}$ & $1.11\pm0.05$ & $13\pm5$\\
					\bottomrule
				\end{tabular}
			\end{center}
			\caption{Comparison of data symmetry exploits in terms of the optimisation metrics and time impact. The best values for each metric are shown in bold, and the setup chosen is also indicated in bold.}
			\label{tab:data_symmetry}
		\end{table}
		
	\FloatBarrier
		\subsection{Stochastic gradient descent with warm restarts}\label{sec:SGDR}
	\textit{Testing of learning-rate annealing schedules. Whilst this results in moderate improvements it eventually becomes obsolete by the schedule in \autoref{sec:one_cycle}.}\\
	
	According to Ref.\cite{Bengio_recommendations}, the learning rate (LR) is one of the most important parameters to set when training a neural network. Whilst initial, optimal values can be found using a LR Range Test \cite{Smith_2015}, it is unlikely that this will remain optimal throughout training. A common approach is to decrease the LR in steps \cite{step_decay} during training (either at preset points or when the loss reaches a plateau).
	
	Reference~\cite{SGDR} instead proposes a \textit{cosine-annealing} approach in which the LR is decreased as a function of mini-batch iterations according to half a cycle of a cosine function. This means that initial training is performed at high LRs, where the model can quickly move across the loss surface towards a minima, and the final training is spent at lower LRs allowing the model to converge to a minima, with a smooth, approximately linear transition between these states. Additionally, by repeating this cosine decay, the paper suggests that the rate of convergence can be further improved, a process described as \textit{warm restarts}.
	
	Reference~\cite{snapshot_ensemble} suggests these sudden changes in the LR allow the model to explore multiple minima. This exploration of minima could in turn lead to the discovery of wider, less steep minima, which should generalise better to unseen data since changes in the centre of the minima will have less of an effect on the loss than if the minima were steeply sided.
	
\begin{figure}[ht]
		\begin{center}
			\begin{subfigure}[t]{\sfMid\textwidth}
				\begin{center}
					\includegraphics[width=\textwidth]{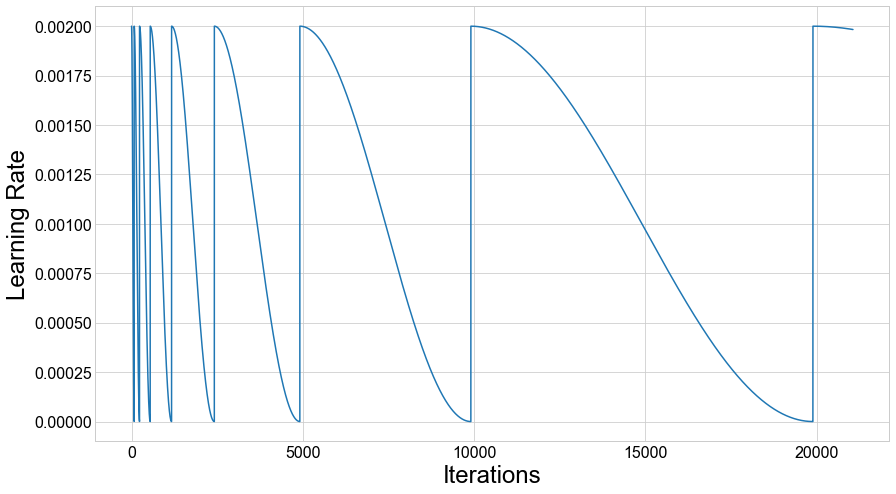}
					\caption{Learning-rate schedule}
				\end{center}
			\end{subfigure}
			\begin{subfigure}[t]{\sfMid\textwidth}
				\begin{center}
					\includegraphics[width=\textwidth]{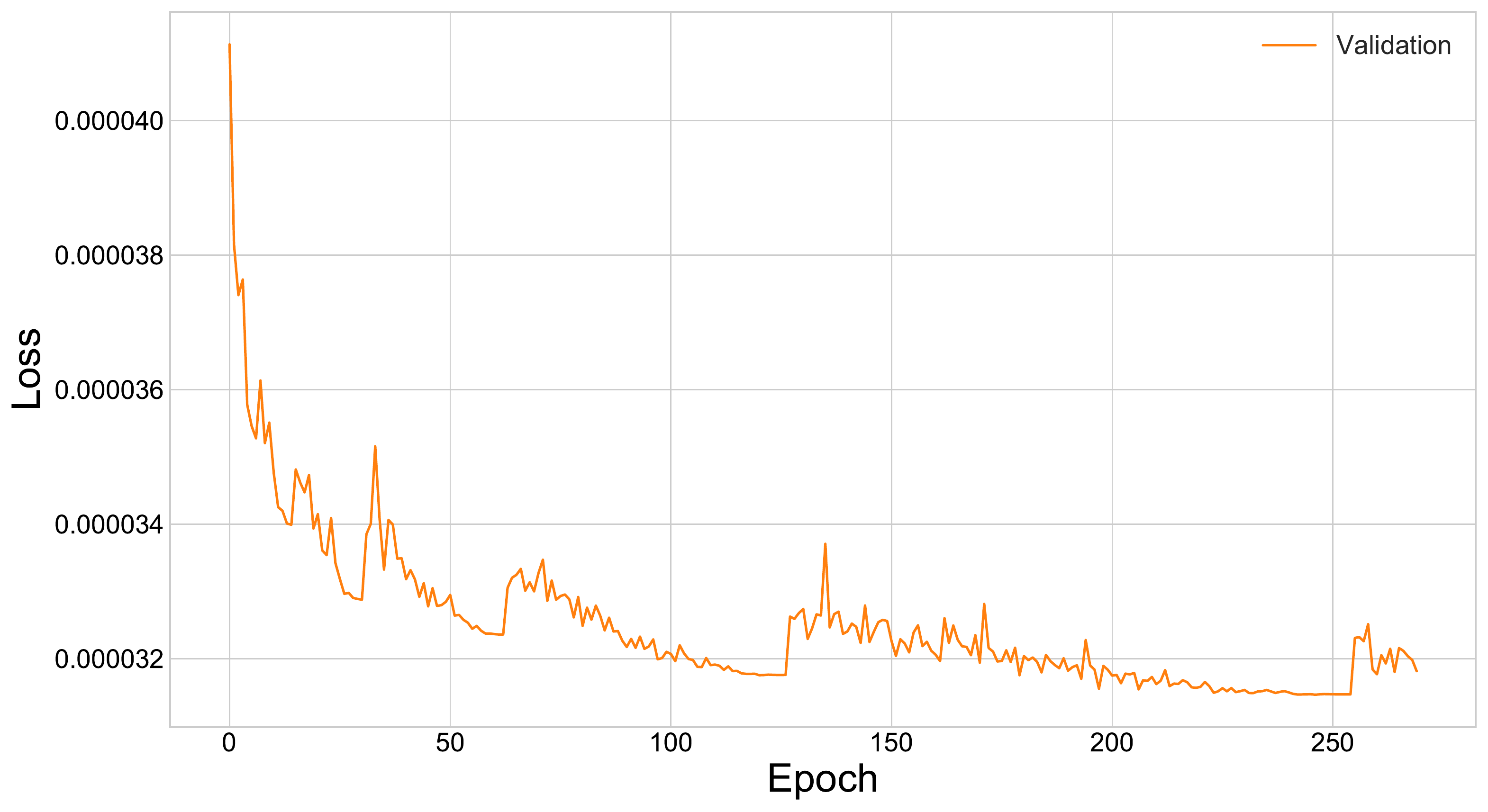}
					\caption{Validation loss evolution}
				\end{center}
			\end{subfigure}
			\caption{Cosine annealed learning-rate schedule with increasing cycle length and example of the corresponding loss evolution for a model trained with such a schedule}
			\label{fig:cosine_lr}
		\end{center}
	\end{figure}

	Figure~\ref{fig:cosine_lr} illustrates the LR schedule and an example of the validation-loss evolution.
	\begin{enumerate}
		\item We begin at an initial LR, $\gamma_0$, of \num{2e-3} and an initial cycle length, $T$, of one complete use of the training folds.
		\item During the cycle, the LR is decreased according to $\gamma_t = 0.5\,\gamma_0\times\left(1+\cos\left(\pi\,t/T\right)\right)$, where $t$ is incremented after each minibatch update
		\item Once the LR would become zero ($t=T$), $t$ is reset to zero, resulting in a sudden increase in the LR (the warm restart). Additionally, the cycle length $T$ is doubled allowing the model to eventually spend longer and longer time at lower LRs, improving convergence.
	\end{enumerate}

	The warm restarts are visible in the validation loss, appearing as small increases in the loss. Initially, the early stopping criterion was modified such that training stopped once the model went one entire cycle without an improvement in the validation loss, however since testing showed models always reached this point in the same annealing cycle, the number of training epochs was instead fixed to avoid unnecessary training time. 
	
	The results (\autoref{tab:solution_evolution} Row ``+SGDR'') show a minor improvement in both MVAC and MAPA, and a decrease in MMVA, indicating a less overly optimistic solution with better generalisation.
		
	\FloatBarrier
		\subsection{Activation function}\label{sec:activation}
	\textit{The inclusion of a newer activation function, which provides slight improvements.}\\

\begin{figure}[ht]
		\begin{center}
			\includegraphics[width=\sfMid\textwidth]{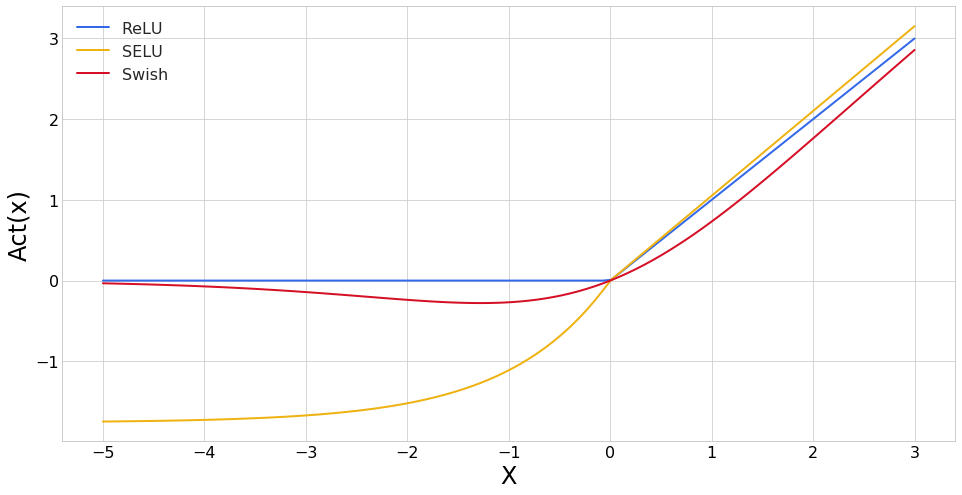}
			\caption{Responses ($Act(x)$) for several activation functions for a range of input values, $x$. PReLU (not shown) is the same as ReLU in the positive region, $Act(x)=x$, and with a constant, non-zero gradient in the negative region, $Act(x)=\alpha x$, where $\alpha$ is a parameter of the model.}
			\label{fig:activations}
		\end{center}
	\end{figure}

	Whilst ReLU is a good default choice for an activation function it does exhibit several problems, which more recent functions attempt to address:
	\begin{itemize}
		\item The Parametrised ReLU (PReLU) \cite{He} is similar to ReLU, but can feature a constant, non-zero gradient for negative inputs, the coefficient of which is learnt via backpropagation during training;
		\item The \textit{Scaled Exponential Linear Unit} (SELU) \cite{selu} uses careful derived scaling coefficients to keep signals in the network approximately unit Gaussian, without the need for manual transformation via batch normalisation \cite{BatchNorm}. It is recommended to use the \texttt{lecun} initialisation scheme \cite{lecun_init} for neurons using the SELU activation function;
		\item The Swish function \cite{swish} was found via a reinforcement leaning based search, and features the interesting characteristic of having a region of negative gradient, allowing outputs to decrease as the input increases: $\mathrm{Swish}\left(x\right)= x\cdot\mathrm{sigmoid}\left(\beta x\right)$, where $\beta$ can either be kept constant or be learnt during training. The recommended weight initialisation scheme for Swish is the same as the one for ReLU, i.e. He.
	\end{itemize}
	A comparison of these functions is shown in \autoref{fig:activations}.
	
	Previous testing, documented in Ref.~\cite{amva_d1.4}, found that the Swish activation function provided superior performance, and so its improvements were again tested here. We use the Swish-1 version, in which $\beta=1$ and is not changed during training. As shown in Row ``+Swish'' of \autoref{tab:solution_evolution}, its inclusion provides slight improvements to MAPA, and so its use was accepted. Note that the small time increase is due to the exponentials in Swish being more complicated to compute than the \texttt{max} operation of ReLU. 
	
	\FloatBarrier
		\subsection{Advanced ensembling}\label{sec:advanced_ensembling}
	\textit{A discussion and experimental test of several more complicated methods of building an ensemble of models. Additionally a new implementation of Stochastic Weight Averaging is introduced which does not rely on a predefined starting point. Whilst this provides slight improvements in training time it is eventually made obsolete by the schedule in \autoref{sec:one_cycle}.}\\

	As mentioned in \autoref{sec:SGDR}, Ref.~\cite{snapshot_ensemble} suggests that Ref.~\cite{SGDR}'s process of restarting the LR schedule allows the training to discover multiple minima in the loss surface. The paper builds on this by introducing a method in which an ensemble of networks results from a single training cycle by saving copies of the networks just before restarting the LR schedule, i.e. when the model is in the minima. This process of \textit{Snapshot Ensembling} (SSE) is further refined into \textit{Fast Geometric Ensembling} (FGE) in Ref.~\cite{FGE}, by forcing the model evolution along curves of constant loss between connected minima (simultaneously and independently discovered by Ref.~\cite{FGE_simul}).
	
	The problem with these methods is that whilst they allow one to train an ensemble in a much reduced time, one still has to run the data through each model individually, so the application-time is still increased. Reference~\cite{SWA} instead finds an approach which leads to an approximation of FGE in a single model. This is done by averaging the models in \textit{weight space}, rather than in \textit{model space}. The general method involves training a model as normal, until the model begins to converge. At this point the model continues to train as normal, but after each epoch a running average of the model's parameters is updated.
	
	In the training, one has to decide on when to begin collecting the averages: \textit{too early}, and the model average is spoiled by ill-performing weights; \textit{too late}, and the model does not explore the loss surface enough to allow SWA to be of use. Additionally, SWA may be combined with a cyclical learning rate, in which case weight averaging should take place at the end of each cycle.
	
	\subsubsection{Implementation}
		\paragraph{SSE}
			SSE was tested by training ten networks as usual, with a cosine-annealed LR (initial LR = \num{2e-3}) with a constant cycle length of 50 folds. Training continued until the validation loss failed to decrease for two complete cycles. Snapshots were then loaded, starting with the best performing set of weights, and then up to four previous snapshots. A weighting function of the form $w=n^{-1}$ was used, where $n$ is the 1-ordered number of weight loads (i.e. best performing weights = 1, first previous snapshot = 2, et cetera). This means that snapshots later in the training are weighted higher than earlier ones, in order to balance the trade off between greater generalisation due to a larger ensemble being used and the poorer performance of earlier snapshots. Additionally snapshots loaded from different model trainings were not reweighted according to validation loss as they were in \autoref{sec:baseline}.
			
		\paragraph{FGE}
			Since FGE expects to send models along curves of near constant loss between minima, it employs a higher frequency saving of snapshots. A linearly cycled LR \cite{Smith_2015} is used, moving from LR = \num{2e-4} to \num{2e-3} and back again over the course of five folds. Training continued until the validation loss had not decreased for nine cycles. Snapshots were loaded in a similar fashion to SSE except up to 8 cycles, as well as the best performing weights, were loaded, and no cycle weighting was used (the losses of the cycles were approximately equal, as expected).
			
		\paragraph{SWA}
			Rather than having to pick a starting point to begin SWA (which would require running the training once beforehand without SWA), SWA was begun early during training and later a second SWA model was started. At set points during training the SWA averages were compared. If the older average performed better, then the younger average was reset to the current model parameters and the time to the next comparison was doubled. If the younger average was better, then the older average was reset to the current model parameters and the time to the next comparison was set back to its initial value (its \textit{renewal period}). Effectively, a range of possible start points for SWA models are tested, and the optimum start position is automatically selected.
			
			Additionally, three LR schedules were tested:
			\begin{itemize}
				\item A constant LR of \num{2e-3} running with a patience of 50 folds with SWA beginning on the fifth fold, and comparisons between averages taking place with an initial separate of five folds.
				\item A Cosine annealed LR between \num{2e-3} and \num{2e-4} over five folds, a patience of 9 cycles, and SWA beginning on the second cycle with a renewal period of two cycles.
				\item A linearly cycled LR between \num{2e-4} and \num{2e-3} over five folds, a patience of 9 cycles, and SWA beginning on the second cycle with a renewal period of two cycles.
			\end{itemize}
		
	\subsubsection{Comparison}
\begin{figure}[ht]
			\begin{center}
				\includegraphics[width=\sfWide\textwidth]{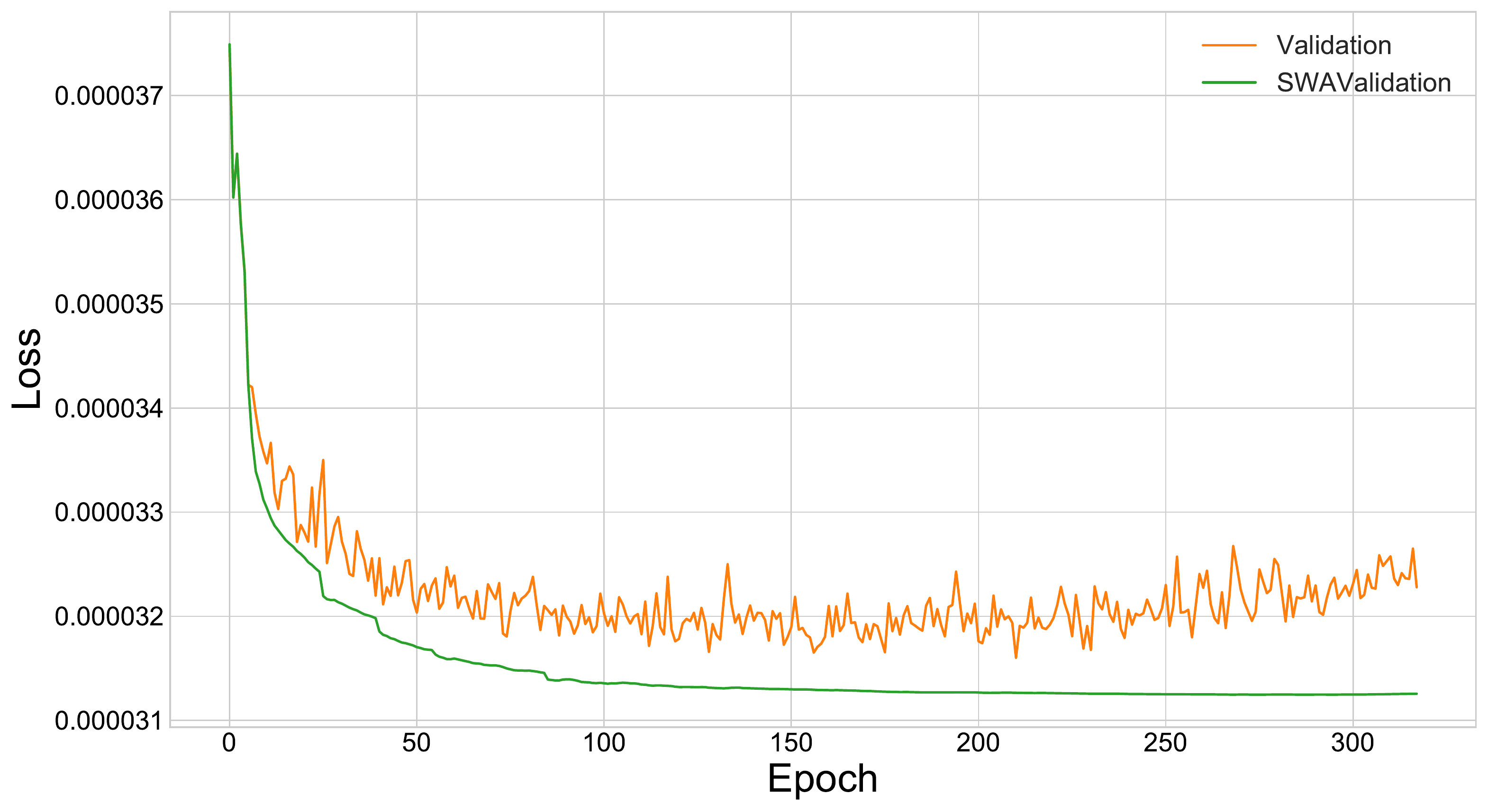}
				\caption{Example evolution of validation loss over training time for both model being trained and the stochastic weight-average of the model states.}
				\label{fig:swa_loss_history}
			\end{center}
		\end{figure}

		Comparing the approaches to the current best solution in \autoref{tab:advanced_ensembling}, we can see that no approach is able to improve the current MAPA, however SWA with a constant LR achieves the same score in a shorter train time. From an example plot of the validation loss over time (\autoref{fig:swa_loss_history}) we can see that SWA not only demonstrates better performance, but it also shows a heavy suppression of loss fluctuations and converges to a loss plateau, where as the non-averaged model eventually over-fits (signified by an increasing loss in the later stages of training). The sharp drops in SWA loss are due to the old average being replaced with the newer average. 
		
		\begin{table}
			\begin{center}
				\begin{tabular}{lccccc}
					\toprule
					Setup & MMVA & MVAC & MAPA & \multicolumn{2}{c}{Fractional time-increase}\\
					& & & & Training & Inference\\
					\midrule
					Current solution & $3.96\pm0.06$ & $3.86\pm0.05$ & $\mathbf{3.81\pm0.02}$ & - & \textbf{-}\\
					SSE & $3.95\pm0.05$ & $3.85\pm0.05$ & $\mathbf{3.81\pm0.02}$ & $0.15\pm0.03$ & $1.5\pm0.3$\\
					FGE & $3.95\pm0.06$ & $3.86\pm0.05$ & $3.80\pm0.02$ & $\mathbf{-0.13\pm0.03}$ & $6\pm1$\\
					\textbf{SWA constant} & $3.96\pm0.06$ & $3.88\pm0.05$ & $\mathbf{3.81\pm0.02}$ & $-0.09\pm0.01$ & \textbf{-}\\
					SWA linear cycle & $\mathbf{3.97\pm0.06}$ & $3.87\pm0.04$ & $\mathbf{3.81\pm0.02}$ & $0.25\pm0.03$ & \textbf{-}\\
					SWA cosine & $3.95\pm0.05$ & $\mathbf{3.89\pm0.05}$ & $\mathbf{3.81\pm0.02}$ & $0.22\pm0.04$ & \textbf{-}\\
					\bottomrule
				\end{tabular}
			\end{center}
			\caption{Comparison of the various advanced ensembling approaches. ``Current solution" refers to the solution as of \autoref{sec:activation}. The best values for each metric are shown in bold, and the setup chosen is also indicated in bold.}
			\label{tab:advanced_ensembling}
		\end{table}
			
	\FloatBarrier
		\subsection{Super-convergence}\label{sec:one_cycle}
	\textit{The application of a schedule for both learning rate and momentum, which provides a significant reduction in training time.}\\
	
\begin{figure}[ht]
		\begin{center}
			\includegraphics[width=\sfMid\textwidth]{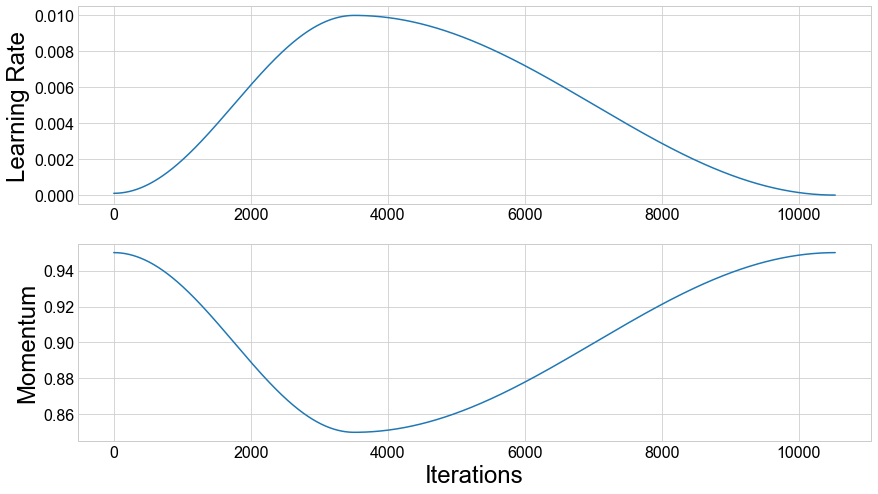}
			\caption{Illustration of the evolution schedule of the learning rate and momentum ($\beta_1$) used during testing. Iterations are minibatch updates.}
			\label{fig:onecycle}
		\end{center}
	\end{figure}

	Reference~ \cite{Smith_2017} introduces the concept of \textit{super convergence}, in which a specific LR schedule, \textit{1cycle}, is used to achieve convergence much quicker (between five and ten times) than traditional schedules allow. This is further discussed in Ref.~\cite{Smith_2018}. The 1cycle policy combines both cyclical LR and cyclical momentum \footnote{Momentum here refers to optimiser momentum and not physical momentum.} (or $\beta_1$ in case of ADAM), and evolves both hyper-parameters simultaneously and continuously from their starting values to another value and then back again over the course of training in a single cycle. Hyper-parameter evolution takes place in opposite directions, with the LR initially increasing and the momentum initially decreasing, i.e. when the LR is at its maximum, the momentum is at its minimum, as illustrated in \autoref{fig:onecycle}. In this way the two parameters help to balance each other, allowing the use of higher learning rates without the network becoming unstable and diverging. In the final stages of training, the LR is allowed to decrease several orders of magnitude below its initial value.

	One of the key points of Ref.~\cite{Smith_2018} is that super-convergence can only occur if the amount of regularisation \footnote{Techniques, or aspects of the data, model, or task, which limit overfitting of the model to the training data or aim to improve the score on the testing dataset. See Ref.~\cite{regularisation_taxonomy} for an in depth discussion.} present in the setup is below a certain amount, but that once super-convergence can be achieved, other forms of regularisation can then be tuned accordingly. In our current setup, we have avoided explicitly adding tuneable forms of regularisation \footnote{Such as $L_2$~\cite{l2} or Dropout~\cite{dropout}.} for this reason. However \autoref{sec:data_symmetry}'s choice to use data augmentation rather than data fixing equates to choosing not to reduce regularisation through data complexity. Because of this, we will be conservative in our testing of super-convergence and attempt to only halve our training time.
	
	Implementation of 1cycle follows the suggestion of fast.ai \cite{fastai_docs}, in which the halves of the cycles are half-cosine functions, rather than linear interpolations. This can be expected to offer the same advantages of the cosine annealing schedule of Ref.~\cite{SGDR} and also provide a smooth transition between the directions of parameter evolution. The total length of the cycle was set to 135 folds (approximately half that required for the solutions of \autoref{sec:SGDR} and \autoref{sec:advanced_ensembling}). The lengths of the first and second parts of the cycle were set to 45 folds and 90 folds, respectively, i.e. a ratio of 1:2, allowing a longer time for final convergence. Based on a LR range test performed at a constant $\beta_1$ of 0.9 and the knowledge that $\beta_1$ would be at a minimum when the LR is maximised, a slightly higher maximum LR of \num{1e-2} and an initial value of \num{1e-4} were chosen.
	
	As shown in \autoref{fig:onecycle}, during the first part of the cycle, the LR is increase from its initial value (\num{1e-4}) to its maximum of \num{1e-2} while $\beta_1$ is decreased from 0.95 to 0.85. In the second half, $\beta_1$ returns back up to 0.95, but the LR is allowed to tend to zero. As shown in Row ``-SWA +1cycle'' of \autoref{tab:solution_evolution}, although 1cycle provides the same level of MAPA performance as SWA, it is able to reach this level of performance in half the time.

		
	\FloatBarrier
		\subsection{Densely connected networks}\label{sec:densenet}
	\textit{An investigation into different connections between layers, which is found to provide a mild improvement in performance whilst reducing the number of free parameters in the networks.}\\

	The fully-connected architectures used so far aim to learn a better representation of the data at each layer, however this means that potentially useful information is lost after each layer if it has not been sufficiently well represented by the latest layer. Being able to do so requires that the layers have a sufficient number of free parameters to avoid losing useful information. Due to limited training data, this limits the depth of the networks which can be trained and so potentially limits the performance of the model.

\begin{figure}[ht]
		\begin{center}
			\includegraphics[width=0.4\textwidth]{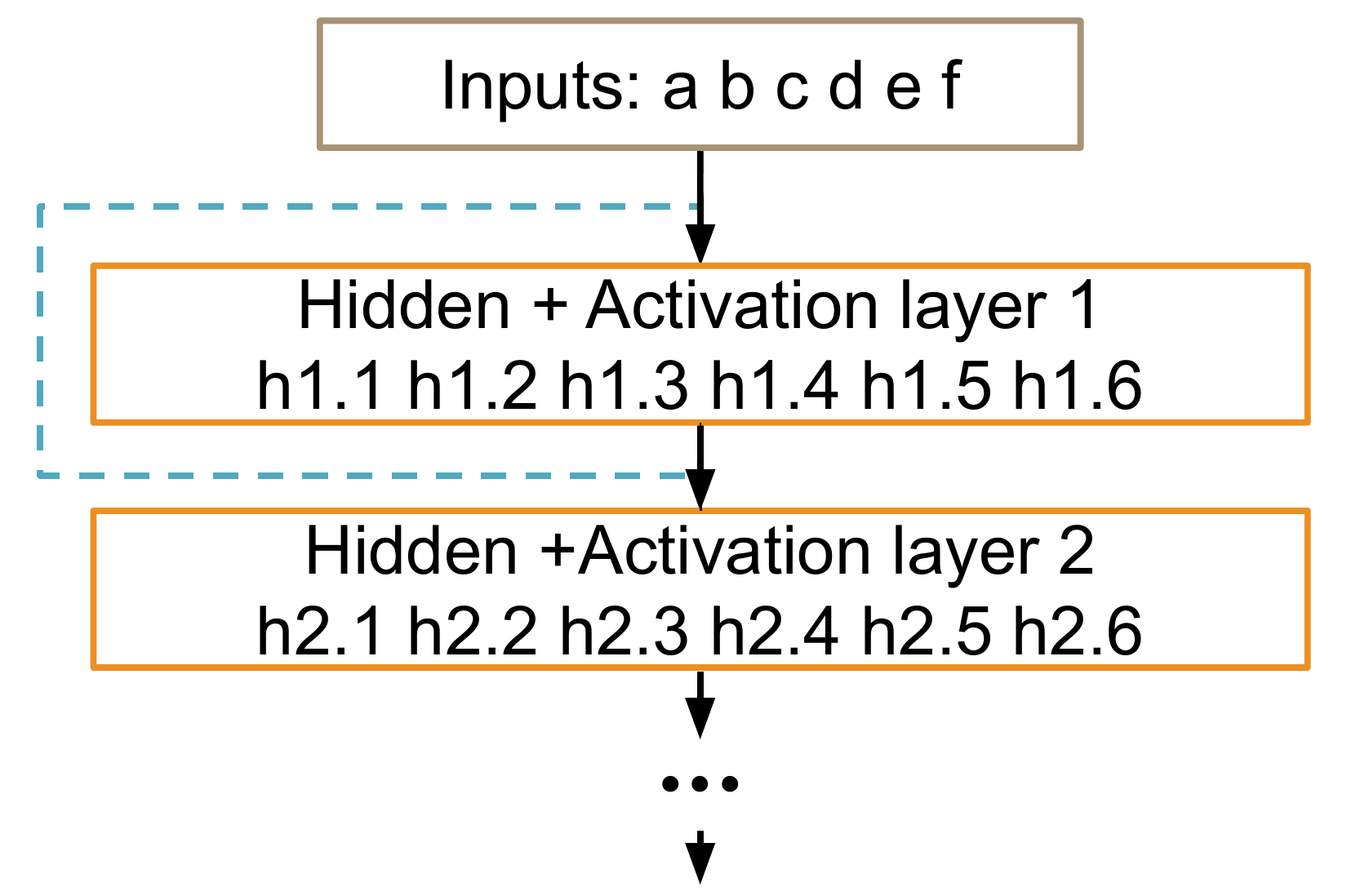}
			\caption{A hypothetical scenario in which it is optimal for hidden layer 2 to have access to both a learned representation of the inputs from hidden layer 1 and one or more of the original inputs. Hidden layer 1 must now learn an identity representation of the required inputs in order for hidden layer 2 to have access to the information it requires. A more efficient approach is for hidden layer 2 to access both the hidden state from layer 1 and the original inputs, as indicated by the dashed line.}
			\label{fig:identity_maps}
		\end{center}
	\end{figure}

	Consider an example as illustrated in \autoref{fig:identity_maps} for a set of inputs $a,b,c,d,e,f$ in the case in which it is optimal to learn first a new feature based on inputs $a,b,c,d,e$ using hidden layer 1 (e.g. the angle between the tau and the electron or muon) and then use hidden layer 2 to combine this new feature with input $f$ (e.g. the angle multiplied by the amount of missing transverse energy). Hidden layer 1 must now learn either an identity representation of input $f$ or a monotonic transformation of it, which requires ``budget'' from both the available free parameters and the amount of training data. It would be more efficient if instead hidden layer 2 were to have direct access to input $f$. Consider further the case in which it is hidden layer 3 that requires access to input $f$, now both hidden layer 1 and 2 must learn identity representations of $f$.

	In their 2016 paper, ``Densely Connected Convolutional Networks" \cite{densenet}, Huang and Weinberger present an architecture in which, within subgroups of layers, the outputs of all previous layers are fed as inputs into all subsequent layers via channel-wise concatenation. This means that lower-level representations of the data are directly accessible by all parts of the block. This potentially allows both a more diverse range of features to be learnt, and for layers to be trained via \textit{deep supervision} \cite{deep_supervision} due to more direct connections when back-propagating the loss gradient. Additionally, it means that the weights which were previously required to encode the low-level information may no longer be necessary.
	
\begin{figure}[ht]
		\begin{center}
			\includegraphics[width=0.25\textwidth]{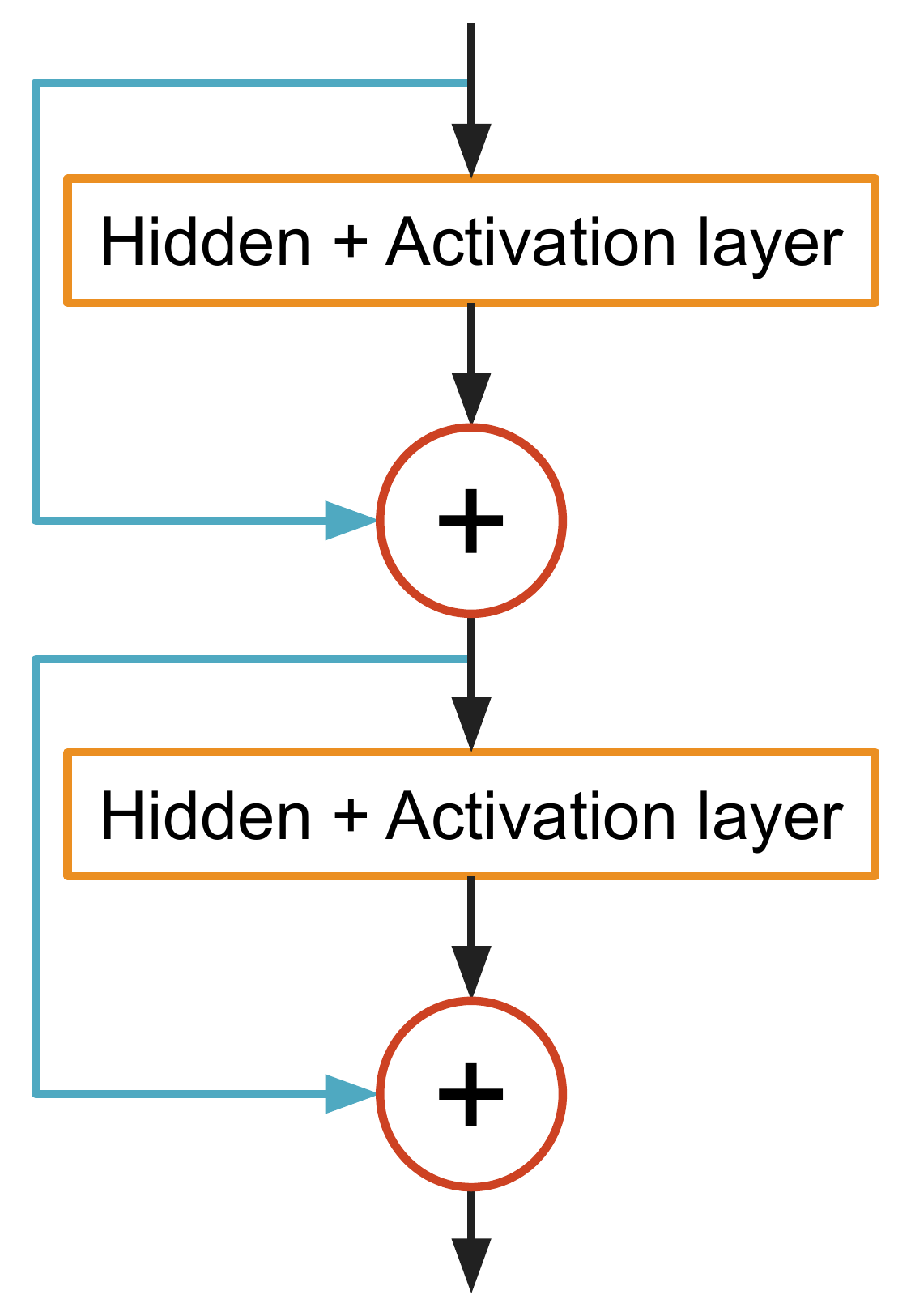}
			\caption{Illustration of a densely-connected, non-convolutional network. $\oplus$ indicates concatenations of hidden states.}
			\label{fig:dense_connections}
		\end{center}
	\end{figure}

	Whilst the paper presents this dense connection in terms channel-wise concatenation of the outputs of convolutional filters, the same idea can be applied to fully connected networks by concatenating width-wise the output tensor of each linear layer with the that layer's input tensor; i.e. $x_{i+1} = D_i\!\left(x_i\right) \oplus x_i$, where $x_i$ is the hidden state before the $i^{\text{th}}$ linear layer ($D_i$), as illustrated in \autoref{fig:dense_connections}.
	
\begin{figure}[ht]
		\begin{center}
			\includegraphics[width=\sfMid\textwidth]{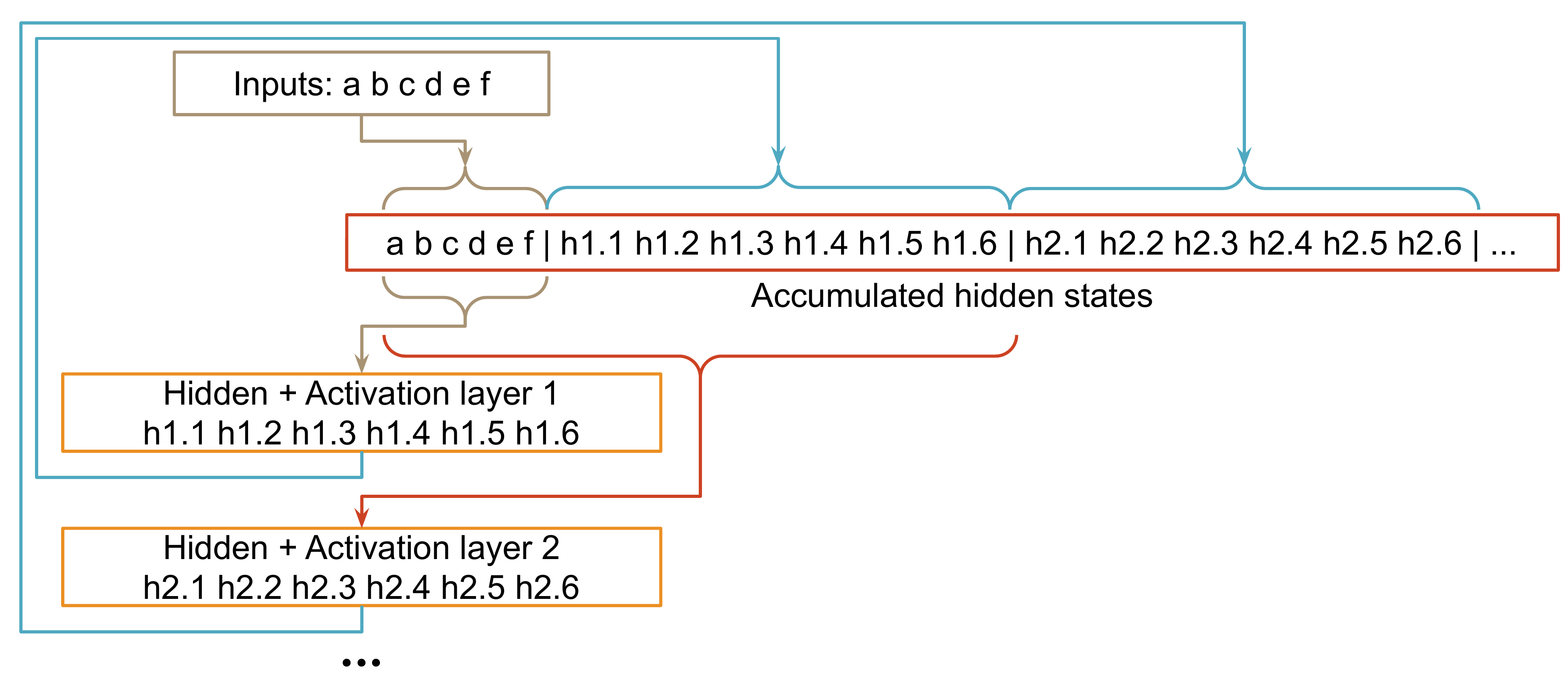}
			\caption{An alternative visualisation of a densely-connected network in which hidden layers append to an ever-growing accumulation of hidden states, taking as input the current the current version of the accumulation.}
			\label{fig:identity_maps_dense}
		\end{center}
	\end{figure}

	This concept can also be thought of as accumulating a tensor of hidden states, as illustrated in \autoref{fig:identity_maps_dense}, in which each hidden layer takes as input the current state of the accumulated tensor and then concatenates its output to the tensor. In this way every subsequent hidden layer has direct access to all previous representations of the inputs, meaning that no identity representations need to be learnt and hidden layers are more easily optimised due to the more direct paths for the loss gradient during back-propagation. As discussed in Ref.~\cite{step_decay}, in the context or residual connections, such ``skip connections'' also help mitigate the effects of over-parameterising a model; rather than forcing redundant layers to learn identity mappings, they can instead be skipped (ignored) or pushed towards a zero output. This can be expected to offer some level of leniency when choosing the number of hidden layers to used in the network.

	\subsubsection{Testing}\label{sec:dense_test}
\begin{figure}[ht]
			\begin{center}
				\includegraphics[width=0.4\textwidth]{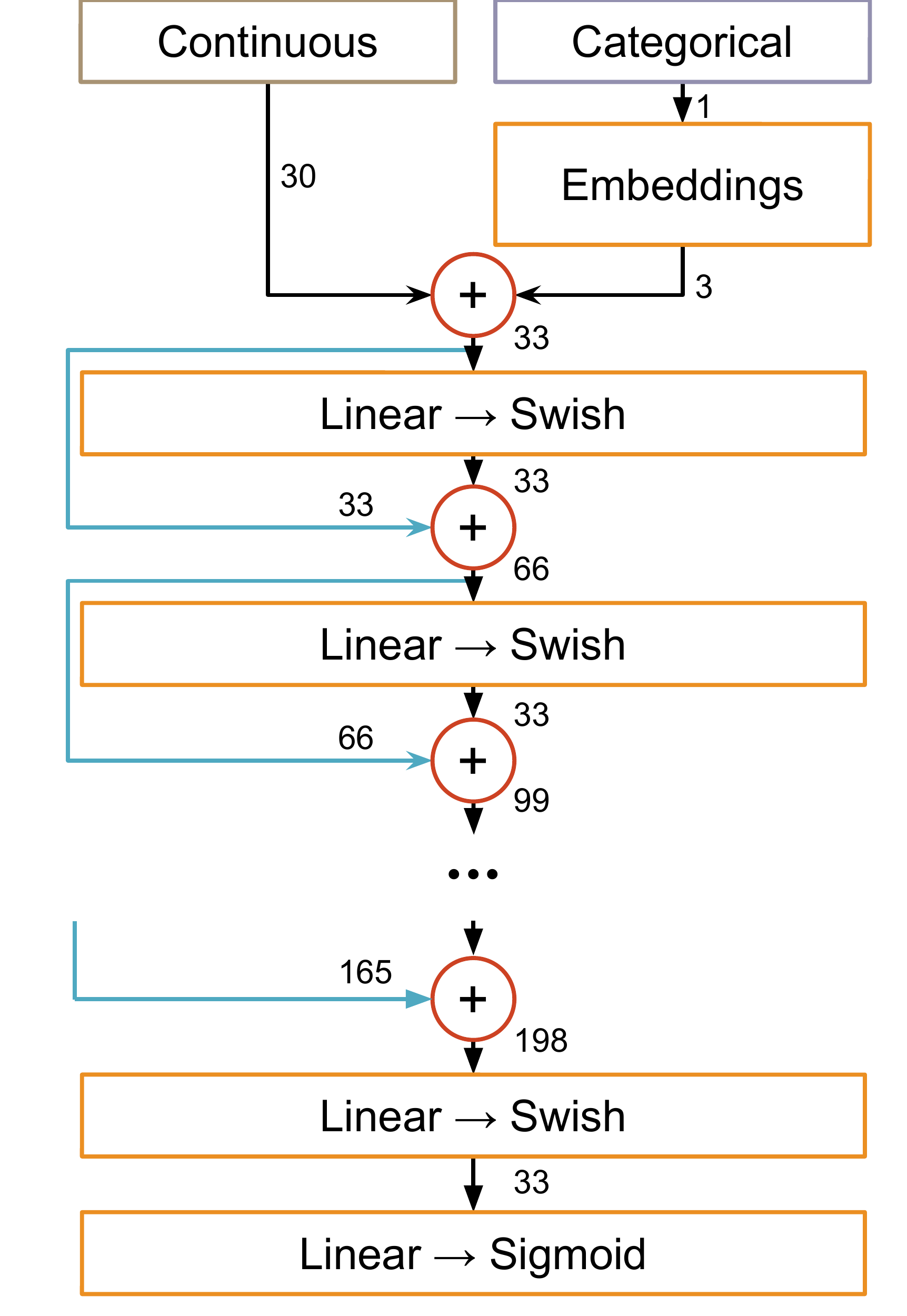}
				\caption{Illustration of the architecture described in \autoref{sec:dense_test}. $\oplus$ indicates concatenations of hidden states and the numbers show the widths of the hidden states at each point. A total of six linear layers are used, each with 33 neurons. Two of these layers are compacted into ``\textbf{\ldots}". As a point of reference, this is the architecture that is used to get the final results of this paper.}
				\label{fig:densenet}
			\end{center}
		\end{figure}

		The architecture up to now has used four hidden layers, each with 100 neurons. Including the embedding and output layers, the total number of free parameters is \num{33813} \footnote{Computable thusly: Embedding matrix $(4\times3)=12$ + first fully-connected hidden layer of 100 neurons and 33 inputs $((33\times100)+100=\num{3400})$ + three subsequent 100 neuron fully-connected hidden layers $(3\times((100\times100)+100)=\num{30300})$ + single-neuron output layer $(100+1)=101$. \num{33813} parameters total.}. For the densely connected architecture we reduce the width of linear layers to 33 (the number of continuous inputs plus the width of the categorical embedding output), and increase the depth of the network to six hidden layers. The outputs of each hidden layer except the last are concatenated with their inputs to compute the input tensor to the next hidden layer. This results in a total of \num{23113} free parameters \footnote{Computable thusly: Embedding matrix $(4\times3)=12$ + six fully-connected hidden layers of 33 neurons with input multiplicities scaling according to $33\times n$: $\sum_{n=1}^6\left[((33\times n)\times33)+33\right]=23067$ + single-neuron output layer $(33+1)=34$. \num{23113} parameters total.}, i.e. about two thirds of the original size. This architecture is illustrated in \autoref{fig:densenet}.

		Training of the networks used the same 1cycle schedule as presented in \autoref{sec:one_cycle}. As shown in Row ``+Dense'' of \autoref{tab:solution_evolution}, the densely connected architecture provides mild improvements in both MVAC and MAPA, despite being significantly smaller, however due to the extra depth and concatenation operations, training and inference timing is approximately the same as the previous architecture.
		
		Unfortunately, a mistake entered into the code during this experiment that led to the ensemble of trained models being uniformly weighted, rather than performance weighted (as per Sec.~\ref{sec:weighted_ensemble}) as intended. This mistake was only discovered during an ablation study performed at the end of the analysis after the private AMS values had been computed. In order to reflect the information seen during the course of the study, and not to draw conclusions from the private AMS, \autoref{tab:solution_evolution} shows the performance of the uniformly-weighted ensemble, which achieves a MAPA of $3.82\pm0.02$. A retraining using performance-weighting achieves a MAPA of $3.81\pm0.01$ (comparable with the $3.81\pm0.02$ achieved by the solution of \autoref{sec:one_cycle}), which might have caused us not to use dense connections: use of dense connections is slightly slower to train, but the slight reduction in variance of performance would have been seen as favourable. This mistake remained in the code throughout \autoref{sec:arch_opt} and App.~\ref{sec:widedeep}, meaning that the comparisons of metrics to what was seen in this section are still valid.
		
	\FloatBarrier
		\subsection{Architecture optimisation}\label{sec:arch_opt}
	In this section we perform a search over the remaining hyper-parameters to finalise our solution. Interestingly, though, the existing architecture is found to work best and the model displays only a weak dependence on its hyper-parameters.

	\subsubsection{Overview of remaining hyper-parameters}
		The choices made so far equate to rough optimisations of hyper-parameters and training-method, but the underlying architecture of the network has remained relatively constant. For the final stage of solution optimisation, we explore the impact of adjusting the remaining hyper-parameters: number of hidden layers (depth), number of neurons per hidden layer (width), dropout rates (DO)  \cite{dropout}, and the amount of $L_2$ regularisation \cite{l2}.
		
		In order to reduce the dimensions of the architecture space the following choices were made:
		\begin{itemize}
			\item Due to the depth of networks explored, batch normalisation \cite{BatchNorm} was not considered.
			\item Having already performed feature selection (\autoref{sec:feature_selection}), $L_1$ regularisation \cite{l1}, which encourages feature sparsity, was not considered.
			\item Rather than adjusting the widths of each layer individually, the widths were allowed to increase or decrease at a constant rate according to the depth of each layer (the \textit{growth rate}), i.e. $w_d=w(1+dg)$, where $w_d$ is the width of the hidden layer at depth $d$ (zero-ordered), $w$ is the nominal layer width, and $g$ is the growth rate. A growth rate of zero corresponds to constant widths for all layers.
			\item Dropout layers, when used, were placed after every activation function expect the output, and a single rate was used for all dropout layers. 
			\item The same value of $L_2$ was used for all weight updates.
			\item Training will use the 1cycle schedule with the same settings used in \autoref{sec:one_cycle}. 
			\item All hidden layers are densely connected as per \autoref{sec:densenet}.
		\end{itemize}
		Whilst other forms of layer-width growth (quadratic, cubic, \textit{et cetera}) could be considered, we assume that the main issue is to have \textit{some} ability to change the layer widths, rather than the exact parametric form; linear growth offers the simplest form, introducing only one extra hyper-parameter. It should also be mentioned that in initial testing of various architectures, the optimal learning rates were all found to be at about the same point ($\num{1e-2}$), so the same LR was used for all parameter sampling, in order to reduce the search space by one dimension. The LR can of course be adjusted later for any promising parameter points found.
		
	\subsubsection{Parameter scan}
		While some choices have allowed a reduction of the number of dimensions of the parameter search space, the remaining space still has five dimensions (depth, width, dropout rate, $L_2$, and growth rate), meaning a full grid-search of all possible combinations is unfeasible; instead a random search will be used. For each architecture, the following parameter-sampling rules were used:
		\begin{itemize}
			\item Depth sampled from $[2,9)\in\mathbb{Z}$
			\item Width sampled from $[33,101)\in\mathbb{Z}$
			\item Growth rate sampled from $[-0.2,1)\in\mathbb{R}$
			\item $L_2$ sampled from $\{0,\num{e-2},\num{e-3},\num{e-4},\num{e-5},\num{e-6}\}\times\num{e-5}$
			\item Dropout rate sampled from $\{0,0.05,0.1,0.25,0.5\}$
			\item The total number of free parameters must be in the range $(\num{5e3},\num{1e5})$
			\item In the case of negative growth rate, the width of the last hidden layer before the output layer must be greater than one.
		\end{itemize}
		
\begin{figure}[ht]
			\begin{center}
				\includegraphics[width=\sfMid\textwidth]{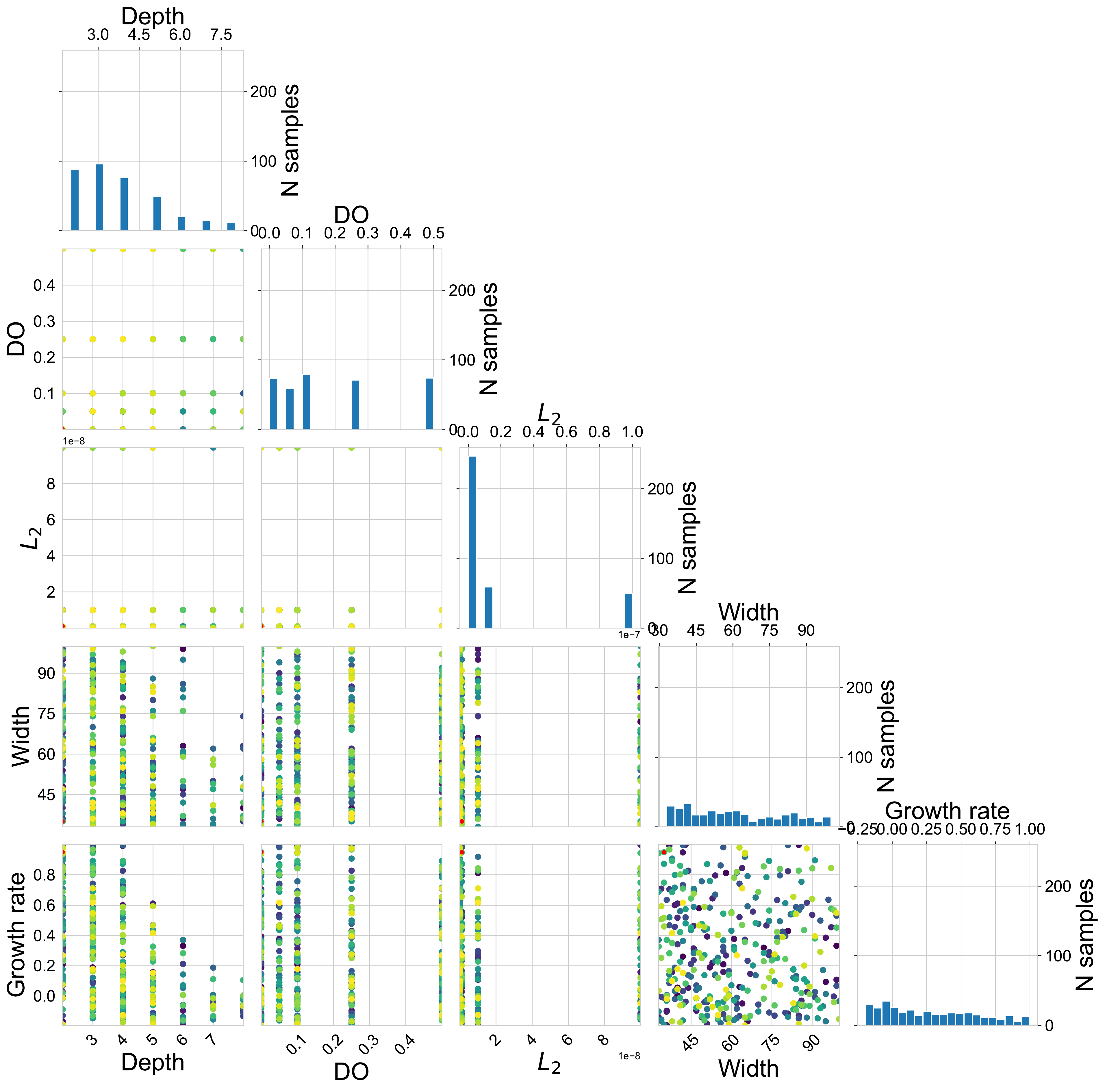}
				\caption{Illustration of the parameter sets sampled during testing.}
				\label{fig:arch_opt_evals}
			\end{center}
		\end{figure}

\begin{figure}[ht]
			\begin{center}
				\includegraphics[width=\sfMid\textwidth]{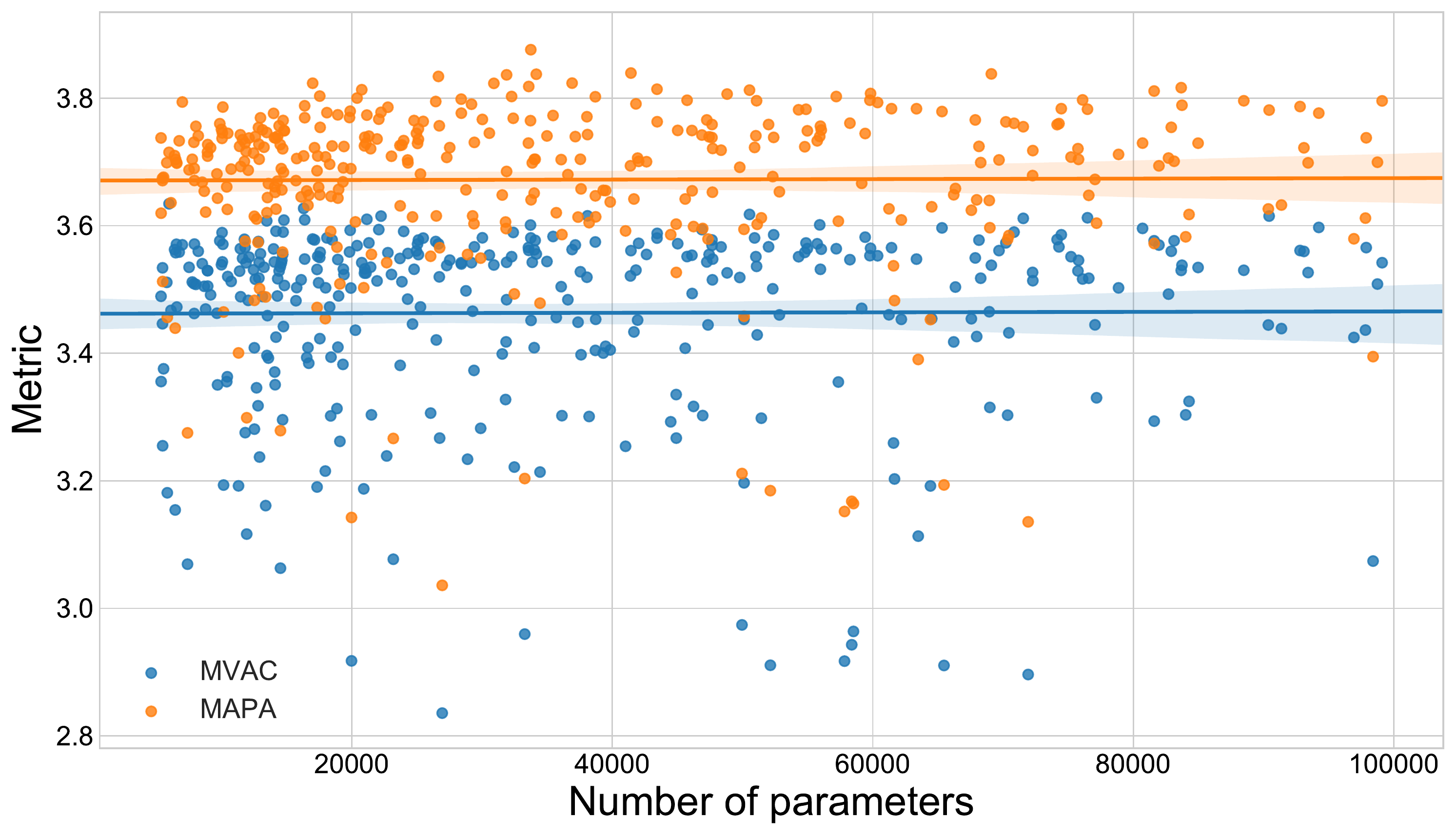}
				\caption{Dependence of the optimisation metrics on the total number of free parameters of the models sampled during testing.}
				\label{fig:arch_opt_complexity}
			\end{center}
		\end{figure}

		Just over 350 parameter sets were tested. Each test consisted of training an ensemble of three networks with the specified architecture. The MVAC and MAPA were then calculated and recorded. Figure~\ref{fig:arch_opt_evals} illustrates the distributions of the sampled parameters. Figure~\ref{fig:arch_opt_complexity} shows MAPA and MVAC as a function of the total number of free parameters in the networks. From the flat linear-fit, we can see that performance is unrelated to model complexity, and that whilst the data and training methodologies are sufficient for optimising large models, such models are unnecessary to reach  better performance. This is possibly due to the leniency on over-parameterised networks discussed in \autoref{sec:densenet}.
		
\begin{figure}[ht]
			\begin{center}
				\begin{subfigure}[t]{0.68\textwidth}
					\begin{center}
						\includegraphics[width=\textwidth]{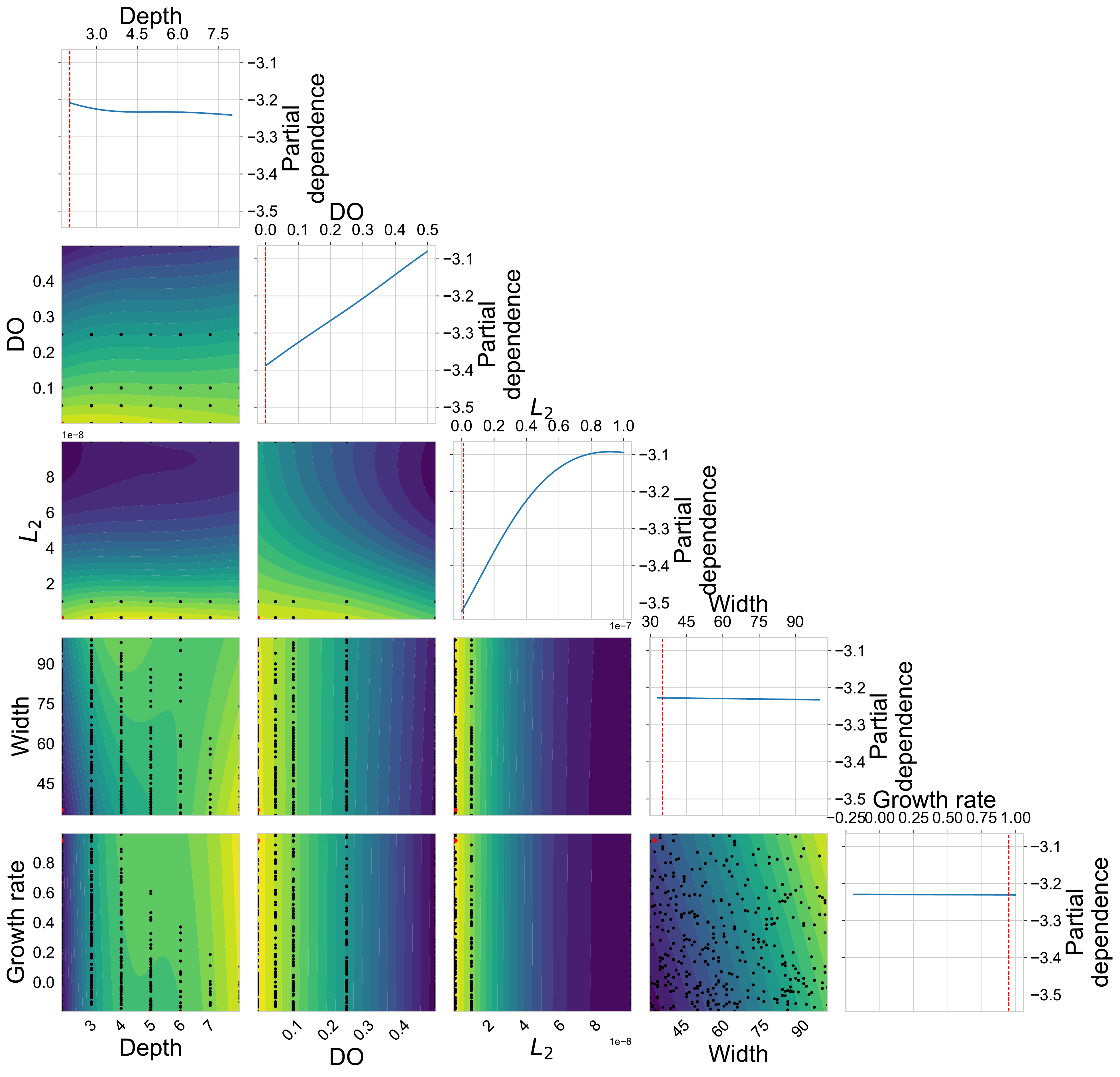}
						\caption{MVAC}
					\end{center}
				\end{subfigure}
				\begin{subfigure}[t]{0.68\textwidth}
					\begin{center}
						\includegraphics[width=\textwidth]{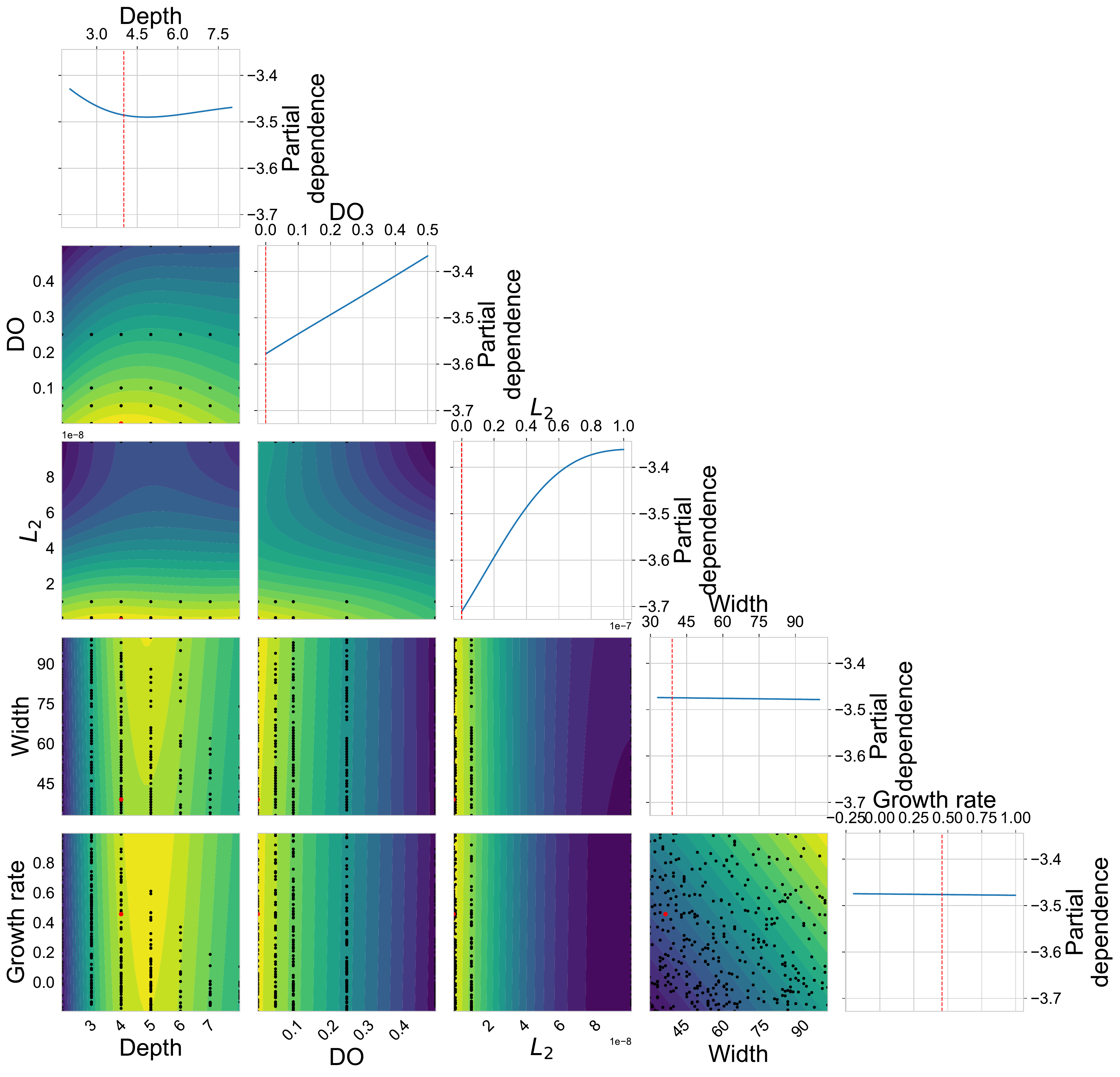}
						\caption{MAPA}
					\end{center}
				\end{subfigure}
				\caption{Partial dependencies of negative MVAC and negative MAPA on network hyper-parameters. Red indicates the position of the best parameter set found during sampling.}
				\label{fig:arch_opt}
			\end{center}
		\end{figure}

		The testing results were then used to fit Gaussian process models in order to evaluate the partial dependencies of the metrics on the hyper-parameters. Figure~\ref{fig:arch_opt} illustrates the one- and two-dimensional partial dependencies of MVAC and MAPA on the hyper-parameters. Note that the machinery used expects a minimisation problem, hence why the negatives of the metrics are shown.

		From the results, the trends in the partial dependencies of both MVAC and MAPA appear to favour minimal explicit regularisation. In terms of depth and width, MVAC shows weak dependence on width and growth rate, and slight dependence on depth, with networks with fewer than four hidden layers performing poorly compared to deeper networks. MAPA again shows weak dependence on width and growth rate, but demonstrates a bounded minimum for depth, favouring networks with around five hidden layers.
		
	\subsubsection{Further testing}
		Based on the scan results and partial dependencies, a number of promising parameter sets were further explored with complete trainings, and the LRs set using LR range tests. The performances of these architectures are compared to the current solution (depth 6, width 33, growth rate 0, and no dropout or weight decay) in \autoref{tab:architecture_search}. As can be seen, none of the new architectures improve on the current architecture in terms of MAPA and although some are able to reach the same level of performance they do not provide significant reductions in terms of training or inference time.
		
		\begin{table}
			\begin{center}
				\begin{tabular}{lccccc}
					\toprule
					Setup & MMVA & MVAC & MAPA & \multicolumn{2}{c}{Fractional time-increase}\\
					& & & & Training & Inference\\
					\midrule
					\textbf{Current} & $3.95\pm0.04$ & $3.89\pm0.05$ & $\mathbf{3.82\pm0.02}$ & - & - \\
					4 100 0.1 & $3.9\pm0.05$ & $3.83\pm0.05$ & $3.74\pm0.01$ & $0.23\pm0.09$ & $0.6\pm0.1$\\
					4 40 0.45 & $3.96\pm0.05$ & $3.90\pm0.05$ & $\mathbf{3.82\pm0.01}$ & $\mathbf{-0.02\pm0.05}$ & $\mathbf{-0.01\pm0.03}$\\
					5 100 -0.1 & $\mathbf{4.03\pm0.05}$ & $\mathbf{3.91\pm0.05}$ & $3.80\pm0.02$ & $0.15\pm0.07$ & $0.6\pm0.1$\\
					5 33 0.9 & $3.95\pm0.05$ & $3.89\pm0.04$ & $3.81\pm0.02$ & $0.19\pm0.08$ & $0.6\pm0.2$\\
					8 33  -0.2 & $3.95\pm0.06$ & $3.89\pm0.06$ & $\mathbf{3.82\pm0.02}$ & $0.4\pm0.1$ & $1.2\pm0.4$\\
					\bottomrule
				\end{tabular}
			\end{center}
			\caption{Comparison of the various explored architectures. ``Current" refers to the solution as of \autoref{sec:densenet}. Architectures names are of the form: [depth] [width] [growth rate]. The best values for each metric are shown in bold, and the setup chosen is also indicated in bold.}
			\label{tab:architecture_search}
		\end{table}
	
\FloatBarrier
	\section{Solution evaluation}\label{sec:testing}
	\subsection{Test results}
\begin{figure}[ht]
			\begin{center}
				\includegraphics[width=\sfWide\textwidth]{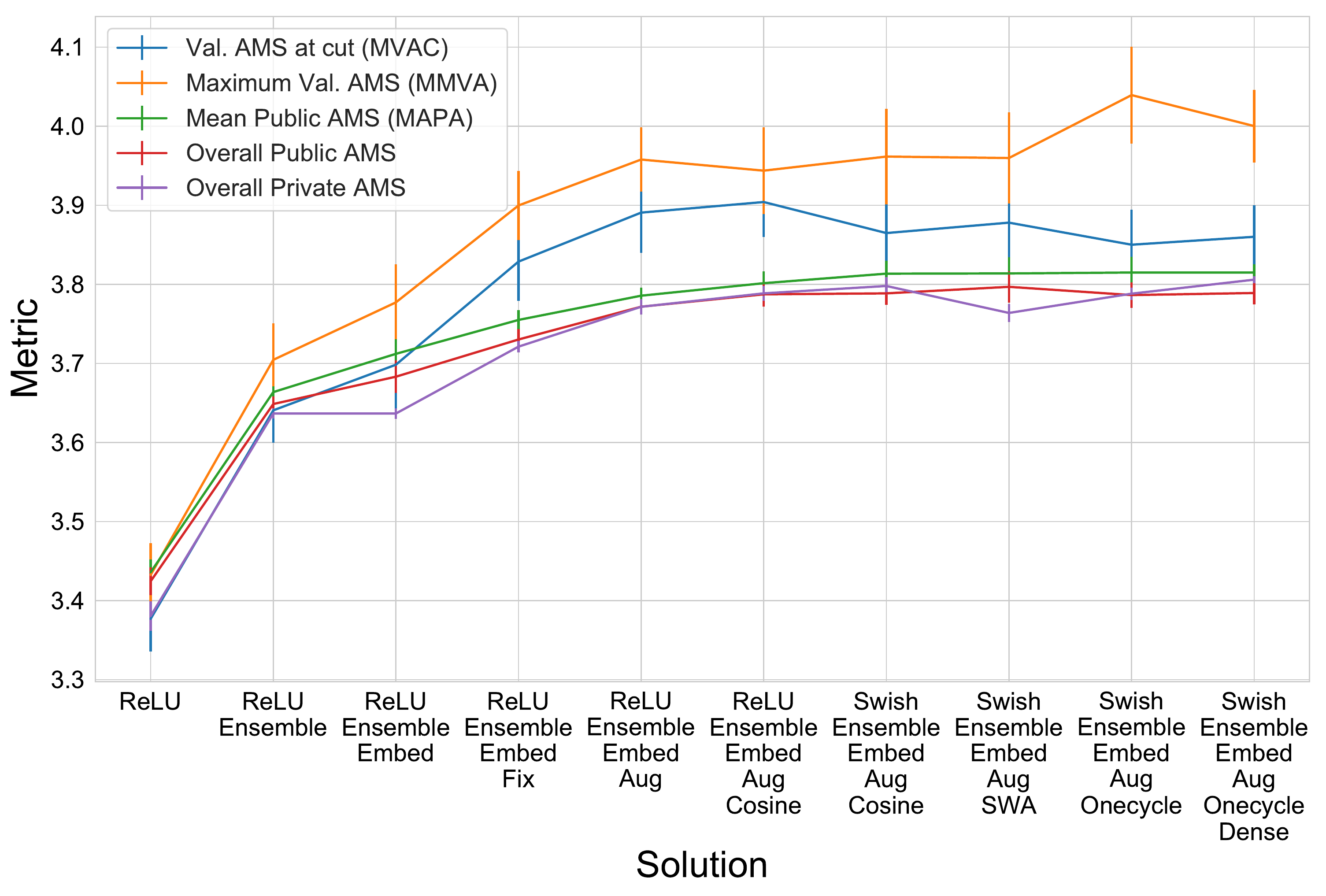}
				\caption{Illustration of the performance metrics for each solution developed. The final model shows the metrics for the weighted ensemble, and not the uniform ensemble that was used during model development (see \autoref{sec:weighted_ensemble} for more details).}
				\label{fig:ams_evo}
			\end{center}
		\end{figure}

		Having built up a solution using several optimisation metrics and timing changes as guides, we now come to checking the performance of the solution using the remaining two metrics: The overall AMS on the public and private portions of the testing data (reminder: MAPA was only an average over subsamples of the public portion of the testing data). In \autoref{fig:ams_evo} we show all metrics as a function of solutions developed, averaged over the six trainings that were run per solution. From the results, we can see that whilst MAPA consistently over-estimated actual performance, its general trend is nonetheless representative of changes in the public and private AMS.
		
	\subsection{Solution timing}
\begin{figure}[ht]
			\begin{center}
				\begin{subfigure}[t]{\sfMid\textwidth}
					\begin{center}
						\includegraphics[width=\textwidth]{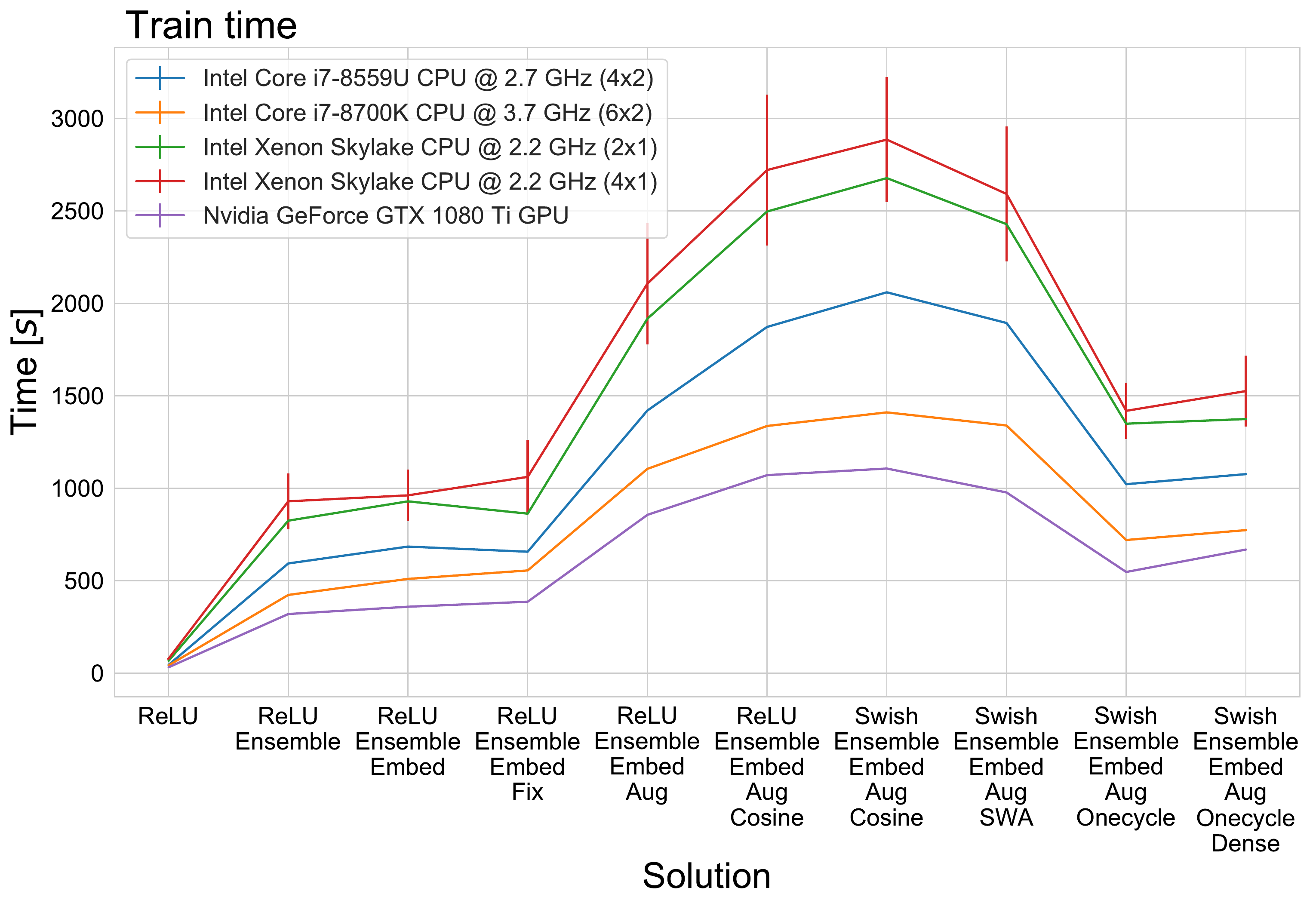}
						\caption{Absolute train-time}
					\end{center}
				\end{subfigure}
				\begin{subfigure}[t]{\sfMid\textwidth}
					\begin{center}
						\includegraphics[width=\textwidth]{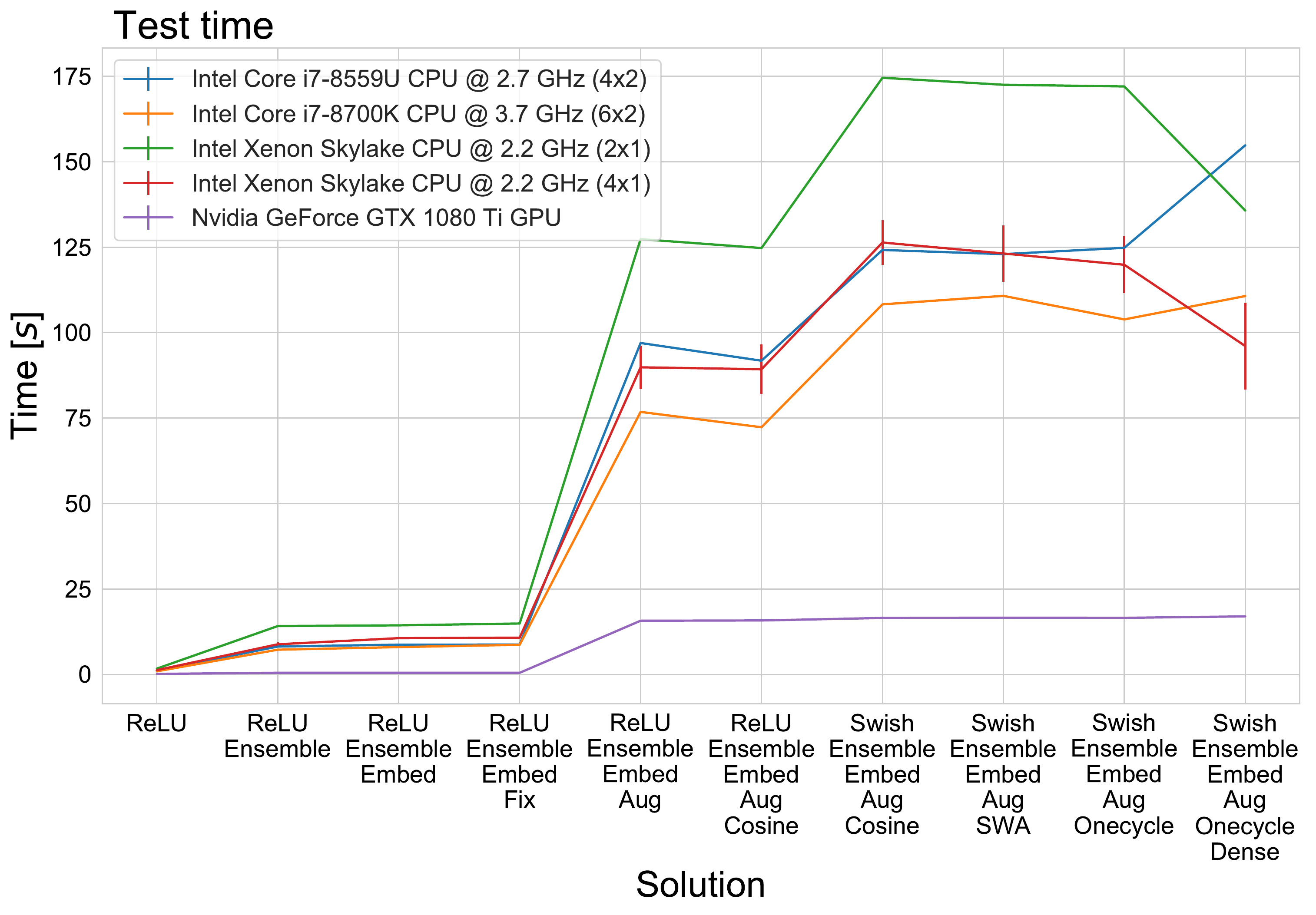}
						\caption{Absolute test-time}
					\end{center}
				\end{subfigure}
				\caption{Absolute (wall-clock) timings for solutions on a variety of hardware. Note that the (4x1) Xenon line is an average of two machines, and hence has error bars; only single machines were available for the other hardware configurations and so their associated timing uncertainties cannot be evaluated.}
				\label{fig:abs_sol_timings}
			\end{center}
		\end{figure}

\begin{figure}[ht]
			\begin{center}
				\begin{subfigure}[t]{\sfMid\textwidth}
					\begin{center}
						\includegraphics[width=\textwidth]{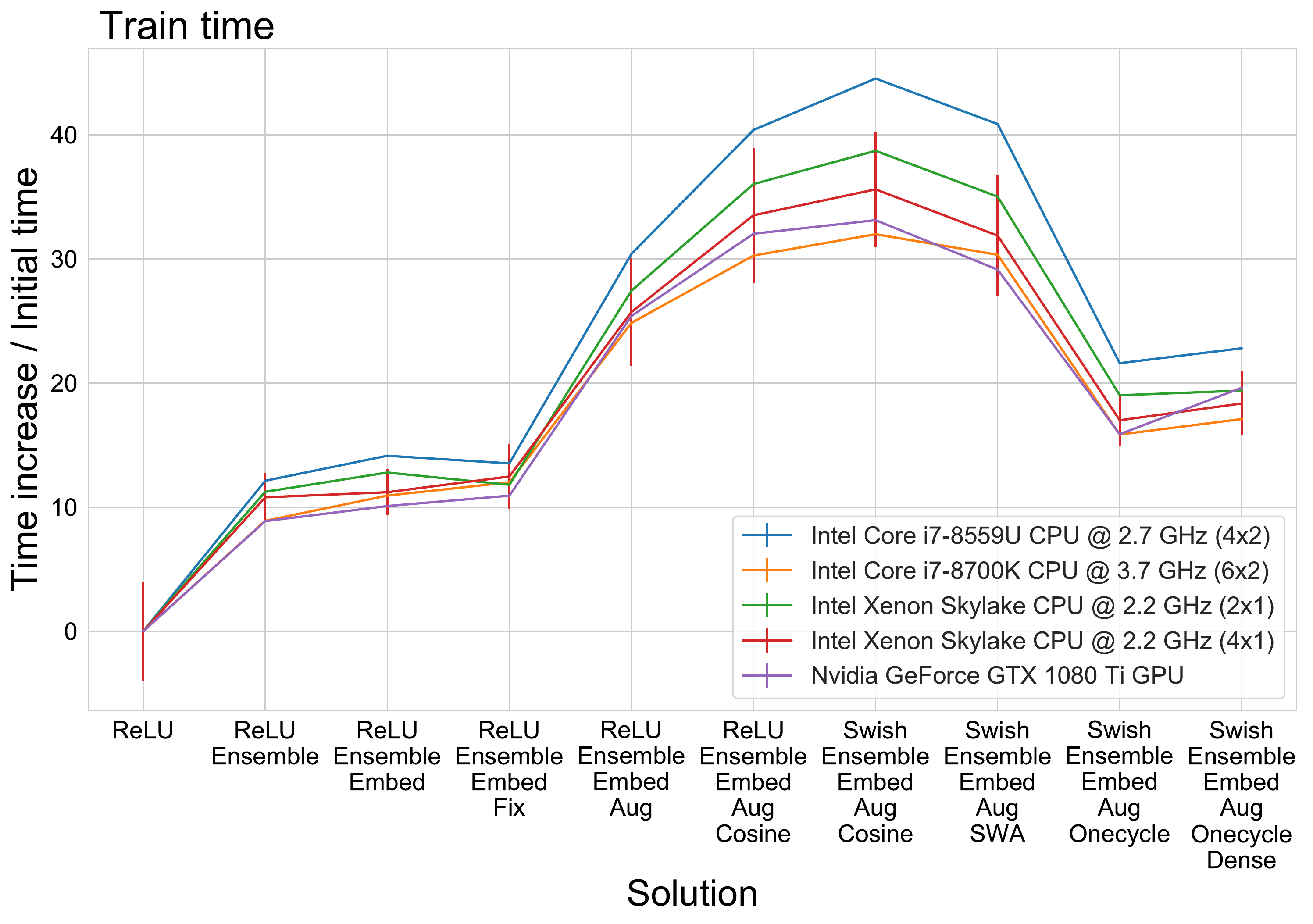}
						\caption{Fractional train-time increase}
					\end{center}
				\end{subfigure}
				\begin{subfigure}[t]{\sfMid\textwidth}
					\begin{center}
						\includegraphics[width=\textwidth]{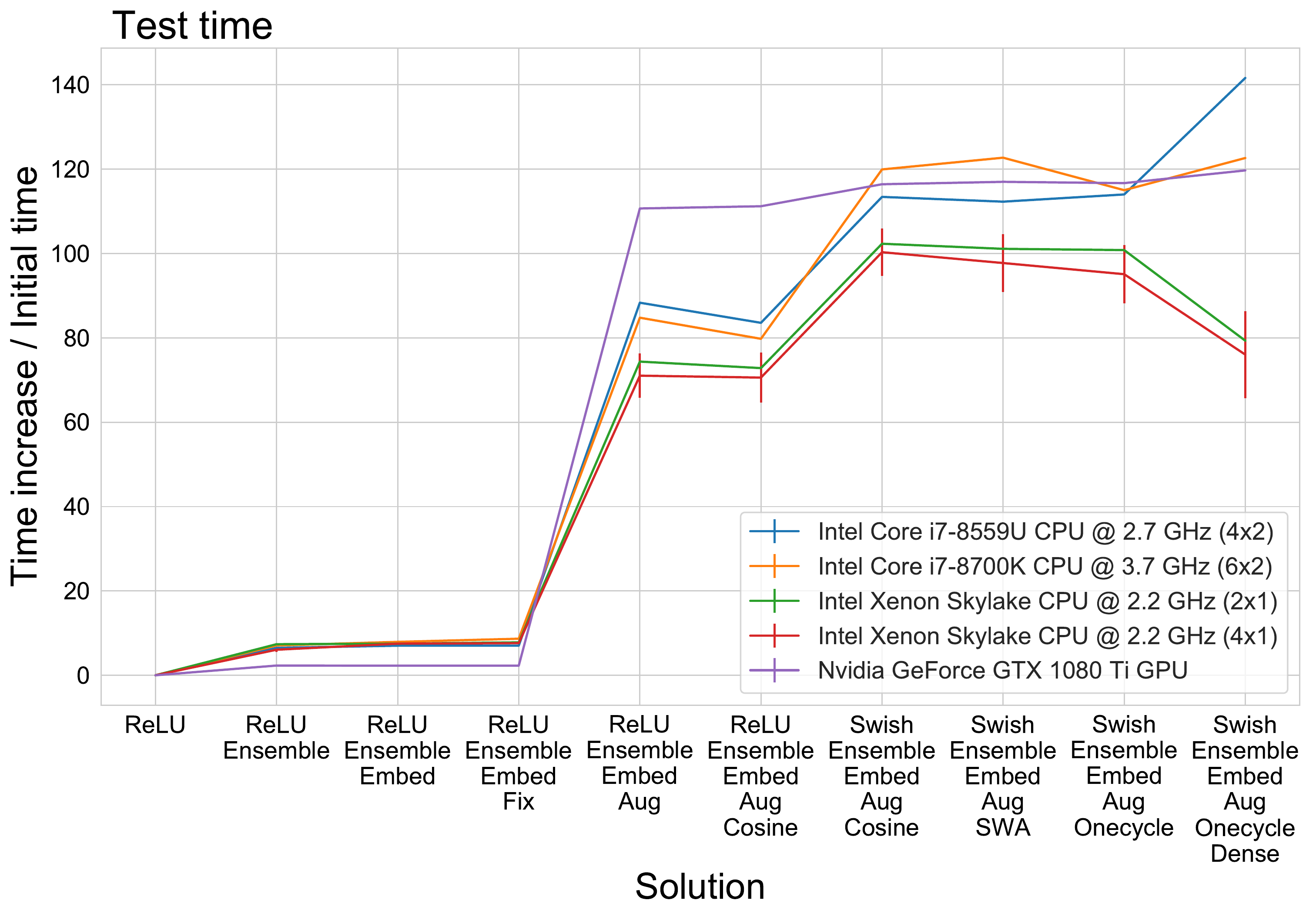}
						\caption{Fractional test-time increase}
					\end{center}
				\end{subfigure}
				\caption{Fractional increases in timings of solutions for a variety of hardware, relative to ``ReLU''. Note that the (4x1) Xenon line is an average of two machines, and hence has error bars; only single machines were available for the other hardware configurations and so their associated timing uncertainties cannot be evaluated.}
				\label{fig:frac_sol_timings}
			\end{center}
		\end{figure}

		Also of interest for real-world application is the time taken by each solution. Plotted in Figs.~\autoref{fig:abs_sol_timings} and \autoref{fig:frac_sol_timings} are the absolute (wall clock) and relative timings for each solution for the variety of hardware tested, as computed by \texttt{timeit}. Note that one solution was run per hardware configuration, except for the quad-core Xenon setup, which shows the mean of two different machines with the same hardware specification. The numbers in parentheses indicate the number of cores and number of threads per core, respectively, except for the Xenon machines, which are cloud-based and are allocated virtual CPUs. From these results, we can see that whilst the GPU is clearly superior, the architectures run sufficiently well on CPU to be viable for use in a particle-physics analysis.
		
		It is also interesting to note the fractional test-time increase for the GPU, which appears to show sub-linear scaling when ensembling, but poor scaling when running inference-time augmentation. This is because \lumin currently loads each data-fold from the hard-drive into either RAM or VRAM sequentially, and the timing includes this. Once in memory, however, all models in the ensemble are run over the fold. Since the load time for VRAM is non-negligible compared to the prediction time, the fractional increase when evaluating the ten models is much less than nine. Additionally, \lumin currently applies data-augmentation when loading the data fold, rather than augmenting the data in-place, meaning that the load time is incurred an extra seven times when running TTA, hence the apparent poor scaling when running TTA on a GPU.
		
	\subsection{Discussion}
		Of the solutions developed, the one from \autoref{sec:densenet} (densely connected network with 1cycle training scheme) offered the highest value for MAPA seen during solution development, and is also much quicker to train compared to solutions that used constant or cosine-annealed learning-rate schedules. Its performance with unweighted ensembling is public : private AMS ($3.80\pm0.01$ : $3.803\pm0.005$), and with weighted ensembling ($3.79\pm0.01$ : $3.806\pm0.005$), see \autoref{sec:ensemble-weighting} for further discussion on this.
		
		As can be seen from the architecture-search results, no explicit regularisation was needed for this problem, and performance was only slightly dependent on the widths of the hidden layers. The rule-of-thumb used to set the initial layer width for the densely connected network (width number = of inputs features (continuous + categorical embeddings)), seems to be appropriate, but further work will be necessary to see how well this rule generalises to other problems and applications.
		
		A comparison to the solutions developed during the running of the challenge is presented in \autoref{tab:sol_comp}. In terms of hardware, for the solutions developed here, the GPU results are for a Nvidia GeForce GTX 1080 Ti, and the CPU result is for a Intel Core i7-8559U (2018 MacBook Pro). RAM and VRAM usage for each setup was found to be less than \SI{1}{\giga\byte}. Details of the other solutions are as follows:
		\begin{enumerate}
			\item The \nth{1} place solution (available at \url{https://github.com/melisgl/higgsml}) primarily used a setup with a Nvidia Titan GPU and \SI{24}{\giga\byte} RAM, but Ref.~\cite{melis} also details timings for an Amazon EC2 \texttt{m2.4xlarge} instance (8-vCPU \& \SI{15}{\giga\byte} RAM);
			\item The \nth{2} place solution (\url{https://github.com/TimSalimans/HiggsML}) was run on an Amazon EC2 \texttt{m2.4xlarge} instance (8-vCPU \& \SI{64}{\giga\byte} RAM) and notes the high RAM requirements of the training;
			\item And the \nth{3} place solution used a 2012 quad-core laptop (\url{https://www.kaggle.com/c/higgs-boson/discussion/10481}).
		\end{enumerate}
		The difference in hardware makes a direct comparison of solution timing difficult, but we can try to scale the timings: The \nth{1} place solution used a Nvidia Titan GPU (2013, \num{1500} gigaflops processing power in double precision) and our solution uses a Nvidia 1080 Ti (2017, \num{10609} gigaflops processing power in single precision). Accounting for the changes in processing power, the \nth{1} place solution should take around \SI{100}{\minute} to train on our GPU. Our solution therefore provides an estimated speedup of \SI{92}{\%} on GPU and \SI{86}{\%} on CPU (comparing to GPU). Given the uncertainties involved in the metric it would be reasonable to say that the solution developed here achieves comparable performance to winning solutions of the challenge in just a fraction of the time.

		Whilst the Kaggle solutions only report the private AMS scores they achieved (single numbers with no uncertainties) it is of interest to know the general performance offered by each of the solutions. The summary paper for the challenge (Ref.~\cite{higgsml_discussion}) estimates the statistical significance of the solution ranking by bootstrap resampling of the scored data. The authors find that at a \SI{95}{\%} confidence level, the \nth{1}-place solution does indeed outperform the rest, but that the \nth{2} and \nth{3}-place solutions are indistinguishable in terms of performance. Ideally such a test should also be performed to compare our solution to the \nth{1}-place solution, however this requires access to the scored data that was submitted during the competition, which is not publicly accessible.
		
		\begin{table}
			\begin{center}
				\begin{tabular}{ll|lll}
					\toprule
					 & Our solution & \nth{1} place & \nth{2} place & \nth{3} place\\
					\midrule
					Method & 10 DNNs & 70 DNNs & Many BDTs & 108 DNNs\\
					Train-time (GPU) & \SI{8}{\minute} & \SI{12}{\hour} & N/A & N/A\\
					Train-time (CPU) & \SI{14}{\minute} & \SI{35}{\hour} & \SI{48}{\hour} &  \SI{3}{\hour}\\
					Test-time (GPU & \SI{15}{\second} & \SI{1}{\hour} & N/A & N/A\\
					Test-time (CPU) & \SI{3}{\minute} & ??? & ??? & \SI{20}{\minute}\\
					Score & $3.806\pm0.005$ & $3.80581$ & $3.78913$ & $3.78682$\\
					\bottomrule
				\end{tabular}
			\end{center}
			\caption{Comparison with challenge solutions. `N/A' indicates that the solution was not run on a GPU and so no timing information is available. `???' indicates that the solution was run on CPU but no timing information was reported. Timings for ``Our solution" are rounded to the nearest minute, or five seconds. Note that the training times differ from those shown in \autoref{fig:abs_sol_timings}, which were trained whilst showing a live feedback of the loss to help diagnose any problems, but equated to a constant increase in time per epoch that was independent of the architecture. For the final timings shown here, the live plotting was turned off.}
			\label{tab:sol_comp}		
		\end{table}

\FloatBarrier
	\section{Benefit and ablation studies}\label{sec:ablation}
    \subsection{Individual benefits}
        Due to all possible permutations of new methods not being considered, it is plausible that some methods would show a greater or worse relative improvement if testing had been performed in a different order. This could happen for a number of reasons, such as: overlaps in improvements in methods; certain pairs of methods working well together; and the non-linear nature of the scoring metric. Table~\ref{tab:add_test} addresses some these reasons by detailing results from retrainings of solutions in which only one new method was added. Whilst all new methods provide at least minor improvements in isolation, it is interesting to see that data augmentation actually provides a greater improvement than ensembling when applied individually, in comparison to \autoref{fig:summary} (which shows the relative contributions of each new method to the overall improvement in performance), indicating an overlap in the two methods. Also the use of dense connections on their own provides a smaller benefit than they do in the full solution, implying that they are most useful when applied in conjunction with other methods.

        \begin{table}
            \begin{center}
                \begin{tabular}{lc}
                    \toprule
                    Setup & Mean improvement in private AMS\\
                    \midrule
                    Baseline + Data augmentation & $7.98\pm0.05\%$\\
                    Baseline + Ensembling & $7.58\pm0.05\%$\\
                    Baseline + 1cycle & $4.00\pm0.03\%$\\
                    Baseline + Swish & $2.22\pm0.03\%$\\
                    Baseline + Embedding & $1.63\pm0.03\%$\\
                    Baseline + Dense connections & $1.25\pm0.03\%$\\
                    \bottomrule
                \end{tabular}
            \end{center}
            \caption{Individual percentage improvements in mean private AMS offered by each new method over a baseline ReLU model. In contrast to \autoref{fig:summary} the model setups here do not include other methods, e.g. ``Baseline + Embedding'' does not also include ensembling.}
            \label{tab:add_test}
        \end{table}
    
    \subsection{Ensemble weighting}\label{sec:ensemble-weighting}
        In \autoref{sec:weighted_ensemble}, performance-weighted ensembling was used. The more simple form of ensembling would have been to instead consider the predictions of each model equally (uniform weighting). The choice to use weighted ensembling was made based on previous experience with similar HEP classification tasks, and was not tested at the time so as to reduce the number of times the performance metrics were computed (reducing the chance of over-fitting to the validation dataset). Having completed the study, though, it is of interest to check the impact of weighted versus unweighted ensembling. As discussed in \autoref{sec:densenet}, the final ensembles were accidentally built using uniform weighting due to a mistake in the code, achieving a private AMS of $3.803\pm0.005$. The mistake was found when performing this ablation study of the ensemble weighting and retraining the ensembles using performance-weighted ensembling, as intended, results in a slightly higher private AMS of $3.806\pm0.005$.
	\section{Conclusion} \label{sec:conclusion}
	In this \whatAmI we have examined several recent advances in deep learning for neural network architectures and training methodologies, and their applicability to a typical problem in the field of high-energy particle-physics, using data from a popular data-science competition. Whilst the solutions developed were unable to consistently improve on the performance of the winning solutions, they were able to reproduce their performance in a significantly shorter time period using less computational resources.
	
\begin{figure}[ht]
		\begin{center}
			\includegraphics[width=\sfMid\textwidth]{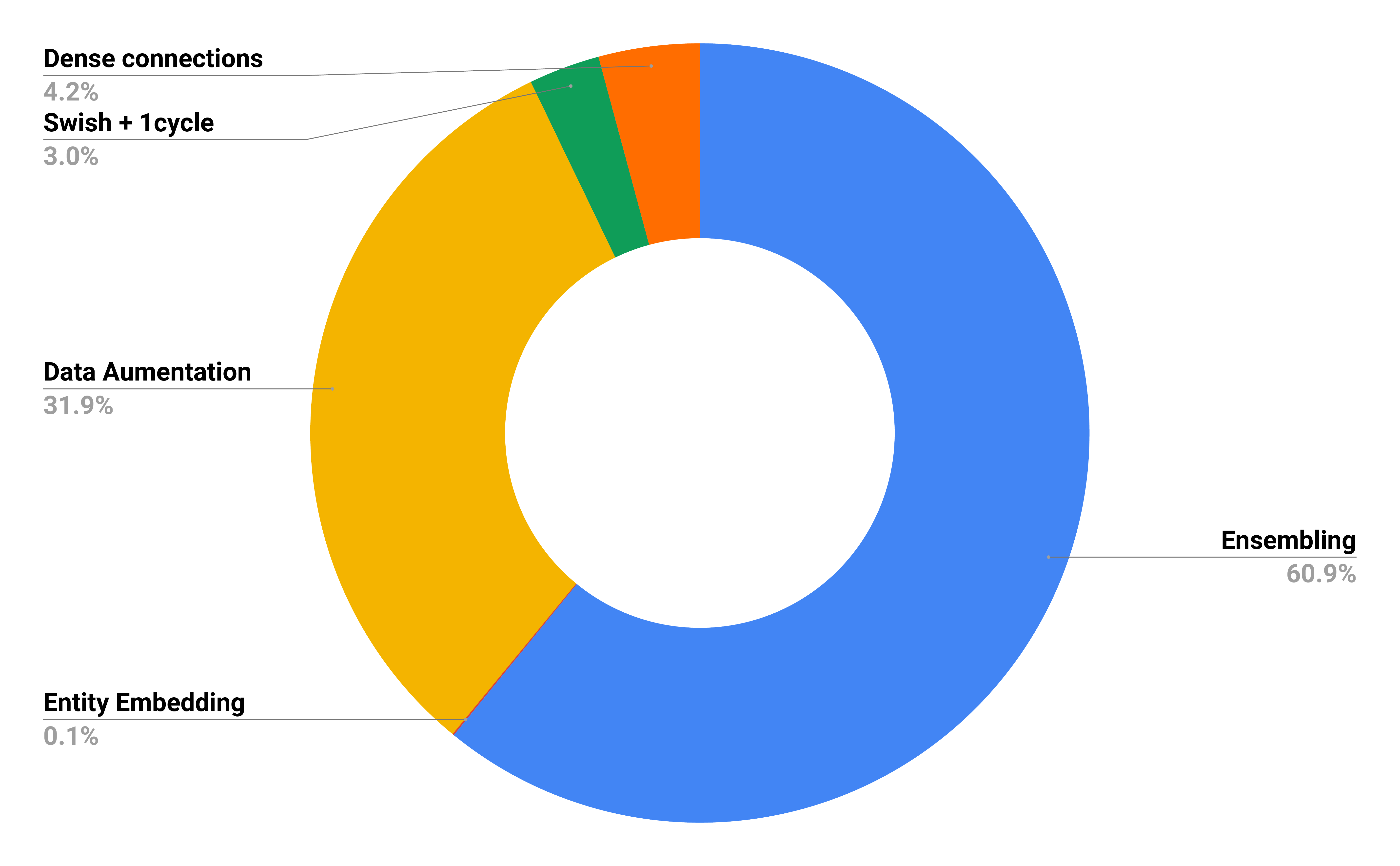}
			\caption{Illustration of the relative contributions of each new method to the overall improvement in mean private AMS over a baseline ReLU model. Note that contributions are calculated sequentially in the order tested, i.e. the entity embedding result includes ensembling, the augmentation result includes embedding and ensembling, \textit{et cetera}.}
			\label{fig:summary}
		\end{center}
	\end{figure}

	The solutions were developed in a systematic manner, rather than considering all possible permutations of hyper-parameters, using a variety of metrics (one of which was unavailable during the competition) and considering the time requirements. The improvements to the performance metric are broken down in \autoref{fig:summary}, where we can see that most of the improvement beyond ensembling (which the winning solutions already used) is coming from domain-specific data augmentation (applied during training and testing), this is particularly interesting as it implies that the models are able to make good use of the low-level information in the data and learn their own high-level representations of it. Learning-rate scheduling provides not only a moderate reduction in train time, but also a small improvement in performance. Finally, using densely connected layers offers a further small improvement in performance as well as reducing the dependence of performance on the widths of network layers, potentially making it easier to find optimal architectures for similar applications.

	It is interesting to note the benefit of data augmentation on performance; 
	
\FloatBarrier
	\section*{Acknowledgements}
    I am very grateful to Michele Gallinaro, Nuno Castro, and João Pedro de Arruda Gonçalves for their valued feedback and comments in reviewing this \whatAmI, and to the other members of the AMVA4NewPhysics ITN for all the discussions over the years, which no doubt helped to contribute to this work. Additional thanks also to the fast.ai team whose courses introduced me to many of the ideas tested here, as well as inspired me to begin writing my own deep learning framework. I would also like to thank Tommaso Dorigo and David Rousseau for the useful discussions, as well the two reviewers of the manuscript for their thorough and helpful recommendations.

\newpage
\small{\bibliography{pub_higgml_lumin}}

\newpage
\begin{appendices}
	\FloatBarrier
	\section{Parallel networks and feature splitting}\label{sec:widedeep}
	\textit{An investigation into more complex architectures which split the input features along separate processing paths. Whilst this section does not result in an improvement, it does illustrate how these architectures can be used to further interpret the data and so has been included as an appendix to the main report. In the course of solution development, this investigation took place just prior to the architecture search detailed in \autoref{sec:arch_opt}.}\\

	So far we have treated all the input features in the same manner (aside from embedding the jet multiplicity) and passed them all through the networks together. However, in particle physics we have lower-level features, such as final-state momenta, but we often compute several high-level features, which are theory inspired non-linear combinations of low-level features, such as the Higgs mass. Indeed, the Higgs ML dataset contains both high (\texttt{DER}) and low (\texttt{PRI}) features.
	
	As was shown in \autoref{fig:ll_hl_feats}, the high-level features in the data already demonstrate a good degree of class separation compared to the low-level features. Because of this, it is possible that we could reduce the number of free parameters of the model; although the low-level features might require a deeper network to learn a useful representation for them, perhaps the high-level features might simply require a single layer to encode them. Such architectures have been referred to in literature as \textit{Wide \& Deep learning}, e.g. Ref.~\cite{widedeep}.
	
	It is likely, however, that the optimal model will require some interactions between the high- and low-level features. We can consider two extra architectures in addition to completely splitting the features:
	In the first architecture, all of the input features are passed through the deep network as normal, but additionally the high-level features are passed through a single layer which is then concatenated with the output of the deep block. In this case the single layer can encode easily the high-level information, allowing the deep block to learn to represent better the low-level information and to capture the interactions between the low- and high-level features.
	
	In the second architecture, the low- and high-level features are split, but two \textit{bottleneck} layers are used to encode a low-dimensional representation of the sets of features. The outputs of the bottleneck layers are then concatenated with their opposite set of features to provide inputs to each block: i.e. the compressed representation of the high-level features is concatenated with the low-level features and passed to the deep block, and vice versa. In this case, the bottleneck layers should learn the most useful representation of the information which each network block would otherwise not have available, whilst still allowing the number of free parameters in the model to be comparatively small.

	\subsection{Testing}
\begin{figure}[ht]
	    	\begin{center}
	    		\begin{subfigure}[t]{\sfSmall\textwidth}
	    			\begin{center}
	    				\includegraphics[width=\textwidth]{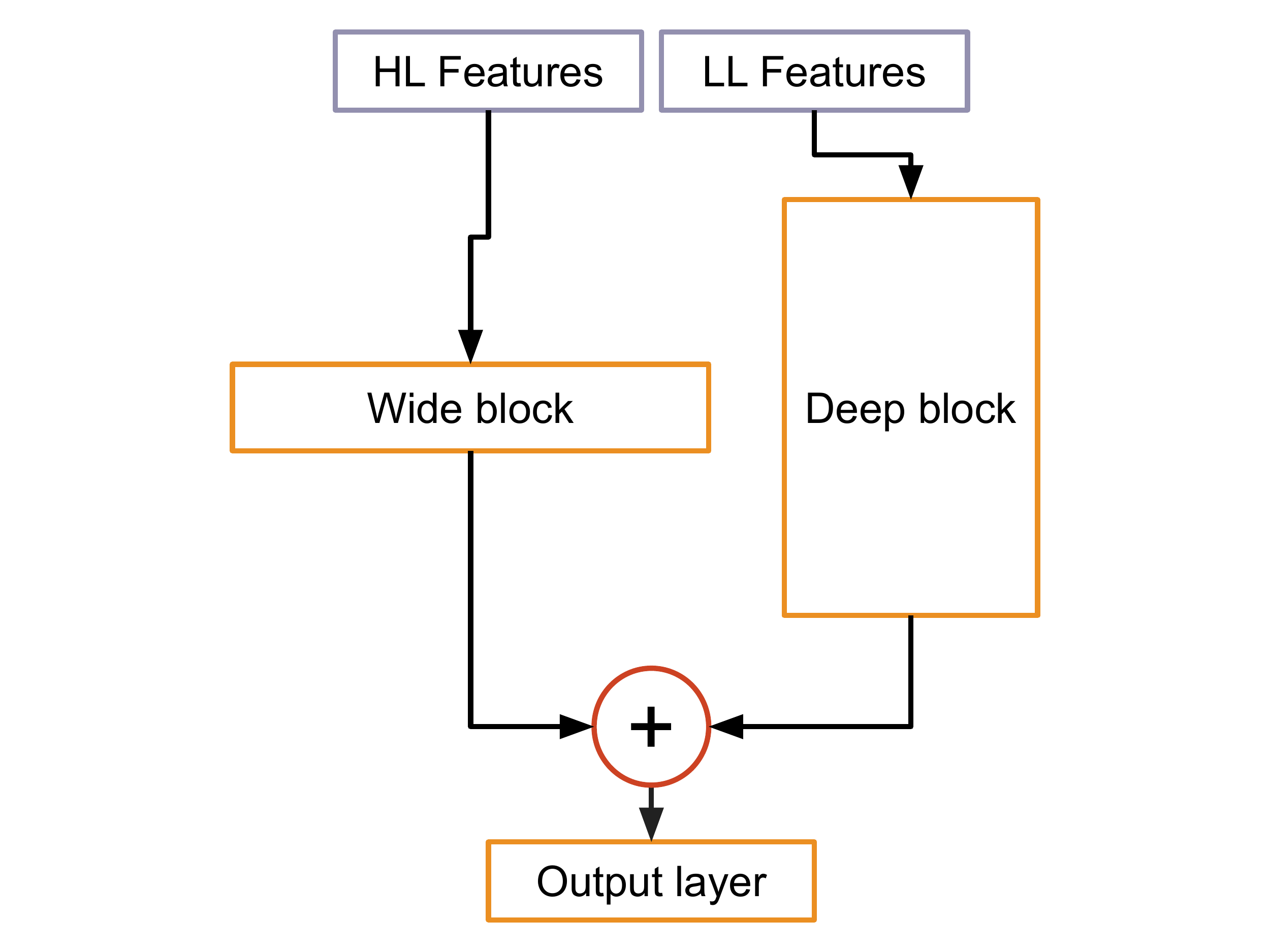}
	    				\caption{Full-split}
	    			\end{center}
	    		\end{subfigure}
	    		\begin{subfigure}[t]{\sfSmall\textwidth}
	    			\begin{center}
	    				\includegraphics[width=\textwidth]{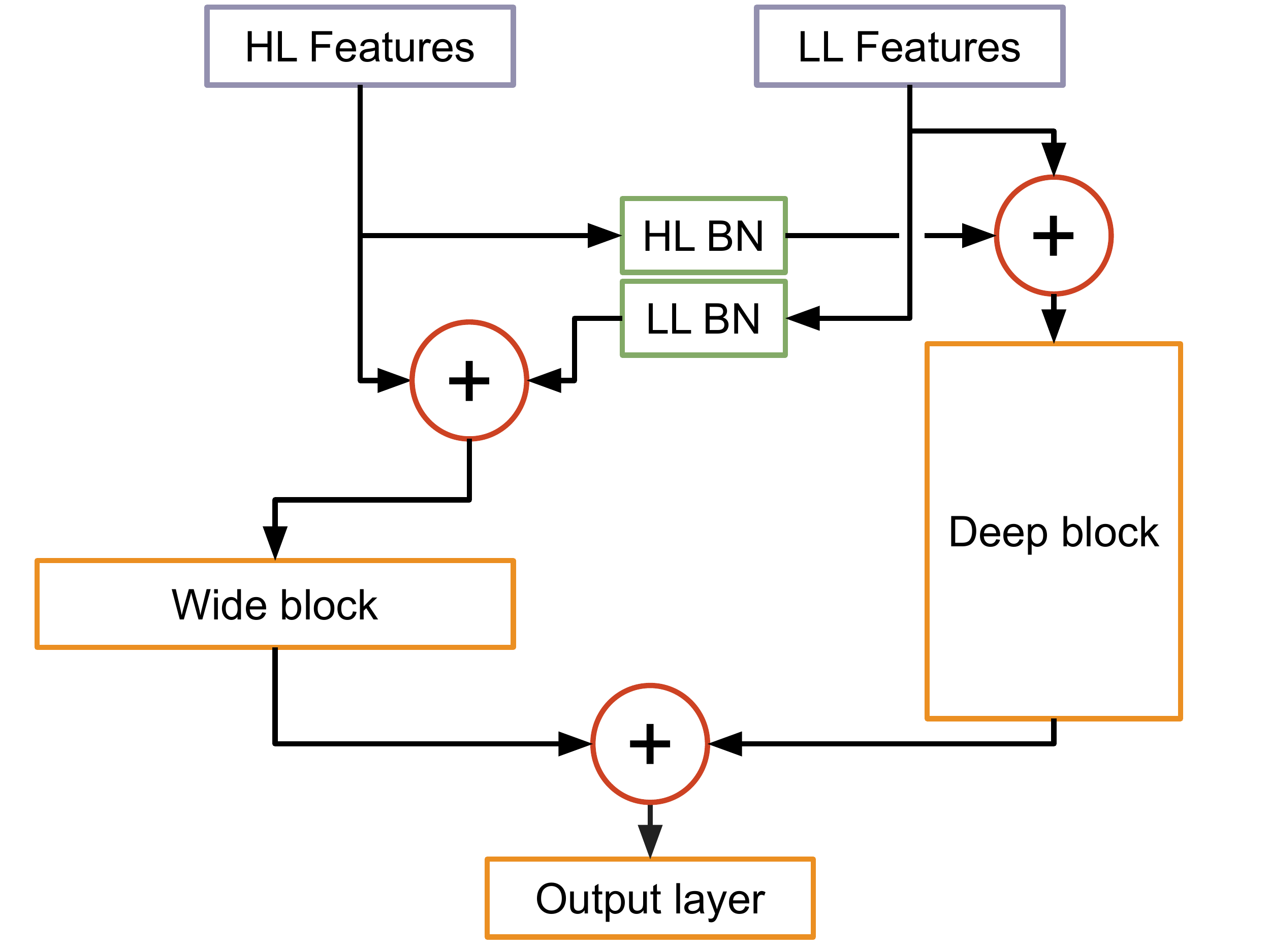}
	    				\caption{Full-split with bottlenecks}
	    			\end{center}
	    		\end{subfigure}
	    		\begin{subfigure}[t]{\sfSmall\textwidth}
	    			\begin{center}
	    				\includegraphics[width=\textwidth]{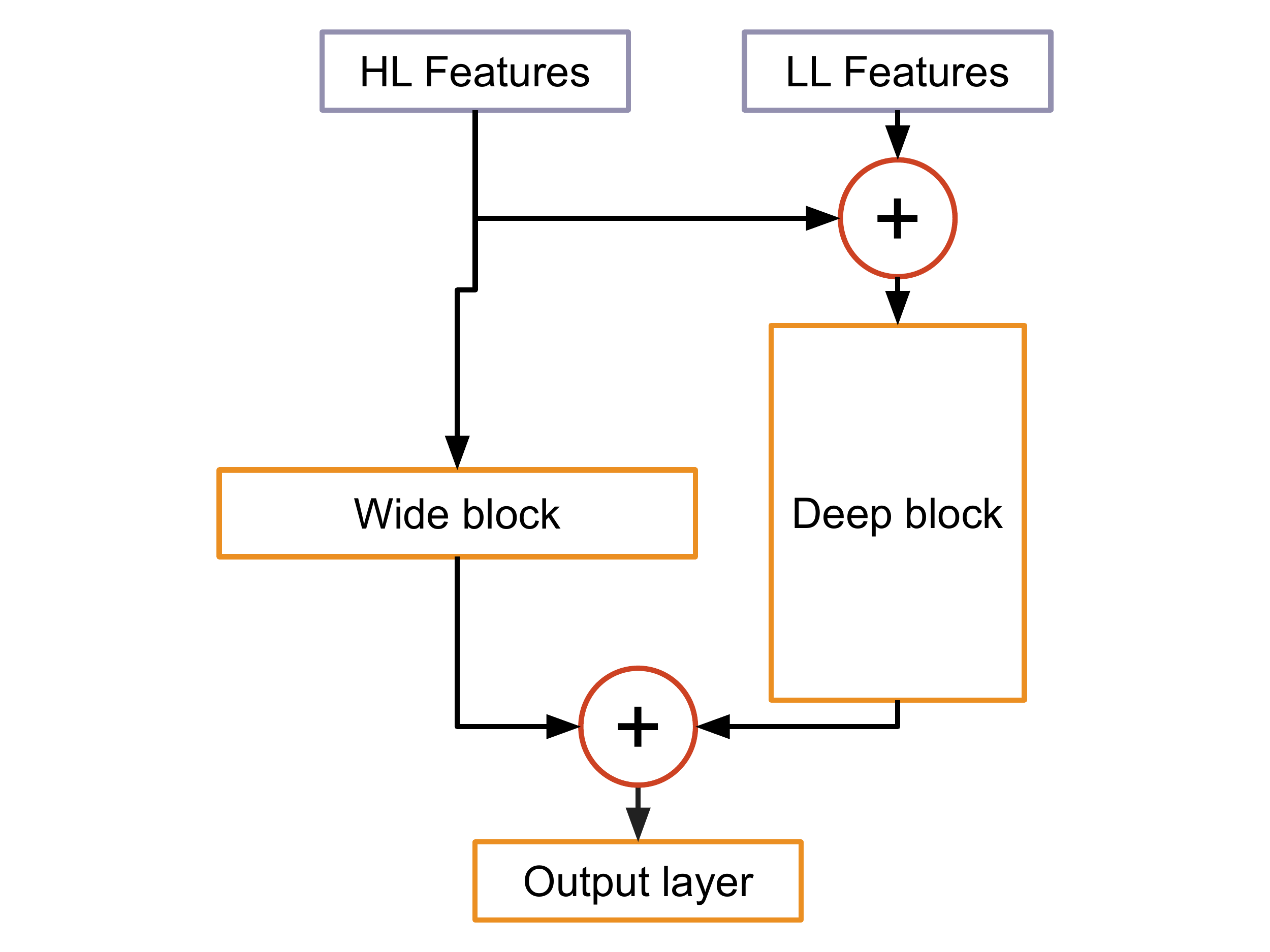}
	    				\caption{Semi-split}
	    			\end{center}
	    		\end{subfigure}
	    		\caption{Architecture diagrams for the parallel networks considered. $\oplus$ indicates concatenations of hidden states, BN stands for ``bottleneck", and HL and LL stand for ``high-level" and ``low-level", respectively.}
	    		\label{fig:split_arch}
	    	\end{center}
	    \end{figure}

		Three architectures were tested:
		\begin{enumerate}
			\item Full-split: The low-level features were passed through a six-layer-deep, 20 neuron-wide, densely connected set of hidden layers, and the high-level features were passed through a single 50-neuron-wide layer.  This results in \num{9303} free parameters.
			\item Full-split+bottlenecks: Each set of low- and -high level features are passed through single neuron bottleneck layers. The low-level features, concatenated with the high-level-bottleneck output, were passed through a six-layer-deep, 21 neuron-wide, densely connected set of hidden layers, and the high-level features were concatenated with the low-level-bottleneck output and passed through a single 51-neuron-wide layer. This results in \num{10272} free parameters.
			\item Semi-split: Both the low and high level features are passed through a six-layer-deep, 33 neuron-wide, densely connected set of hidden layers, and additionally the high-level features were passed through a single 50-neuron-wide layer. This results in \num{23863} free parameters.
		\end{enumerate}
	    In all cases, Swish activation functions were placed after every linear layer, and the outputs of the wide and deep blocks were then concatenated and passed through the output neuron. Figure~\ref{fig:split_arch} illustrates the three network architectures. For reference the current solution as of \autoref{sec:densenet} uses \num{23113} free parameters.
	    
	    From \autoref{tab:widedeep}, we can see that completely splitting the features severely reduces model performance, but that allowing the network to learn compressed representations of the features goes some way to recover performance. We can also see that the addition of the linear embedding for high-level features, offered by the semi-split architecture does not provide any improvement over the current solution, and only serves to increase the model size and timings.
	    
	    \begin{table}
	    	\begin{center}
	    		\begin{tabular}{lccccc}
	    			\toprule
	    			Setup & MMVA & MVAC & MAPA & \multicolumn{2}{c}{Fractional time-increase}\\
	    			& & & & Training & Inference \\
	    			\midrule
	    			\textbf{Current} & $\mathbf{3.95\pm0.04}$ & $\mathbf{3.89\pm0.05}$ & $\mathbf{3.82\pm0.02}$ & \textbf{-} & - \\
	    			Full-split & $3.69\pm0.06$ & $3.64\pm0.06$ & $3.596\pm0.006$ & $0.02\pm0.03$ & $ \mathbf{-0.10\pm0.04}$ \\
	    			Full-split+BN & $3.80\pm0.05$ & $3.72\pm0.05$ & $3.647\pm0.006$ & $0.14\pm0.05$ & $0.00\pm0.05$ \\
	    			Semi-split & $3.93\pm0.05$ & $3.86\pm0.05$ & $\mathbf{3.82\pm0.02}$ & $0.12\pm0.02$ & $0.1\pm0.1$ \\
	    			\bottomrule
	    		\end{tabular}
	    	\end{center}
	    	\caption{Comparison of the various parallel network architectures. ``Current" refers to the solution as of \autoref{sec:densenet}. `BN' in this case stands for BottleNeck. The best values for each metric are shown in bold, and the setup chosen is also indicated in bold.}
	    	\label{tab:widedeep}
	    \end{table}
    
    \subsection{Interpretation}
    	Whilst it was found that none of the parallel architectures provided improvements over the current solution, they do allow for several informative interpretation methods due to the fact that the features are being split into subgroups.
    	
    	Since the output layer in each model consists of a single neuron which directly takes its inputs from the wide and deep block outputs, we can compute the absolute values of the dot products of slices of the output neuron's weights ($\bar{w}$) with the corresponding inputs ($\bar{x}$) from each of the wide and deep blocks for a range of data points. This then gives an idea of the reliance of the model's prediction on particular subsets of inputs.
    	
\begin{figure}[ht]
    		\begin{center}
    			\begin{subfigure}[t]{\sfMid\textwidth}
    				\begin{center}
    					\includegraphics[width=\textwidth]{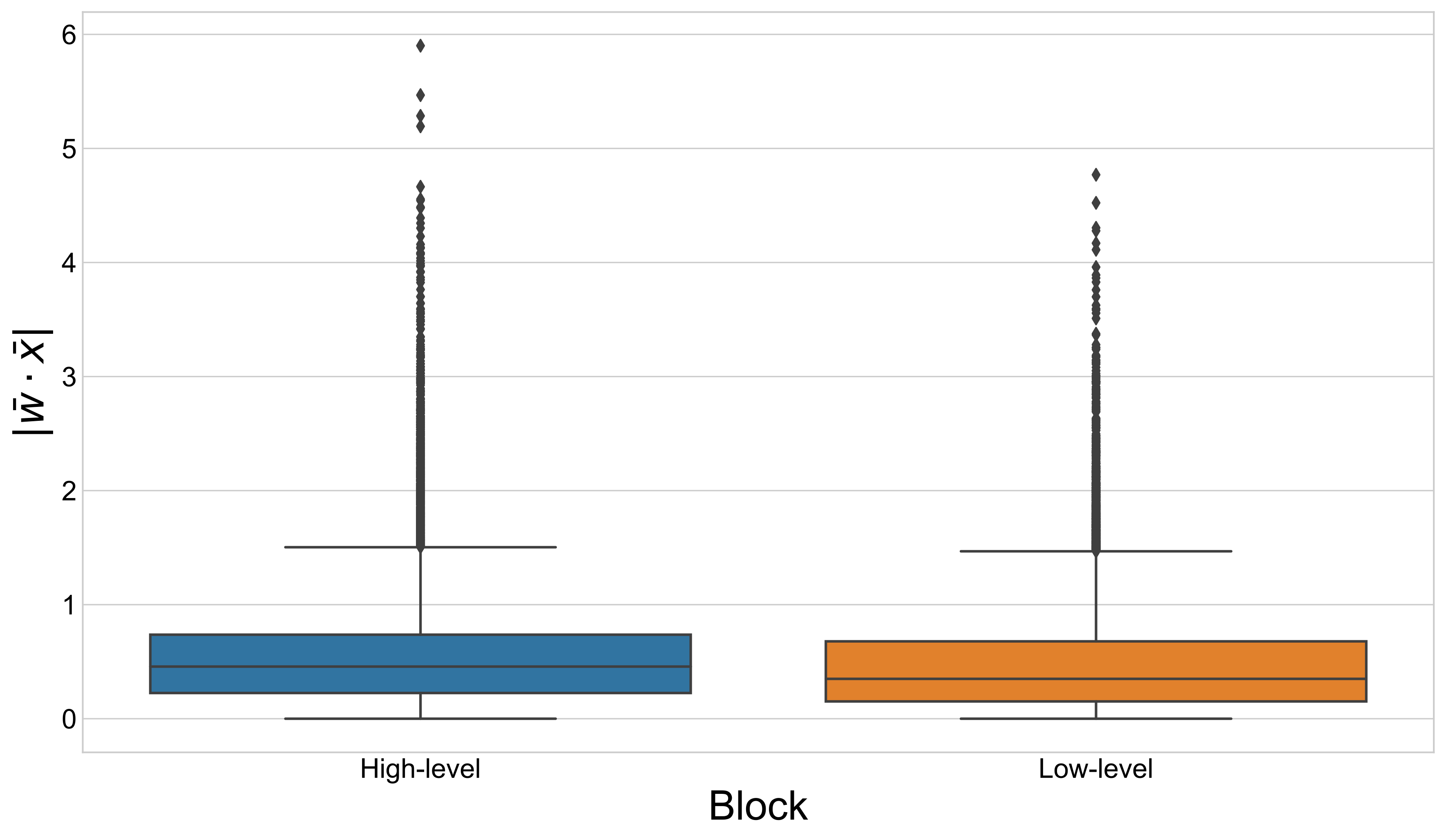}
    					\caption{Model reliance on high-level and low-level features as processed by the ``Full-split" architecture.}
    				\end{center}
    			\end{subfigure}
    			\begin{subfigure}[t]{\sfMid\textwidth}
    				\begin{center}
    					\includegraphics[width=\textwidth]{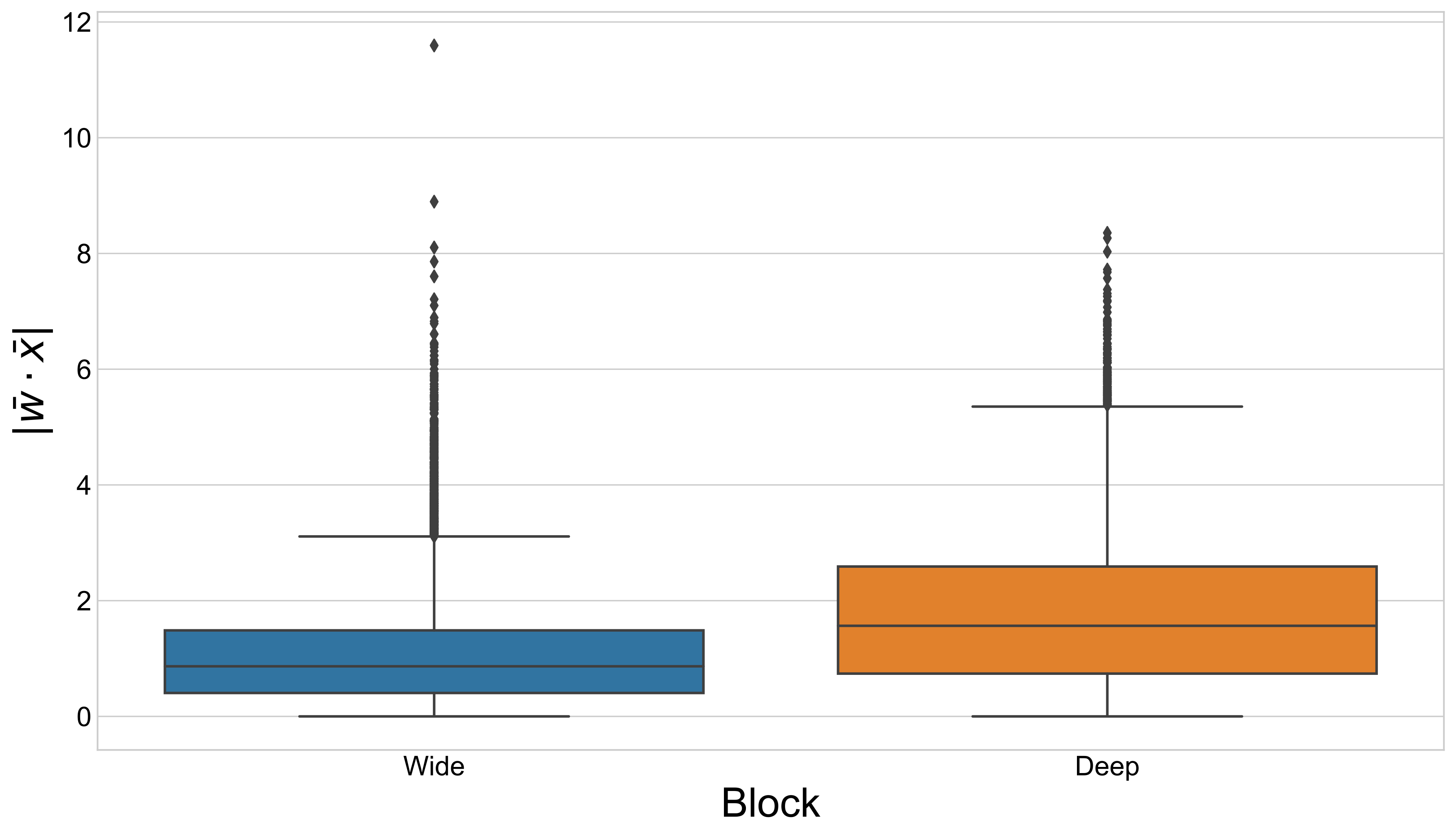}
    					\caption{Model reliance on high-level and high+low-level features as processed by the ``Semi-split" architecture.}
    				\end{center}
    			\end{subfigure}
    			\caption{Model reliance on subsets of features}
    			\label{fig:split_interp}
    		\end{center}
    	\end{figure}

    	From \autoref{fig:split_interp} we can see that the ``Full-split" architecture shows a slightly higher dependence on the single layer fed with the high-level features, but the fact that this difference in dependence is only minor suggests that the deep block is learning a powerful representation of the  low-level features. The ``Semi-split" model, on the other hand, shows a strong reliance on the deep block (which contains both high and low features), confirming the result that the single-layer embedding of the high-level features does not add much to the model performance.
    	
\begin{figure}[ht]
    	\begin{center}
    		\begin{subfigure}[t]{\sfMid\textwidth}
    			\begin{center}
    				\includegraphics[width=\textwidth]{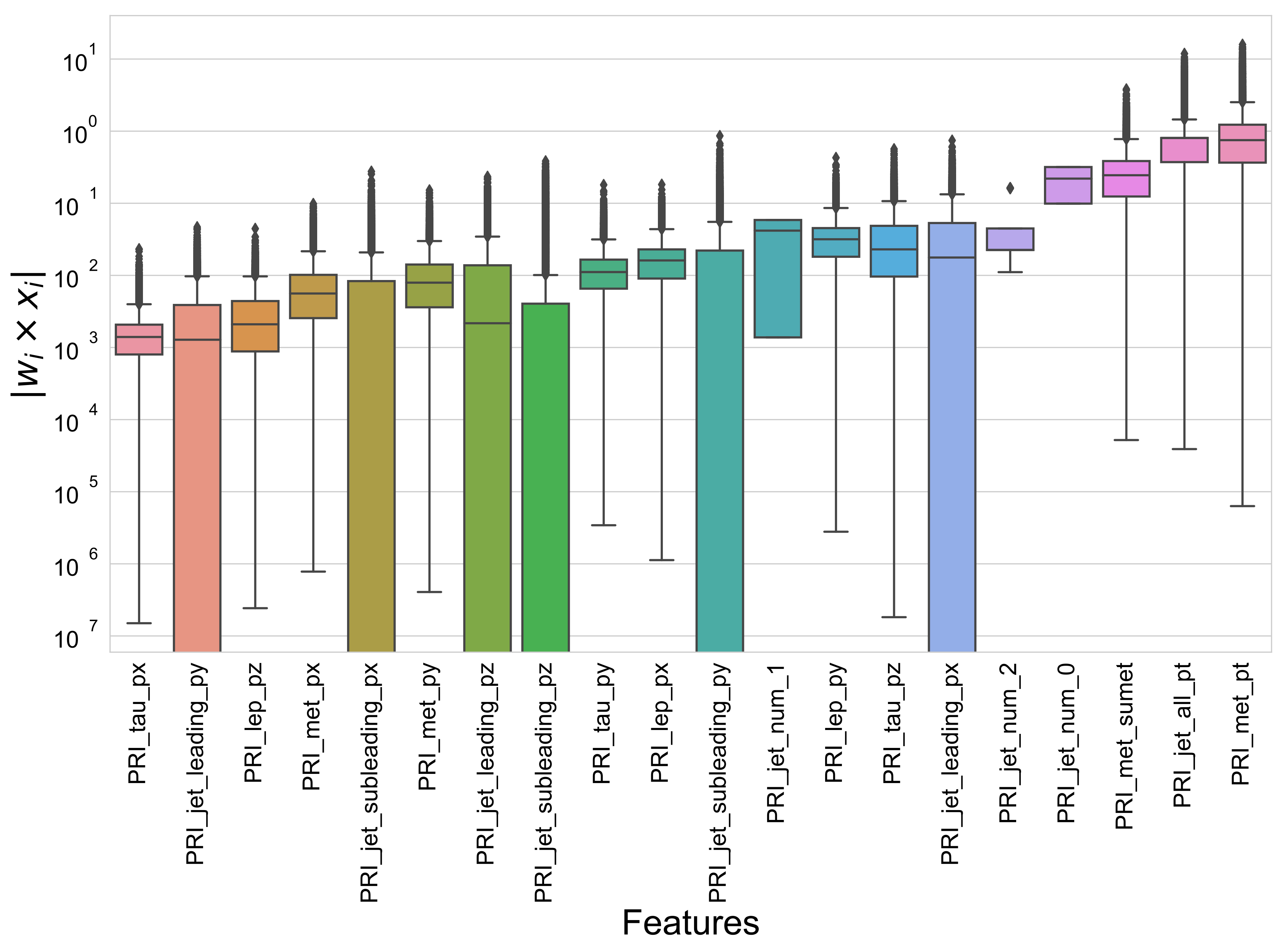}
    				\caption{Low level}
    				\label{fig:a:bn_imp_interp}
    			\end{center}
    		\end{subfigure}
    		\begin{subfigure}[t]{\sfMid\textwidth}
    			\begin{center}
    				\includegraphics[width=\textwidth]{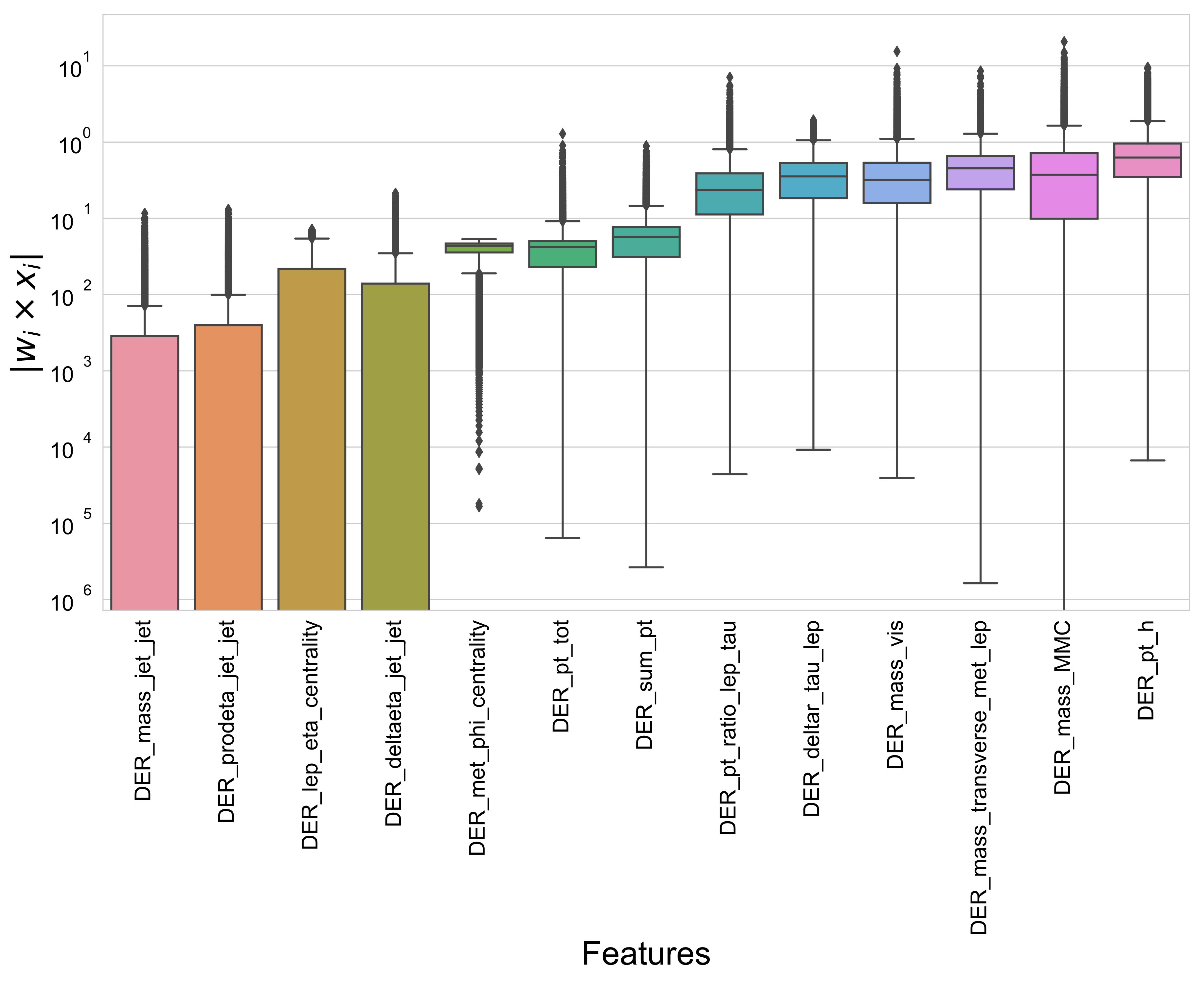}
    				\caption{High level}
    				\label{fig:b:bn_imp_interp}
    			\end{center}
    		\end{subfigure}
    		\caption{Bottleneck reliance on individual features.}
    		\label{fig:bn_imp_interp}
    	\end{center}
    \end{figure}

    In a similar fashion, the bottleneck layers of the ``Full-split+bottlenecks" architecture can be interpreted by computing the absolute values of the weights ($w_i$)times the feature values ($x_i$, for feature $i$ )for a range of data points. Comparing \autoref{fig:b:bn_imp_interp} to \autoref{fig:hl_imp}, we can see that whilst the high-level bottleneck representation does rely heavily on the two most important features (the MMC mass and the transverse mass), the most used feature is the $p_{t,h}$, which has comparatively lower importance (as shown in \autoref{fig:hl_imp}). This is perhaps an indication that the deep block is being used to refine this weaker feature by combining it with low-level information, or the response of the deep block is being parametrised according to the Higgs \pt. It is also interesting to note in \autoref{fig:a:bn_imp_interp} that the low-level bottleneck appears to be used for passing information about jet multiplicity and missing transverse energy; features which could potentially be used to parametrise the response of the wide layer to the jet-related features.
    
\begin{figure}[ht]
		\begin{center}
			\begin{subfigure}[t]{\sfMid\textwidth}
				\begin{center}
					\includegraphics[width=\textwidth]{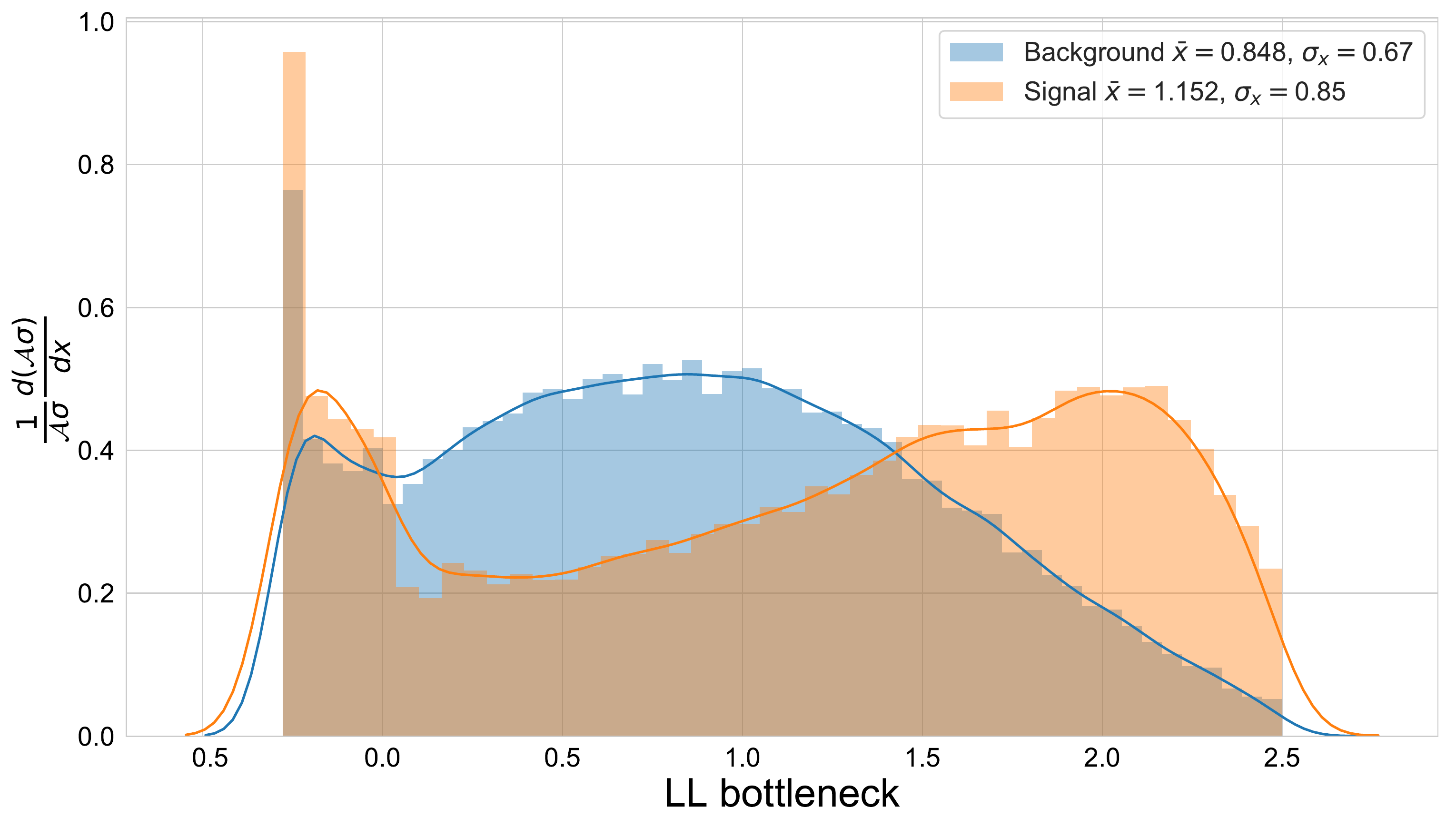}
					\caption{Low level}
				\end{center}
			\end{subfigure}
			\begin{subfigure}[t]{\sfMid\textwidth}
				\begin{center}
					\includegraphics[width=\textwidth]{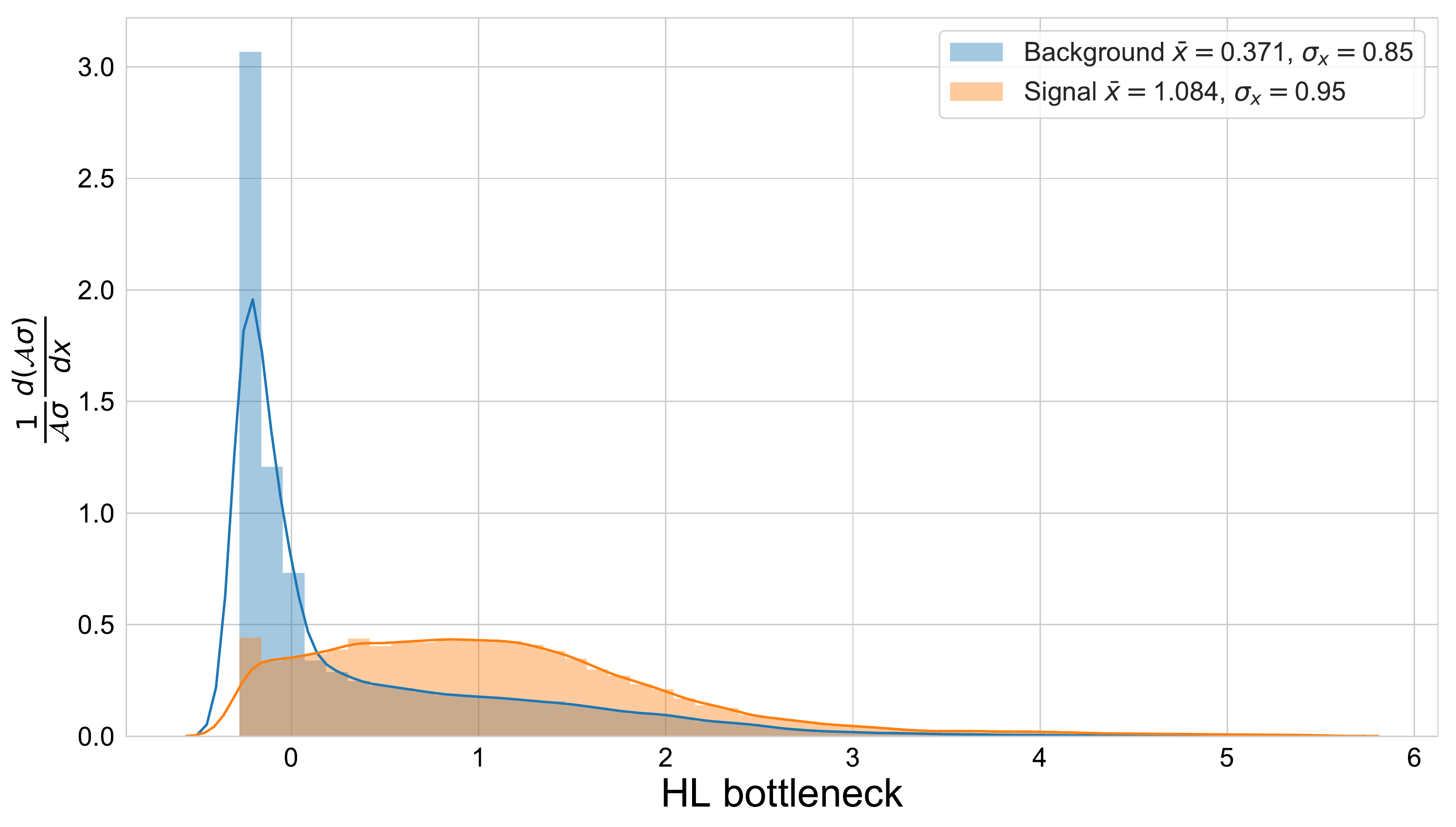}
					\caption{High level}
				\end{center}
			\end{subfigure}
			\caption{Output distribution of the bottleneck layers separated by event class. These are the outputs after the activation layers of the bottlenecks. It is interesting to note the ``double-peak'' distribution of the low-level bottleneck output; further inspection revealed that the peak close to zero was only present in events containing at least one jet, further supporting the hypothesis that the low-level bottleneck learns to pass jet-multiplicity information to the wide-block.}
			\label{fig:bn_output_interp}
		\end{center}
	\end{figure}

	As a final stage of interpretation of the bottleneck layers, we can plot the distributions of the their outputs for a range of data points (\autoref{fig:bn_output_interp}). We can see that both bottlenecks learn features which display decent separation between the data classes.
	
	\newpage
	\FloatBarrier
	\section{A study of the SuperTML result}\label{sec:super_tml}
    As mentioned in \autoref{sec:higgsml_sota}, Ref.~\cite{super_tml} claims to achieve an AMS of \num{3.979} through the use of pretrained models. If true, this would potentially be revolutionary for the HEP field, and so it is necessary to confirm and investigate further. 
 
    \subsection{Pretraining and transfer learning}
        \subsubsection{Overview}
            Pretraining is a method in which a model is trained to perform a task on a large dataset. The trained model can be used as a starting point for refinement on another dataset to perform a different task. This allows more powerful models to be used in situations where such models would otherwise not be able to be trained, or simply to provide a better initialisation than random weighting. This is known as transfer learning (see e.g. Ref.~\cite{transfer_learning_1} and Ref.~\cite{transfer_learning_2}).
            
            One example is the pretraining of models on large collections of images, such as ImageNet~\cite{ImageNet}, which allows the model to learn common and transferable features maps such as corners and edges, and more complex features like materials and objects. The large model can now be used as a starting point for learning on a much smaller dataset (e.g. R-CNN~\cite{r_cnn}), or for different tasks such as image analysis of historical texts~\cite{historical_texts} where only a small number of texts have human annotations due to the time and expertise required to label the data. Another example is the pretraining of models on large corpuses of texts, such as WikiText~\cite{wikitext}, in order to learn the basic grammar and vocabulary of a language, the trained model can then be further refined for other tasks \cite{ulmfit}.

            Reference~\cite{rethinking_imagenet} notes, however, that the benefits of pretraining can be limited by the differences in target data-space, or task, notably in image data when the target task is sensitive to the spacial locations of objects in the image.

        \subsubsection{Transfer learning in tabular data}
            Model pretraining for computer vision and natural-language processing is achieved by using architectures which have some insensitivity to the dimensions of the input data; recursive and convolutional layers allow the model to be applied to images of different sizes, or sentences of different lengths, to those of the training data. In tabular data, however, the number of features can change considerably from one dataset to another, with no (current) architecture capable of dealing with this variation in an efficient manner. Even if there were such an architecture, the main benefit of pretraining comes from learning methods of feature extraction from unstructured data, whereas tabular data already has features extracted and these carry meanings specific to the dataset.
            
            Imagine for example employee data for an international business. One could train a classification model to predict which of the employees leave during the data-taking period based on employee details. Now imagine similar data for employees working at a small startup company. The same features are potentially present for both datasets, so it is possible that the classifier trained for the international business might offer some transfer-learning benefits when fine-tuned on the smaller company, but severe differences in company and employee type will likely limit the benefits of using the pretrained model. Taking this idea further, the pretrained model is now fine-tuned to predict whether it rained on a given day using data which happens to have the same number as input features as the original training data. Here no transfer learning can take place; the input features do not share similar meanings and the target task is far too different to the original task.

    \subsection{The SuperTML method}
\begin{figure}[ht]
            \begin{center}
                \begin{subfigure}[t]{\sfSmall\textwidth}
                    \begin{center}
                        \includegraphics[width=\textwidth]{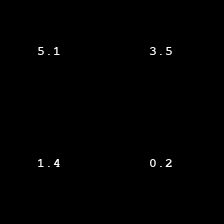}
                        \caption{Iris}
                    \end{center}
                \end{subfigure}
                \begin{subfigure}[t]{\sfSmall\textwidth}
                    \begin{center}
                        \includegraphics[width=\textwidth]{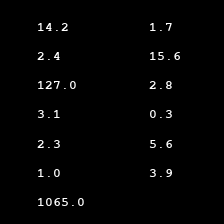}
                        \caption{Wine}
                    \end{center}
                \end{subfigure}
                \begin{subfigure}[t]{\sfSmall\textwidth}
                    \begin{center}
                        \includegraphics[width=\textwidth]{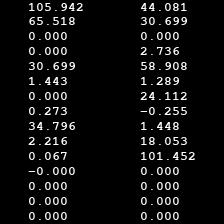}
                        \caption{HiggsML}
                    \end{center}
                \end{subfigure}
                \caption{Example images as encoded for the SuperTML method.}
                \label{fig:supertml_image}
            \end{center}
        \end{figure}

        The Super Tabular data Machine Learning method (SuperTML) is a way of allowing pretrained models to be applied to tabular data, such as the HiggsML dataset and potentially other HEP data. It avoids restrictions on the data dimensionality (number of features) by transforming each datapoint into its own image. Each image is black and the values of the features are written in white text on the image, as shown in \autoref{fig:supertml_image}.

        Having transformed the data into images, it uses the data to fine-tune large models that have been pretrained on ImageNet. The authors demonstrate their method on four classification datasets: Iris~\cite{iris}, Wine~\cite{wine}, Adult~\cite{adult}, and HiggsML.

    \subsection{Re-implementation of the method}
        The authors do not provide a publicly accessible distribution of their method, and did not offer to provide their code during our correspondence; we are therefore left with re-implementing their method based on the paper description. In correspondence, the authors did clarify some technical points of their method, but not all. Our implementation and experimental results are available at Ref.~\cite{my_supertml}, and are documented below.

        \subsubsection{Image preparation}
            Images are prepared as 224x224 pixel three-channel (RGB) images. Pixels are initially set to $(0,0,0)$ (black). Feature values are written in columns in the order as listed in the datasets in 13pt \texttt{FreeMonoBold} text in white ($255,255,255)$ and saved in png format. The use of three-channel images, despite being greyscale, is due to the pretrained models expecting full-colour images. This method of data preparation is compatible with that of a later re-implementation to which the authors directed us (\url{https://github.com/EmjayAhn/SuperTML-pytorch}).

        \subsubsection{Model fine-tuning}
            There are several choices when fine-tuning models. A two-stage process can be adopted:
            \begin{enumerate}
                \item An initial training in which the pretrained model is frozen and not updated and extra linear layers are added at the end of the model and trained, allowing the new layers to adjust to the outputs of the pretrained model;
                \item The full model is unfrozen and a second stage of training is performed, allowing more optimal feature maps to be learnt.
            \end{enumerate}
            In correspondence, the authors confirmed that they did not add any extra layers to the end of the model, and that training was performed in a single stage on the unfrozen model. We perform the fine-tuning using the fast.ai library v1 \cite{fastai}, with 1cycle scheduling and no forms of data augmentation.

        \subsubsection{Implementation confirmation tests}
            To show that our implementation works we attempt to recover the authors' scores of \SI{93.33}{\%} accuracy on the Iris dataset and \SI{97.30}{\%} accuracy on the Wine dataset. We follow the authors' data split of \SI{80}{\%} training and \SI{20}{\%} validation. For the model we use ResNet-34~\cite{step_decay}, instead of SE-net-154~\cite{senet} due to it being quicker to train for these confirmation tests. Despite the smaller model, our implementation achieves validation accuracies of \SI{100}{\%} and {97}{\%} on the Iris and Wine datasets, respectively. From these results we believe that our implementation of the method matches the authors'. Note that we do not test on the Adult dataset due to the authors' using one of their own methods to provide text embeddings for certain features in the data and we do not wish to introduce a source of uncertainty by implementing another of their techniques.

    \subsection{HiggsML tests}
        From the paper we read that the open-data version of the dataset is used, allowing the authors' to compute the AMS metric themselves on the testing data. They also say they use the same \num{250000} events for training and the same \num{550000} events for testing as were used in the Kaggle competition \footnote{The authors confirmed a typographical error in the paper where at one point the dataset sizes were quoted to be ten times smaller.}. To monitor overtraining we randomly sample \SI{20}{\%} of the training data for validation. The paper gives no mention of cut optimisation at all and the authors did not give any comment when asked, so we are left to assume that they assign events to the class with the highest prediction, i.e. a cut of 0.5 in the binary classification sense of the problem. The data images are prepared as normal, and we do not apply any preprocessing to them, (Fig.~4(a) in the paper gives no indication that the data are preprocessed). Transforming the dataset into images increases the size of the data from \SI{187}{\mega\byte} to \SI{9.8}{\giga\byte} (50 times larger).

        For the model we fine-tune SE-net-154, and in testing we found that the model converges within four epochs. The batch size was set to 20, which was the largest size that could safely fit in the \SI{11}{\giga\byte} VRAM of the GPU (Nvidia 2080 Ti). Each epoch takes just over one hour 50 minutes; total training time is about seven and a half hours. A validation accuracy of around \SI{83}{\%} is reached.
        
        The trained model is then applied to the testing dataset, which takes just over an hour. Computing the AMS for a cut of 0.5 gives public and private scores of 2.83 and 2.84, respectively, in disagreement with the authors' score of 3.98 \footnote{We indeed ran many experiments and were unable to recover the claimed score. These are all included in our GitHub repository.}. One possibility is that the authors do optimise a cut, but computing the maximum possible AMS on both public and private datasets, results in AMS scores of 3.20 and 3.25, respectively.
        
        During correspondence, the authors did not reply to questions about whether or not they split the testing dataset into the public and private samples; they are supplied together and must be manually separated by the user. This gives another possibility that the authors did not split the testing sample and instead computed the AMS on the entire sample. It should be noted from \autoref{eq:ams}, that the AMS contains no self-normalising components, unlike other ML metrics such as accuracy. All normalisation comes from the weights of the samples on which it is computed. Since the public and private samples have the same normalisation (integrated luminosity) computing the AMS on the entire testing set equates to running with double the integrated luminosity and can be expected to increase the AMS by a factor of approximately $\sqrt{2}$: $\sqrt{2}\times 2.84=4.02$. Indeed computing the AMS on the entirety of the testing dataset produces an AMS of 4.01. Allowing for slight changes in model training and other sources of stochasticity, this score is compatible with the claimed score of 3.98. We therefore suspect that this is how the authors arrived at their score.

    \subsection{Closing remarks}
        Although the SuperTML provides a solution to the problem of running pretrained models on tabular data with differing numbers of features, it doesn't solve the problem that the features can have very different meanings. Additionally the target task is extremely different to the pretraining task; whilst the models may learn to recognise numerals during pretraining, the task never explicitly encourages the model to learn to perform mathematical operations on the numerals. The task is also sensitive to the spatial locations of each feature, as warned against in Ref.~\cite{rethinking_imagenet}. It is therefore difficult to see how the pretraining really benefits the method, and indeed further testing showed that the same performance was achievable when starting from random initialisation. 

\begin{figure}[ht]
            \begin{center}
                \includegraphics[width=\sfMid\textwidth]{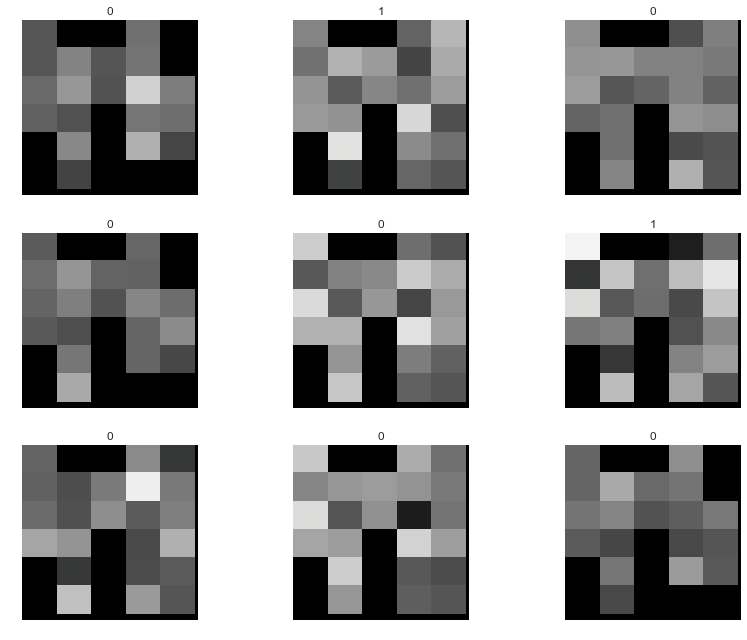}
                \caption{Examples of several datapoints encoded as greyscale blocks. The class of datapoint is indicated above each image: 0 - background, 1 - signal.}
                \label{fig:pixel_encoding}
            \end{center}
        \end{figure}

        We also found that slight improvements were possible by transforming the feature values to the range $[0,255]$ and encoding the data as solid, greyscale, blocks of pixels, as shown in \autoref{fig:pixel_encoding}; this mitigates the necessity of text recognition by the model by providing direct access to the numerical representation of the feature values. Additionally it allows smaller images to be used since font size is no longer a limitation. Smaller images means that the increase in data file size is reduced, and larger batch-sizes may be used, which reduced the training time.

        As it stands, though, the method is perhaps not best suited for data analysis in HEP, where we nowadays work with analysis-level data of at least several gigabytes. Even with the pixel-block encoding and 56x56 images, the increase in storage requirements is still a factor of 17. Additionally, the method is slow to train and to apply, and for final-state analysis the model will normally have to be retrained several times, occasionally at short notice. The final point is that not all analysers will have access to GPUs, and even fewer will have them on-demand. At least for a few years to come it will be necessary that analysis-level algorithms work reasonably on CPU, as well. Further discussion on the requirements and future goals for ML in HEP may be found in Ref.~\cite{hepml_cwp}.

        Certainly, though, the concept of pretraining domain-specific models in HEP is enticing; training data is usually computer simulated, and for analysis-level must be of very high quality, which requires the running of comparatively slow simulations, especially for detector simulations (e.g. \geant~\cite{geant0,geant1}). There are, however, software packages that still provide detector simulation but through the use of rough parameterisations of the detector responses (e.g. \delphes~\cite{delphes0,delphes1,delphes2}). It is possible then that powerful models could be pretrained using large samples of roughly simulated data, and then fine-tuned on smaller samples of high-quality simulated data.
	\newpage
	\FloatBarrier
	\section{Software details}
	The investigation performed in this \whatAmI made use of several packages, which are detailed in \autoref{tab:software}.
	The framework and notebook experiments are made available at Ref.~\cite{higgsml_lumin_git}.
	
\begin{table}[ht]
	\begin{center}
		\begin{tabular}{llll}
			\toprule
			Software & Version & References & Use/Notes\\
			\midrule
			\lumin & 0.0.0-0.3.1 & \cite{lumin} & Wrapping \pytorch to implement new \\
			& & & methods presented\\
			\pytorch & 1.0.0 & \cite{pytorch} & Implementing neural networks\\
			\textsc{Seaborn} & 0.8.1 & \cite{Seaborn} & Plot production\\
			\textsc{Matplotlib} & 3.0.2 & \cite{MatPlotLib} & Plot production\\
			\textsc{Pandas} & 0.23.4 & \cite{Pandas} & Data analysis and computation\\
			\textsc{NumPy} & 1.15.2 & \cite{Numpy} & Data analysis and computation\\
			\textsc{Scikit-Learn} & 0.20.0 & \cite{SKLearn} & Cross-validation and pre-processing\\
			\textsc{RFPImp} & 1.3.4 & \cite{rfpimp} & Random Forest Permutation Importance \\
			& & & and mutual dependency calculations\\
			\textsc{SciKit-Optimize} & 0.5.2 & \cite{skopt} & GP fitting and hyper-parameter\\
			& & &  partial-dependence calculations\\
			\textsc{FastAI} & 1.0.55 & \cite{fastai} & Implementing the SuperTML method\\
			\textsc{Pillow} & 6.0.0 & \cite{pillow} & Image preparation for SuperTML method\\
			\bottomrule
		\end{tabular}
	\end{center}
	\caption{Software used for the investigation}
	\label{tab:software}
\end{table}


\end{appendices}

\end{document}